\documentclass[twocolumn,10pt]{IEEEtran}

\usepackage{verbatim}
\usepackage{epsfig,bbm}
\usepackage[colorlinks,bookmarksopen,bookmarksnumbered,citecolor=red,urlcolor=red,]{hyperref}
\usepackage{CJK}
\usepackage{indentfirst}
\usepackage{multirow}
\usepackage{epstopdf}
\usepackage{graphicx}
\usepackage{footmisc}
\graphicspath{{./figures/}}
\usepackage{amsfonts}
\usepackage{mathrsfs}
\usepackage{setspace}
\usepackage{amsmath}
\usepackage{algorithm,algorithmic,amsbsy,amsmath,amssymb,epsfig,bbm,mathrsfs, bbm} 
\usepackage{amsthm}
\usepackage{verbatim}
\usepackage[noadjust]{cite}
\usepackage[subfigure]{graphfig}

\begin{document}

\title{Wireless Charging Technologies: Fundamentals, Standards, and Network Applications}
 \author{ Xiao Lu$^{\dagger}$, Ping Wang$^{\ddagger}$,  Dusit Niyato$^{\ddagger}$, Dong In Kim$^{\S}$, and Zhu Han$^{\wr}$\\
 ~$^{\dagger}$ Department of Electrical and Computer Engineering, University of Alberta, Canada \\
    ~$^{\ddagger}$ School of Computer Engineering, Nanyang Technological University, Singapore \\
    $^{\S}$  School of Information and Communication Engineering, Sungkyunkwan University (SKKU), Korea \\
    $^{\wr}$  Electrical and Computer Engineering, University of Houston, Texas, USA.
    \thanks{ {\bf Dong In Kim} is the corresponding author.}
   }
   
    \markboth{IEEE Communications Surveys and Tutorials, to appear}%
    {Shell \MakeLowercase{\textit{et al.}}: Bare Demo of IEEEtran.cls for Journals}

\maketitle
\begin{abstract}
 
Wireless charging is a technology of transmitting power through an air gap to electrical devices for the purpose of energy replenishment. The recent progress in wireless charging techniques and development of commercial products have provided a promising alternative way to address the energy bottleneck of conventionally portable battery-powered devices. However, the incorporation of wireless charging into the existing wireless communication systems also brings along a series of challenging issues with regard to implementation, scheduling, and power management. In this article, we present a comprehensive overview of wireless charging techniques, the developments in technical standards, and their recent advances in network applications. In particular, with regard to network applications, we review the static charger scheduling strategies, mobile charger dispatch strategies and wireless charger deployment strategies. Additionally, we discuss open issues and challenges in implementing wireless charging technologies. Finally, we envision some practical future network applications of wireless charging.
 
\end{abstract}

\emph{Index terms- Wireless Charging, wireless power transfer, inductive coupling, resonance coupling, RF/Microwave radiation, energy harvesting, Qi, PMA, A4WP, simultaneous wireless information and power transfer (SWIPT), energy beamforming, wireless powered communication network (WPCN), Magnetic MIMO, Witricity}.

\section{Introduction}

Wireless charging~\cite{A.2014Costanzo,J.2013Garnica}, also known as wireless power transfer, is the technology that enables a power source to transmit electromagnetic energy to an electrical load across an air gap, without interconnecting cords. This technology is attracting a wide range of applications, from low-power toothbrush to high-power electric vehicles because of its convenience and better user experience.  Nowadays, wireless charging is rapidly evolving from theories toward standard features on commercial products, especially mobile phones and portable smart devices. In 2014, many leading smartphone manufacturers, such as Samsung, Apple and Huawei, began to release new-generation devices featured with built-in wireless charging capability. IMS Research \cite{imsresearch} envisioned that wireless charging would be a $4.5$ billion market by 2016. Pike Research \cite{pikeresearch} estimated that wireless powered products will triple by 2020 to a $15$ billion market.

Compared to traditional charging with cord, wireless charging introduces many benefits as follows.
\begin{itemize}
\item Firstly, it improves user-friendliness as the hassle from connecting cables is removed. Different brands and different models of devices can also use the same charger.
 
\item  Secondly, it renders the design and fabrication of much smaller devices without the attachment of batteries. 
 
\item  Thirdly, it provides better product durability (e.g., waterproof and dustproof) for contact-free devices. 
  
\item  Fourthly, it enhances flexibility, especially for the devices for which replacing their batteries or connecting cables for charging is costly, hazardous, or infeasible (e.g., body-implanted sensors). 
  
\item Fifthly, wireless charging can provide power requested by charging devices in an on-demand fashion and thus are more flexible and energy-efficient. 

\end{itemize} 

Nevertheless, normally wireless charging incurs  higher implementation cost compared to wired charging. First, a wireless charger needs to be installed as a replacement of traditional charging cord. Second, a mobile device requires implantation of a wireless power receiver. Moreover, as wireless chargers often produce more heat than that of wired chargers, additional cost on crafting material may be incurred.

The development of wireless charging technologies is advancing toward two major directions, i.e., radiative wireless charging (or radio frequency (RF) based wireless charging) and non-radiative wireless charging (or coupling-based wireless charging). Radiative wireless charging adopts electromagnetic waves, typically RF waves or microwaves, as a medium to deliver energy in a form of radiation. The energy is transferred based on the electric field of an electromagnetic wave, which is radiative. Due to the safety issues raised by RF exposure~\cite{S.2009Branch}, radiative wireless charging usually operates in a low power region. For example, omni-directional RF radiation is only suitable for sensor node applications with up to 10mW power consumption \cite{L.2013Xie,A.2008Sample}. Alternatively, non-radiative wireless charging is based on the coupling of the magnetic-field between two coils within the distance of the coils' dimension for energy transmission. As the magnetic-field of an electromagnetic wave attenuates much faster than the electric field, the power transfer distance is largely limited. Due to safety implementation, non-radiative wireless charging has been widely used in our daily appliances (e.g., from toothbrush to electric vehicle charger \cite{A.2013Covic}) by far.

In this article, we aim to provide a comprehensive survey of the emerging wireless charging systems with regard to the fundamental technologies, international standards as well as applications in wireless communication networks. 
Our previous work in~\cite{X.LuSurvey} presented a review of research issues in RF-powered wireless networks with the focus on the receiver-side (i.e., energy harvester) designs. This survey differs from~\cite{X.LuSurvey} in the following aspects: this survey i) covers various major wireless charging techniques, namely, inductive coupling, magnetic resonance coupling and RF/microwave radiation, from fundamental principles to their applications, ii) reviews the existing international standards, commercialization and implementations, and iii) emphasizes on the transmitter-side (i.e., wireless charger) strategy designs for different types of network applications. Another recent survey in~\cite{S.2015Ulukus} provides an overview of self-sustaining wireless communications with different energy harvesting techniques, from the perspective of information theory, signal processing and wireless networking. Unlike~\cite{S.2015Ulukus}, this survey focuses on the wireless charging strategies in communication networks with wireless energy harvesting capability, also referred to as wireless powered communication networks (WPCNs)~\cite{S.1408.2335Bi}.

\begin{table}
\footnotesize
\centering
\caption{\footnotesize  Summary of Existing Survey in Related Area.} \label{Survey_comparison}
\begin{tabular}{|p{0.7cm}|p{1.5cm}|p{5.5cm}|} 
\hline
\footnotesize  Survey & Scope & Main Contribution \\
\hline
\cite{X.LuSurvey} &  Wireless network with RF energy harvesting & Review of i) fundamentals and circuit design for RF energy harvesting, ii) resource allocation schemes and communication protocols for various types of RF-powered wireless network, and iii) practical challenges and future directions. \\
\hline
\cite{S.2015Ulukus} & Wireless network with energy harvesting  & Review of i) information-theoretic physical layer performance limits to transmission scheduling policies and medium access control protocols, ii) emerging paradigm of energy transfer and cooperation that occur separately or jointly with information transfer, and iii) energy consumption models for energy harvesting communication systems. \\
\hline
\cite{S.2011Sudevalayam} & Sensor nodes with energy harvesting & Review of architectures, energy sources and storage technologies, as well as applications of sensor nodes with energy harvesting. \\
\hline 
 \cite{V.2014Prasad} & Devices with ambient energy harvesting   & Review of i) various types
 of energy harvesting techniques, ii) different energy harvesting models, and ) power management and networking
 aspects of the energy harvesting devices. \\
\hline
 \cite{R.2014Valenta} & RF/microwave energy harvesting circuit &  Review of i) basics and designs of the RF energy harvesting circuit, and ii) energy conversion efficiency of existing implementations of RF energy harvesting circuits.  \\
 \hline
\end{tabular}
\end{table}

Existing literatures~\cite{S.2011Sudevalayam,R.2014Valenta,V.2014Prasad} also presented relevant reviews in energy harvesting research, mainly from the perspective of device-level techniques and hardware implementations. In~\cite{S.2011Sudevalayam}, the authors gave an overview of the sensor nodes powered by different energy harvesting techniques. Reference~\cite{V.2014Prasad} focused on the techniques of harvesting energy from ambiance. The authors in~\cite{R.2014Valenta} investigated RF/microwave energy harvesting circuit design and surveyed the energy efficiency of the state-of-the-art implementations.   
In Table \ref{Survey_comparison}, we summarize the scope and main contributions of the existing survey papers relevant to the topic of ours.

\begin{figure} 
\centering
\includegraphics[width=0.35\textwidth]{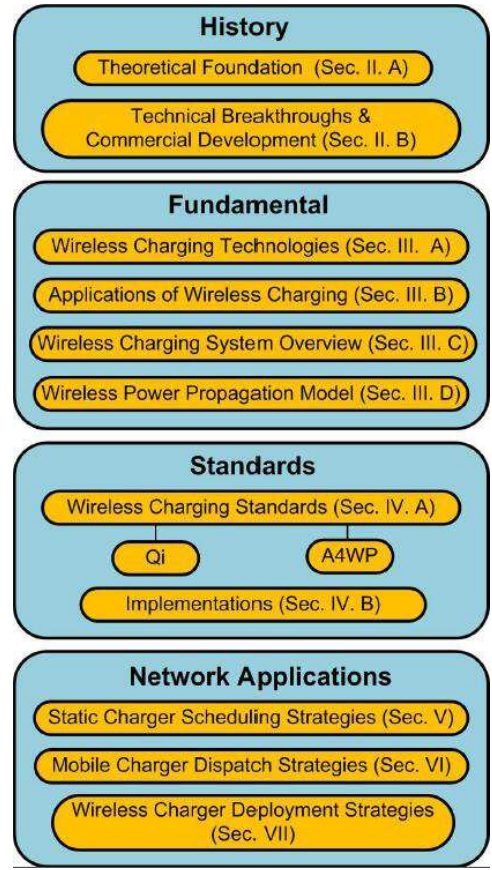}
\caption{An outline of the scope of this survey.} \label{outline}
\end{figure}

\begin{table}
\footnotesize
\centering
\caption{\footnotesize  List of Abbreviations.} \label{Abbreviations}
\begin{tabular}{|l|l|} 
\hline
\footnotesize  {\bf Abbreviation} & {\bf Description} \\
\hline
RF & Radio Frequency \\
\hline
WPCN & Wireless powered communication networks \\
\hline
SPS & Solar Power Satellite\\
\hline
WPC & Wireless Power Consortium \\
\hline
PMA & Power Matters Alliance \\
\hline
A4WP & Alliance for Wireless Power \\
\hline
IPT & Inductive power transfer \\
\hline
RFID & Radio Frequency Identification \\
\hline
SWIPT & Simultaneous wireless information and power transfer  \\
\hline
FCC & Federal Communications Commission \\
\hline
EV & Electric vehicle \\
\hline
PHEV & Plug-in hybrid electric vehicle \\
\hline
LED & Light-emitting diode  \\
\hline
WRSN & Wireless renewable sensor network\\
\hline
AM & Amplitude modulated \\
\hline
GSM & Global System for Mobile Communications \\
\hline
WBAN & Wireless body area network \\
\hline
AC & Alternating current \\
\hline
DC & Direct current \\
\hline
SISO & Single-input single-output \\
\hline
MISO & Multi-input single-output \\
\hline
SIMO & Single-input multi-output \\
\hline
MIMO &  Multi-input multi-output  \\
\hline
PTU & Power transmitter unit \\
\hline
PRU & Power receiver unit \\
\hline
ISM & Industrial Scientific Medical \\
\hline
BLE & Bluetooth low energy \\
\hline
E-AP & Energy access point\\
\hline
D-AP & Data access point \\
\hline
H-AP & Hybrid access point \\
\hline
AWGN & Additive white Gaussian noise \\
\hline
TDMA & Time division multiple access   \\
\hline
CSI   & Channel state information \\
\hline
SNR & Signal to noise ratio \\
\hline
SHC & Shortest Hamiltonian cycle \\
\hline 
TSP & Traveling Salesman's Problem\\
\hline
NP & Non-deterministic polynomial-time \\
\hline
LP  & Linear programming  \\
\hline
NLP & non-linear programming \\
\hline
MINLP & Mixed-integer nonlinear programming \\
\hline
MILP & Mixed integer linear programming  \\
\hline
QoM & Quality of Monitoring \\
\hline
PSO & Particle swarm optimization \\ 
\hline
WISP & Wireless identification and sensing platform \\
\hline
ILP & Integer linear programming  \\
\hline
CMOS & Complementary metal–oxide–semiconductor \\
\hline

\end{tabular}
\end{table}

Figure~\ref{outline} outlines the main design issues for wireless charging systems. We first describe a brief history of wireless power transfer covering the progress in theoretical foundation, technical breakthroughs as well as recent commercialization development in Section~II. Then, in Section~III, we present an overview of existing wireless charging techniques and their applications, followed by the introduction of magnetic-field propagation models. We also review the hardware implementation of these wireless charging technologies. Subsequently, in Section~IV, the specifications of the leading international wireless charging standards are described in details. The existing implementations of those standards are also outlined. We then survey the network applications including static charger scheduling strategies, mobile wireless charger dispatch strategies, and wireless charger deployment strategies in Sections~V, VI and VII, respectively. Furthermore, in Section~VIII, we shed light on some open research directions in implementing wireless charging technologies. Additionally, we envision some future network applications. Finally, Section~IX concludes the survey.  The abbreviations used in this article are summarized in Table~\ref{Abbreviations}.

\section{History and Commercialization}

\begin{figure*} 
\centering
\includegraphics[width=1\textwidth]{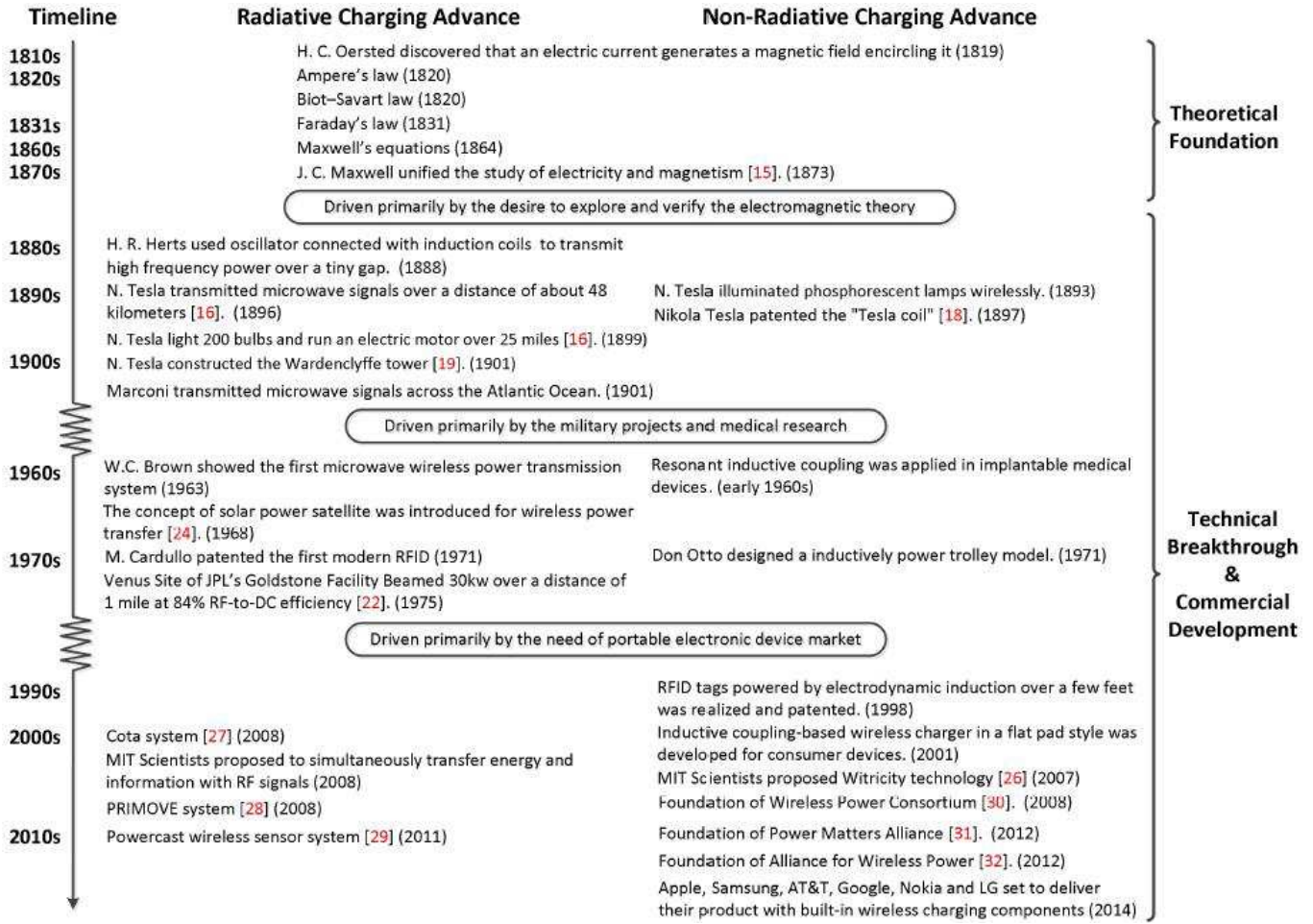}
\caption{A brief development history of wireless power transmission.} \label{timeline}
\end{figure*}

This section provides an overview of the development history of wireless charging research as well as some recent commercializations.  
Figure~\ref{timeline} shows a brief history and major milestones of wireless charging technology.

\subsection{Theoretic Foundation}
The study of electromagnetism originates from 1819 when H. C. Oersted discovered that electric current generates a magnetic field around it. Then, Ampere's Law, Biot-Savart's Law and Faraday's Law were derived to model some basic property of magnetic field. They are followed by the Maxwell's equations introduced in 1864 to characterize how electric and magnetic fields are generated and altered by each other.
Later, in 1873, the publication of J. C. Maxwell's book \emph{A Treatise on Electricity and Magnetism} \cite{C.1873Maxwell} unified the study of electricity and magnetism. Since then, electricity and magnetism are known to be regulated by the same force. These historic progress established the modern theoretic foundation of electromagnetism.

\subsection{Technical breakthroughs and Research Projects}
The history has witnessed a series of important technical breakthroughs, going along with two major research lines on electric field and magnetic field. In 1888, H. R. Herts used oscillator connected with induction coils to transmit electricity over a tiny gap. This first confirmed the existence of electromagnetic radiation experimentally. Nikola Tesla, the founder of alternating current electricity, was the first to conduct experiments of wireless power transfer based on microwave technology. He focused on long-distance wireless power transfer~\cite{N.1914Tesla} and realized the transfer of microwave signals over a distance about 48 kilometers in 1896. Another major breakthrough was achieved in 1899 to transmit $10^8$ volts of high-frequency electric power over a distance of 25 miles to light 200 bulbs and run an electric motor~\cite{N.1914Tesla}. However, the technology that Tesla applied had to be shelved because emitting such high voltages in electric arcs would cause disastrous effect to humans and electrical equipment in the vicinity~\cite{R.2009Bhutkar}. 

\begin{figure*}
\centering
\includegraphics[width=1\textwidth]{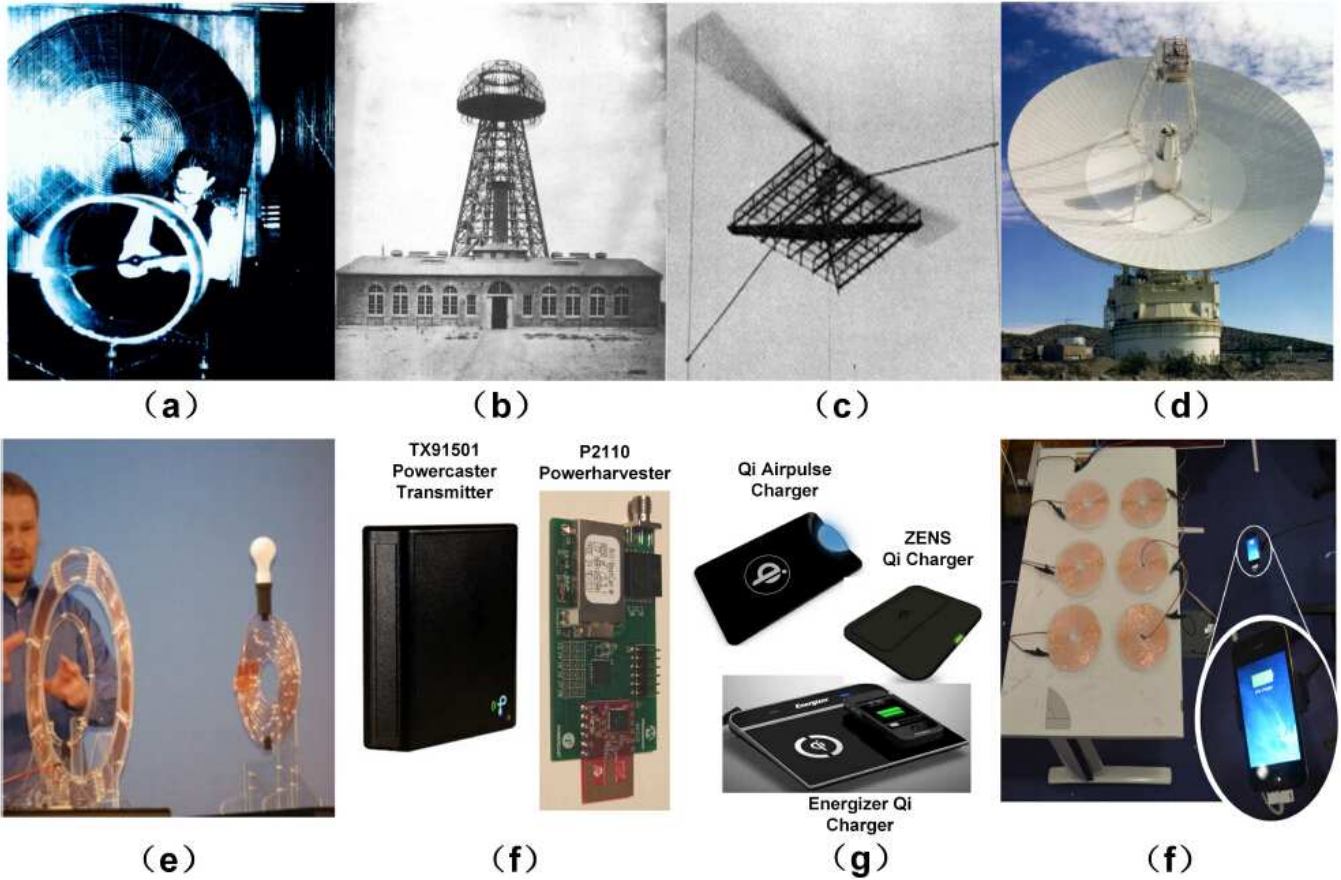}
\caption{Illustrations of wireless power transmission systems. a) Tesla coil~\cite{teslauniverse},  b) Wardenclyffe Tower~\cite{M.1981Cheney}, c) Microwave-powered airplane~\cite{airplanesandrockets}, d) JPL’s Goldstone Facility~\cite{jpl}, e) Witricity system~\cite{topnews}, f) Powercaster transmitter and harvester~\cite{Powercast}, g) Qi charging pads~\cite{Energizer,zens,bitmore}, h) Magnetic MIMO system~\cite{J.2014Jadidian}. (IEEE Copyright)} \label{illustration}
\end{figure*}

Around the same period, Tesla also made a great contribution to promote the magnetic-field advance by introducing the famous ``Tesla coil", illustrated in Figure~\ref{illustration}a. In 1901, Tesla constructed the Wardenclyffe Tower, shown in Figure~\ref{illustration}b to transfer electrical energy without cords through the Ionosphere. However, due to technology limitation (e.g., low system efficiency due to large-scale electric field), the idea has not been widely further developed and commercialized. Later, during 1920s and 1930s, magnetrons were invented to convert electricity into microwaves, which enable wireless power transfer over long distance. However, there was no method to convert microwaves back to electricity. Therefore, the development of wireless charging was abandoned.

\begin{figure*}
\centering
\includegraphics[width=0.9\textwidth]{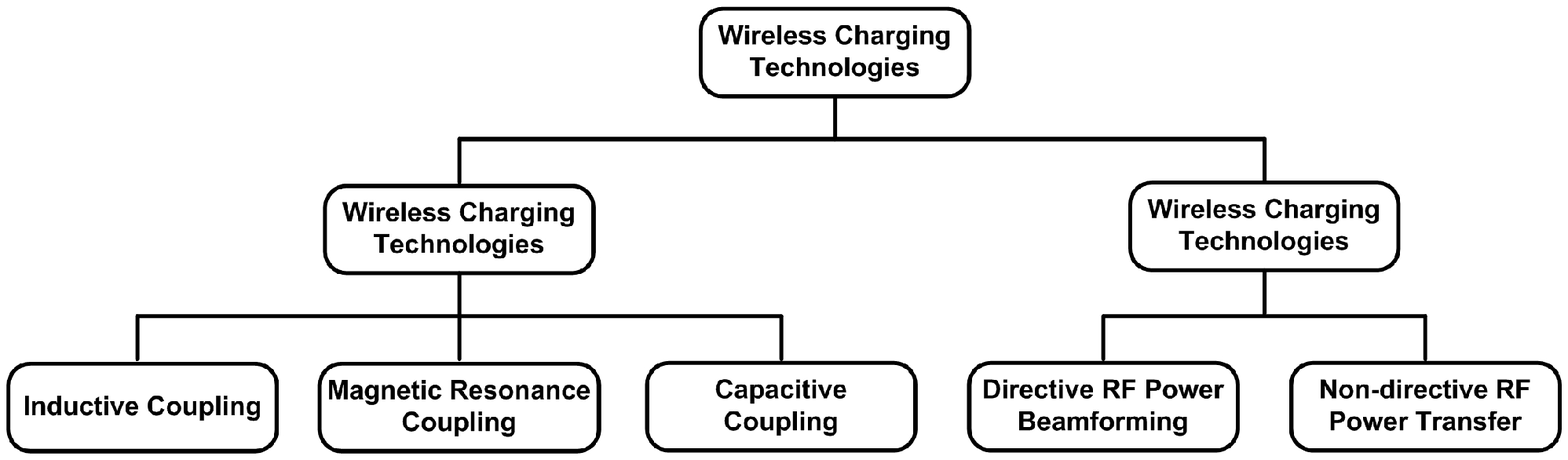}
\caption{Classification of wireless charging technologies.} \label{WPT_Classification}
\end{figure*}

It was until 1964, when W. C. Brown, who is regarded as the principal engineer of practical wireless charging, realized the conversion of microwaves to electricity through a rectenna.
Brown demonstrated the practicality of microwave power transfer by powering a model helicopter, demonstrated in Figure~\ref{illustration}c, which inspired the following research in microwave-powered airplanes during 1980s and 1990s in Japan and Canada~\cite{J.1988Schlesak}. In 1975, Brown beamed 30kW over a distance of 1 mile at $84\%$ with Venus Site of JPL’s Goldstone Facility \cite{B.2013Strassner}, shown in Figure~\ref{illustration}d.
Solar power satellite (SPS), introduced in 1968, is another driving force for long-distance microwave power transfer \cite{O.2002McSpadden}. The concept is to place a large SPS in geostationary Earth orbit to collect sunlight energy, and transmit the energy back to the Earth through electromagnetic beam. NASA's project on SPS Reference System prompted abundant technology developments in large-scale microwave transfer during 1970s and 1980s. 
During the same period, coupling-based technology was developed under slow progress. Though inductive coupling for low-power medical applications was successful and widely applied in 1960s, there were not many technical boosts.     

\subsection{Commercialization}

The recent upsurge of interests in wireless charging research was primarily forced by the need of portable electronic device market. In 1990s, commercialized wireless charging product began to emerge because of the explosive wide spread of portable electronic devices~\cite{A.2006Vanderelli}. Both far-field and near-field based wireless charging approaches are undergoing progress. In 2007, Kurs \emph{et al} proposed Witricity technology, shown in Figure~\ref{illustration}e, which was demonstrated through experiments that mid-range non-radiative wireless charging is not only practical but also efficient. Moreover, radiative wireless charging systems like Cota system~\cite{cota_system}, PRIMOVE~\cite{primove}, and Powercast wireless rechargeable sensor system \cite{Powercast} (illustrated in Figure~\ref{illustration}f) have been commercialized.  

More recently, different consortiums, e.g., Wireless Power Consortium (WPC)~\cite{WPC}, Power Matters Alliance (PMA)~\cite{PMA}, and Alliance for Wireless Power (A4WP)~\cite{WiPower}, have been established to develop international standards for wireless charging. Nowadays, these standards have been adopted in many electronic products available in the market, such as smart phones and wireless chargers demonstrated in Figure~\ref{illustration}g. At the end of 2014, a breakthrough technology, named magnetic MIMO (MagMIMO), illustrated in Figure~\ref{illustration}h, has been designed to perform multi-antenna beamforming based on magnetic waves. This technology has opened an area for the magnetic-field beamforming research. The history of wireless charging then continues. The reader can refer to \cite{C.1984Brown,B.2013Strassner} for a more detailed history of progress.

\section{Fundamentals of Wireless Charging}

In this section, we provide some basic knowledge of wireless charging which covers the principles of charging techniques, existing applications of wireless charging as well as charging system designs in terms of architectures, hardware designs and implementations. In addition, we introduce the wireless power propagation models for non-radiative charging systems.


\subsection{Wireless Charging Technologies}

As illustrated in Figure~\ref{WPT_Classification}, wireless charging technologies can be broadly classified into non-radiative coupling-based charging and radiative RF-based charging. The former consists of three techniques: inductive coupling~\cite{L.2011Ho}, magnetic resonance coupling~\cite{Kurs2007A} and capacitive coupling~\cite{M.2011Kline}, while the latter can be further sorted into directive RF power beamforming and non-directive RF power transfer~\cite{Z2013Popovic}. 
In capacitive coupling, the achievable amount of coupling capacitance is dependent on the available area of the device~\cite{Y2013Hui}. However, for a typical-size portable electronic device, it is hard to generate sufficient power density for charging, which imposes a challenging design limitation. As for directive RF power beamforming, the limitation lies in that the charger needs to know an exact location of the energy receiver. 
Due to the obvious limitation of above two techniques, wireless charging is usually realized through other three techniques, i.e., magnetic inductive coupling, magnetic resonance coupling, and non-directive RF radiation.

The magnetic inductive and magnetic resonance coupling work on near field, where the generated electromagnetic field dominates the region close to the transmitter or scattering object. The near-field power is attenuated according to the cube of the reciprocal of the charging distance~\cite{X.LuSurvey}. 
Alternatively, the microwave radiation works on far field at a greater distance. The far-field power decreases according to the square of the reciprocal of the charging distance~\cite{X.LuSurvey}. Moreover, for the far-field technique, the absorption of radiation does not affect the transmitter. By contrast, for the near-field techniques, the absorption of radiation influences the load on the transmitter~\cite{N.2012Shinohara}. 
This is because, a transmitting antenna and a
receiving antenna are not coupled for the far-field technique. While a transmitting coil and a
receiving coil are coupled for the near-field techniques~\cite{2011NShinohara}.

\begin{figure*} 
 \centering
 \subfigure [Inductive Coupling] {
  \label{IC}
  \centering
  \includegraphics[width=0.36 \textwidth]{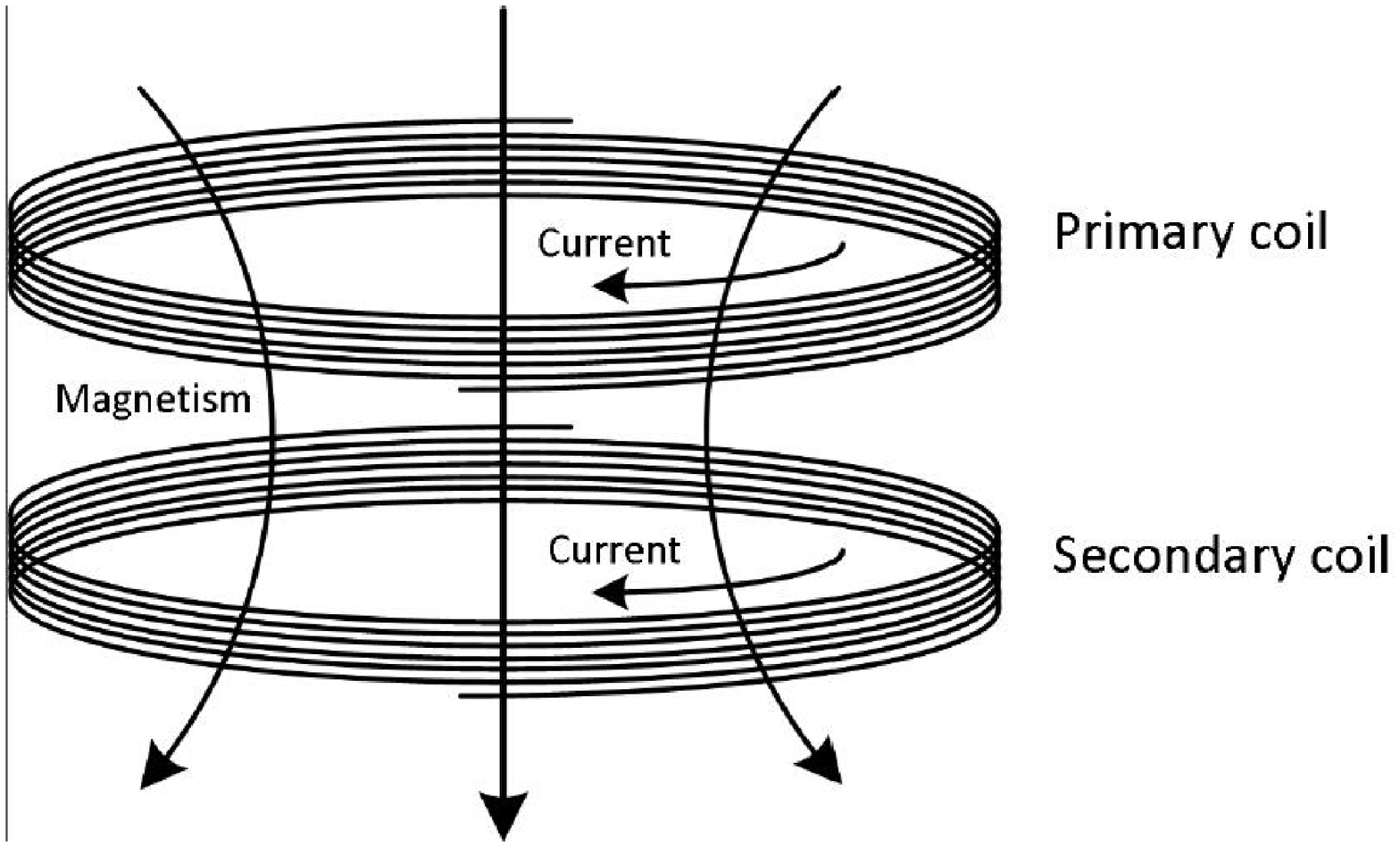}}  
  \centering
  \subfigure [Magnetic Resonance Coupling] {
   \label{MRC}
   \centering
   \includegraphics[width=0.5  \textwidth]{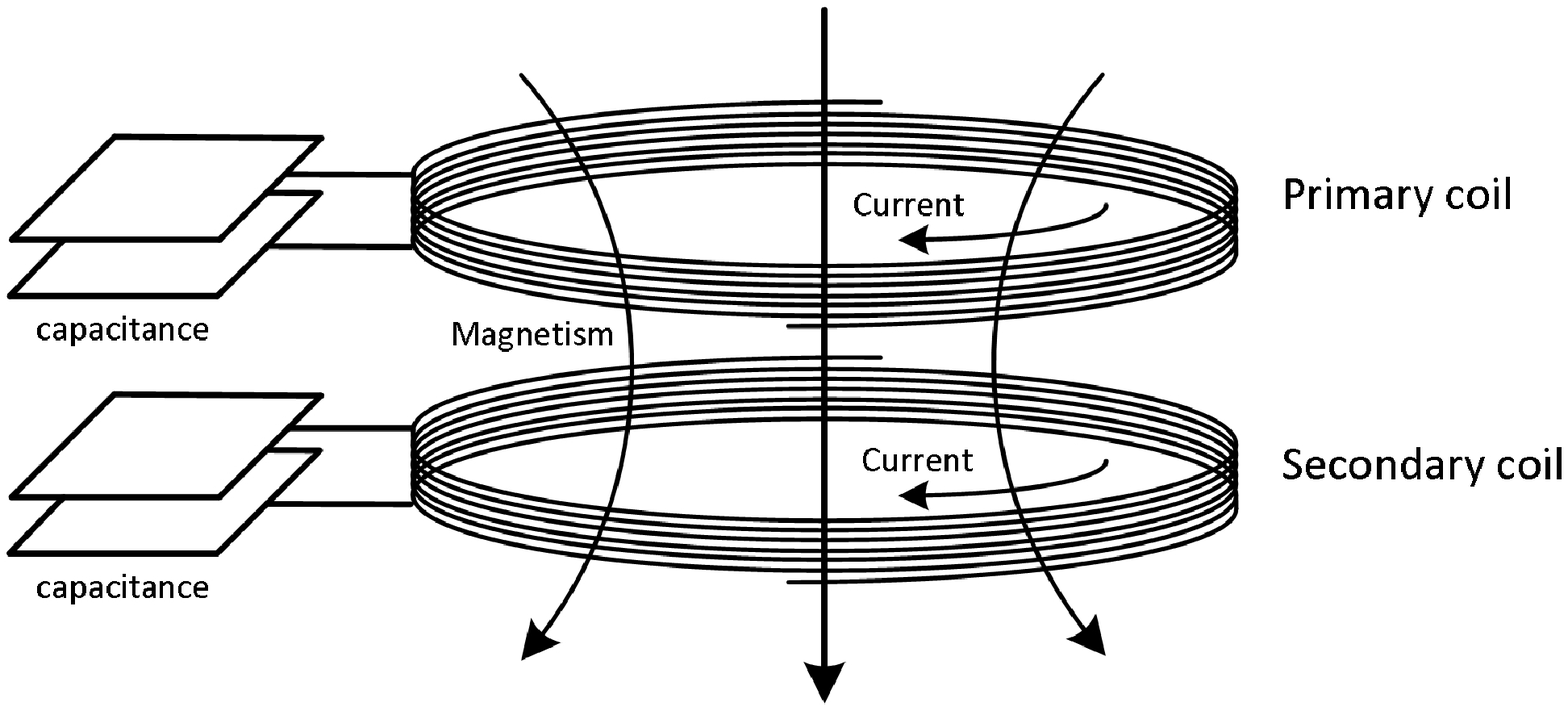}}   
   \centering
 \caption{Models of wireless charging systems for inductive coupling and magnetic resonance coupling.} 
 \label{receiver_designs}
 \end{figure*}

\begin{figure*}
\centering
\includegraphics[width=0.85\textwidth]{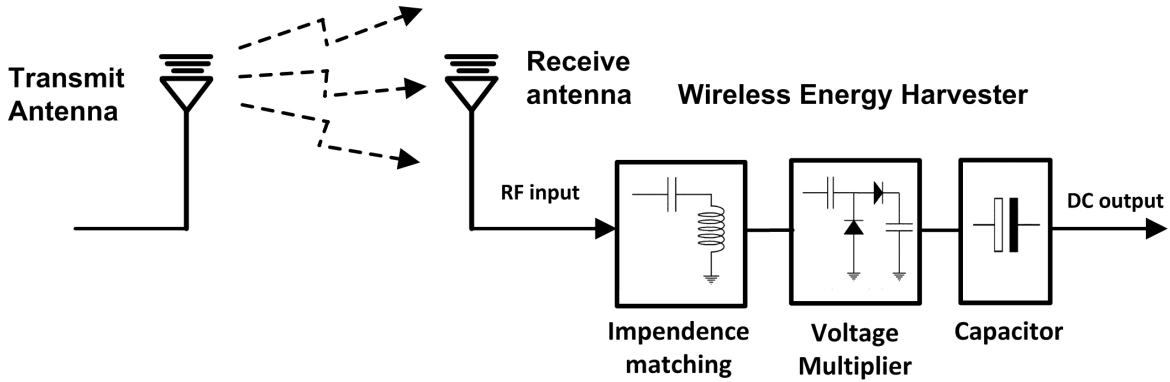}
\caption{Far-field wireless charging.} \label{microwave}
\end{figure*}
 
\subsubsection{Inductive Coupling}

Inductive coupling~\cite{WPC} is based on magnetic field induction that delivers electrical energy between two coils. Figure~\ref{IC} shows the reference model. 
Inductive power transfer (IPT) happens when a primary coil of an energy transmitter generates predominantly varying magnetic field across the secondary coil of the energy receiver within the field, generally less than a wavelength. The near-field magnetic power then induces voltage/current across the secondary coil of the energy receiver within the field. This voltage can be used for charging a wireless device or storage system.  
The operating frequency of inductive coupling is typically in the kilo Hertz range. The secondary coil should be tuned at the operating frequency to enhance charging efficiency~\cite{X.2014Wei}. The quality factor is usually designed in small values (e.g., below 10 \cite{S.2004Wang}), because the transferred power attenuates quickly for larger quality values~\cite{Z.2012Pantic}. Due to lack of the compensation of high quality factors, the effective charging distance is generally within 20cm~\cite{X.2014Wei}. Inductively coupled radio frequency identification (RFID) \cite{S.2008Ahson,L.2013Roselli} is an example that pushes the limit to extend the charging distance to tens of centimeters, at the cost of diminished efficiency (e.g., 1-2$\%$~\cite{A.2008Sample}) with received power in micro watt range. Despite the limited transmission range, the effective charging power can be very high (e.g., kilowatt level \cite{H2012Wu} for electric vehicle re-charging).

The advantages of magnetic inductive coupling include ease of implementation, convenient operation, high efficiency in close distance (typically less than a coil diameter) and ensured safety. Therefore, it is applicable and popular for mobile devices.
Very recently, MIT scientists have announced the invention of a novel wireless charging technology, called MagMIMO \cite{J.2014Jadidian}, which can charge a wireless device from up to 30cm away. It is claimed that MagMIMO can detect and cast a cone of energy toward a phone, even when the phone is put inside the pocket.   

\begin{table*} \small 
\centering
\caption{\footnotesize Comparison of different wireless charging techniques.} \label{WET}
\begin{tabular}{|p{2.5cm}|p{5cm}|p{5cm}|p{3.5cm}|} 
\hline
\footnotesize {\bf Wireless charging technique} & {\bf Advantage} & {\bf Disadvantage} & {\bf Effective charging distance}    \\
\hline
Inductive coupling & Safe for human, simple implementation  & Short charging distance, heating effect, not suitable for mobile applications, needs tight alignment between chargers and charging devices & From a few millimeters to a few centimeters \\ 
\hline
Magnetic resonance coupling & Loose alignment between chargers and charging devices, charging multiple devices simultaneously on different power, high charging efficiency, non-line-of-sight charging  & Not suitable for mobile applications, limited charging distance, complex implementation &  From a few centimeters to a few meters \\ 
\hline
RF radiation & Long effective charging distance, suitable for mobile applications & Not safe when the RF density exposure is high, low charging efficiency, line-of-sight charging  & Typically within several tens of meters, up to several kilometers \\ 
\hline
\end{tabular}
\end{table*}  
 
\subsubsection{Magnetic Resonance Coupling}

Magnetic resonance coupling~\cite{A.2008Karalis}, as shown in Figure~\ref{MRC}, is based on evanescent-wave coupling which generates and transfers electrical energy between two resonant coils through varying or oscillating magnetic fields. As two resonant coils, operating at the same resonant frequency, are strongly coupled, high energy transfer efficiency can be achieved with small leakage to non-resonant externalities.
For instance, an up-to-date prototype~\cite{X.Li2015} was demonstrated to achieve the maximum power transfer efficiency of $92.6\%$ over the distance of 0.3cm. 
Due to the property of resonance, magnetic resonance coupling also has the advantage of immunity to neighboring environment and line-of-sight transfer requirement.  
Previous demonstrations \cite{Kurs2007A,B.2009Cannon,C.2008Zhu,N.2009Low} of magnetically coupled resonators have shown the capability to transfer power over longer distance than that of inductive coupling, with higher efficiency than that of RF radiation approach. Additionally, magnetic resonance coupling can be applied between one transmitting resonator and many receiving resonators. Therefore, it enables concurrent charging of multiple devices~\cite{A.2010Kurs,S.2011Rajagopal,J.2012Choi,Z.2012Kim,B.2009Cannon}.

As magnetic resonance coupling typically operates in the megahertz frequency range, the quality factors are normally high. With the increase of charging distance, the high quality factor helps to mitigate the sharp decrease in coupling co-efficiency, and thus charging efficiency. Consequently, extending the effective power transfer distance to meter range is possible. In 2007, MIT scientists proposed a high-efficient mid-range wireless power transfer technology, i.e., Witricity, based on strongly coupled magnetic resonance~\cite{Kurs2007A,witricity}. It was reported that wireless power transmission can light a 60W bulb in more than two meters with the transmission efficiency around $40\%$. The efficiency increased up to $90\%$ when the transmission distance is one meter. However, it is difficult to reduce the size of a Witricity receiver because it requires a distributed capacitive of coil to operate. This poses an important challenge in implementing Witricity technology in portable devices. Magnetic resonance coupling can charge multiple devices concurrently, by tuning coupled resonators of multiple receiving coils~\cite{A.2010Kurs}. This has been shown to achieve an improved overall efficiency. However, mutual coupling~\cite{G.2014Kim} of receiving coils can result in interference, and thus proper tuning is required.

\subsubsection{RF Radiation}

RF radiation utilizes diffused RF/microwave as a medium to carry radiant energy~\cite{2014X.Lu}. RF/microwave propagates over space at the speed of light, normally in line-of-sight. The typical frequency of RF/microwave ranges from 300MHz to 300GHz~\cite{X.LuIEEENetwork}. The energy transfer can use other electromagnetic waves such as infrared and X-rays. However, due to the safety issue, they are not widely used. Figure~\ref{microwave} shows the architecture of a microwave power transmission system. The power transmission starts with the AC-to-DC conversion, followed by a DC-to-RF conversion through magnetron at the transmitter side. After propagating through the air, the RF/microwave captured by the receiver rectenna are rectified into electricity again, through an RF-to-DC conversion.

The RF-to-DC conversion efficiency is highly dependent on the captured power density at receive antenna, the accuracy of the impedance matching between the antenna and the voltage multiplier, and the power efficiency of the voltage multiplier that converts the received RF signals to DC voltage~\cite{S.2013Ladan}.
An example of the state-of-the-art implementation in~\cite{V.2015Kuhn} demonstrated that the RF-to-DC conversion efficiency was achieved at 62$\%$ and 84$\%$ for a cumulative -10dBm and 5.8dBm input power, respectively. A more detailed review on the conversion efficiency of RF energy harvester implementations can be found in~\cite{X.LuSurvey,R.2014Valenta}.
Furthermore, from the theoretical analysis perspective, the closed-form mathematical characterization of energy-conversion efficiency and maximum output power for an energy-harvesting circuit has been provided in~\cite{R.2015Valenta}.

The RF/microwave energy can be radiated isotropically or toward some direction through beamforming. The former is more suitable for broadcast applications. For point-to-point transmission, beamforming transmits electromagnetic waves, referred to as energy beamforming~\cite{ZhangRuiMIMO}, can improve the power transmission efficiency. A beam can be generated through an antenna array (or aperture antenna). The sharpness of energy beamforming improves with the number of transmit antennas. The use of massive antenna arrays can increase the sharpness. The recent development has also brought commercial products into the market. For example, the Powercaster transmitter and Powerharvester receiver~\cite{Powercast} allow 1W or 3W isotropic wireless power transfer.

Besides longer transmission distance, microwave radiation offers the advantage of compatibility with existing communications system. Microwaves have been advocated to deliver energy and transfer information at the same time~\cite{Varshney2008}. The amplitude and phase of microwave are used to modulate information, while its radiation and vibration are used to carry energy. This concept is referred to as simultaneous wireless information and power transfer (SWIPT)~\cite{ZhangRuiMIMO}. To accommodate SWIPT, advanced smart antenna technologies~\cite{Z.Ding2015} employed at the receiver side have been developed to achieve a favorable trade-off between system performance and complexity.  By contrast, the deployment of dedicated power beacons overlaid with existing communication system has also been proposed as an alternative because of its cost-effectiveness and applicability~\cite{K1207.5640Huang}. However, because of health concern of RF radiations, the power beacons should be constrained following RF exposure regulations, such as the Federal Communications Commission (FCC) regulation \cite{FCC}, and the maximum permissible exposure levels specified in IEEE C95.1-2005 standard~\cite{MPE}. Therefore, dense deployment of power beacons is required to power hand-held cellular mobiles with lower power and shorter distance.

Table~\ref{WET} shows a summary of the wireless charging techniques. The advantage, disadvantage, effective charging distance and applications are highlighted.

\subsection{Applications of Wireless Charging}
In this subsection, to provide a better illustration of the diverse and promising use of wireless charging, we introduce the existing applications of wireless charging with regard to near-field and far-field practices.   

\subsubsection{Near-field Charging}

Near-field charging applications can be realized based on inductive coupling and magnetic resonance coupling. Because of the ease and low-cost of implementation, most of the existing applications have primarily adopted inductive coupling. As aforementioned, IPT is capable of supporting high power transfer above kilowatt level, so it is widely used by industrial automation. The major applications include robot manipulation~\cite{J.2007Gao,A.1996Kawamura}, automated underwater vehicles~\cite{T.2007McGinnis,Z.2015Cheng,F.2012Tang}, induction generators~\cite{H.2014Gorginpour}, and induction motors~\cite{H.2014Guzman,Y.2014Liang,A.2014Mahmoudi}.
High-power IPT has also been adopted to provide real-time power for public transportation~\cite{A2013Covic} such as monorail systems~\cite{S.2008Raabe,J.2006Elliott,A.2010Keeling}, people-mover systems~\cite{M.2008Jufer}, railway-powered electric vehicles~\cite{J.2014Shin,A.2007Covic,Y.2015Choi,A.2013Russer} and high speed trains~\cite{J.2015Kim,S.2012Lee}. The transferred power level ranges from kilowatt to hundreds of kilowatt. 
For example, the online electric vehicle system~\cite{J.2014Shin,Y.2012Chun} realizes 100kW output power with $80\%$ power efficiency over a 26cm air gap.

Another widely adopted high-power charging application is to energize the battery of electric vehicles (EVs) including plug-in hybrid electric vehicles (PHEVs). Inductive coupling has been introduced for EV charging since 1990s~\cite{G.1999Hayes,R.1996Severns}. Inductive chargers for both unidirectional charging~\cite{H.2012Wu,G.2007Egan,N.2015Liu} and bi-directional charging \cite{K.2011Madawala,C.2015Kisacikoglu} that enable vehicle-to-grid power~\cite{W.2013Zhou} have been developed. Typically, the charging efficiency is above 90$\%$ with 1-10kW power across a 4-10mm gap~\cite{A.2013Covic}.
A review on the recent progress of inductive charging for EVs can be found in \cite{H.2011Wureview}. More recently, magnetic resonance coupling-based charging systems for EVs have also been demonstrated and evaluated~\cite{H.Kim2014,W.2014Khan-ngern,Di2012Tommaso,X.2013Wang,S.2012Krishnan}. Compared to inductive chargers, magnetic resonance coupling-based EV charging allows larger charging distance as well as efficiency. For instance, the experiments in \cite{H.Kim2014} achieves over 95$\%$ efficiency over a 22.5cm air gap. 

The medium-power near-field charging (ranging from several watts to tens of watts operating power) has primarily been applied to medical apparatuses and our daily appliances. 
Various biomedical implant designs based on inductive coupling have been shown in~\cite{H.2010Jiang,H.2013Jiang,A.2012Arshad,K.2013RamRakhyani,K.2012RamRakhyani}. The up-to-date implementation can achieve above 50$\%$ overall charging efficiency over 10mm air gap~\cite{H.2013Jiang}. Magnetic resonance coupling based charging for biomedical implants \cite{K.2011RamRakhyani,F.2013Xue,Q.2013Xu,A.2014Qusba,G.2012Yilmaz,D.2014Ahn} exhibits more powerful penetration ability. As the charging distance is much larger than the coil dimension, magnetic resonance coupling enables smaller implanted device size with a normal charging range. As demonstrated in~\cite{D.2014Ahn}, with a 3cm transmit coil and 2cm receive coil, above 60$\%$ charging efficiency can be realized over 20cm distance. The state-of-the-art implementation can result in above 70$\%$ charging efficiency in bio-tissue environments~\cite{G.2012Yilmaz}. 

As for daily appliance powering, the mainstream of the applications is for household devices and portable devices. Inductive toothbrush~\cite{M.2004Stratmann}, TV~\cite{J.2012Kim}, lighting~\cite{S.2011Rajagopal,W.2004Baarman}, wall switch~\cite{A.1993Johnson}, heating system~\cite{Y.2014Xu} are examples for household devices. With regard to portable devices, variant standard compliant wireless chargers, such as  RAVPower's Qi charger~\cite{RAVPower}, Verizon Qi charging pad~\cite{verizonwireless}, Duracell Powermat~\cite{Duracell2011}, Energizer Qi charger~\cite{Energizer}, ZENS Qi charging pad~\cite{zens}, Airpulse charging pad~\cite{bitmore}, have been developed and commercialized for supplying energy to laptops, tablets, and cellphones.

\begin{figure*} 
\centering
\includegraphics[width=0.8\textwidth]{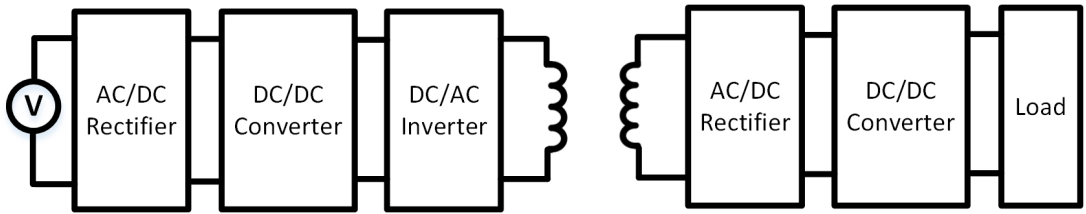}
\caption{A block diagram of non-radiative wireless charging system.} \label{Block_diagram}
\end{figure*}

Furthermore, near-field charging applications have recently expanded to oil well~\cite{X.2013Xin}, off-shore energy harvesting~\cite{Van2009Neste}, coal mine~\cite{Y.2011Zhou}, electric bike~\cite{F.2013Pellitteri}, sensors~\cite{K.2006Fotopoulou}, wearable devices~\cite{O.2013Jonah,N.2014Dobrostomat}, implantable systems~\cite{S.2013Ho,F.2009Zhang}, RFID~\cite{O.2014Mourad}, light-emitting diode (LED) display~\cite{I.2011Cho}, power line communication~\cite{S.2014Barmada} and smart grid~\cite{A.2014Kukde}.

\subsubsection{Far-field Charging}

Far-field charging systems can be realized through either non-directive RF radiation or directive RF beamforming~\cite{J.2013Visser}. 
Non-directive RF radiation can be conducted without line of sight, and is less sensitive to the orientation and position relative to the transmit antenna~\cite{E.2011Falkenstein}. However, the resulted charging efficiency is relatively low. 
Low-power wireless systems, such as wireless renewable sensor networks (WRSNs)~\cite{E.2011Falkenstein}  and RFID systems~\cite{A.2008Sample} have become the most widely adopted applications for non-directive charging. WRSNs with low duty cycles can maintain a perpetual operation with typically RF power densities in the 20-200$\mu$W/cm$^{2}$ range~\cite{Z.2013Popovic}. For example, in \cite{J.2012Hong}, the authors devised an ultra-low power sensor platform with far-field charging. The implemented sensor transmitter and receiver consume the power of 1.79mW and 0.683mW, respectively, to achieve the data rate of 500kbps. Similar system designs with dedicated wireless charger have been reported in \cite{D.2009Mascarenas,S.2014Percy,C.2014Cato} for sensors with batteries and \cite{N.Shinohara2014,L.2014Xia} for battery-less sensors. Instead of relying on dedicated wireless charger, wireless charging systems based on ambient energy harvesting have also been developed. Literature has demonstrated the development of self-recharging sensors platform harvesting environmental RF signals from TV broadcast~\cite{H.2010Nishimoto,P.2013Nintanavongsa,S.2014Kim,R.2013Vyas}, amplitude modulated (AM) radio broadcast~\cite{X2013Wang}, Global System for Mobile Communications (GSM) bands (900/1800)~\cite{B.2013Lim,M.2014Borges}, WiFi routers \cite{Abd2014Kadir,F.2014Alneyadi}, cellular base stations \cite{N.2013Parks} and satellite~\cite{2013A.Takacs,A.2013Takacs,A.Takacs2013}.

RF-powered sensors also appear in other environments, such as wireless body area networks (WBANs)~\cite{H.2013Jiang,S.2015Rao} (e.g., for health-care monitoring). WBANs can be mainly classified into wearable and implanted devices~\cite{F.2010Zhang}, which are put on or inside the human body. Battery-less wearable WBAN designs and implementations have been reported in~\cite{Y.2014Toh,N.2014Desai, J.2012Olivo}.
Typically, the power consumption of the body sensors is tens of milliwatt, and the charging efficiency is around several percent (e.g, 1.2$\%$ in~\cite{N.2014Desai}). 
By contrast, powering implanted sensors deeply inside bodily organs achieves much lower charging efficiency, typically smaller than 0.1$\%$~\cite{S.2014Majerus}.   
Demonstrated in \cite{S.2014Majerus,Y.2011Chow,M.2013Arsalan}, with a micro-watt level RF power source, typical implanted sensors can be powered from tens of centimeters away (e.g., 30cm in~\cite{Y.2011Chow}). Besides, the safety issues regarding RF powering to implantable devices have been investigated in \cite{A.2013Bercich}. RF-powered sensors have also been induced to Internet of Things (IoT)~\cite{K.2012Chen,S.2014Gollakota}, machine-to-machine (M2M) communication systems~\cite{K.2012Ng}, and smart grid systems \cite{Erol-Kantarci2012Suresense,Erol-Kantarci2012DRIFT,U.2014Baroudi}.

Directive RF beamforming can be utilized to support electronic devices with larger power consumption. Ultra-high power transfer systems transmitting on hundreds of kilo-watts have been developed since 1960s with the advance of microwave technology~\cite{K.2015Huang}. Through microwave beamforming techniques, delivering high power across long distance is not rare. For instance, in 1975, the Goldstone microwave power transfer experiment conducted at 2.388GHz managed to deliver 30kW with a 450kW beam power over 1.54 kilometer distance. Far-field microwave beamforming has also propelled the development of a strand of massive wireless charging systems, such as SPS~\cite{H.1996Nansen,E.1977Glaser,C.1992Brown}, unmanned aerial vehicles \cite{A.1996Foote,J.2006Kim}, microwave-driven unmanned vehicles~\cite{J.2013Miyasaka,J.2014Miyasaka,A.2007Oida}, high altitude 
electric motor powered platforms (HAPP)~\cite{K.1982Stefan,B.1982Ernald}, Raytheon Airborne Microwave Platform (RAMP)~\cite{B.2013Strassner},  and stationary high altitude relay program (SHARP)~\cite{J.1985Schlesak, R.1992East}. 

More recently, with the increasing market penetration of EV/PHEV, microwave beamforming has been adopted as a mean to remotely power EVs~\cite{N.2011Shinohara,N2011Shinohara,N.2013Shinohara,Y.2012Kubo,N2013Shinohara}.
A prototype that utilizes roadside transmitter to energize an EV has been implemented and investigated in~\cite{N.2013Shinohara}. The rectenna developed is shown to rectify 10kW power with over 80$\%$ RF-DC conversion efficiency.
During the past decade, directive RF beamforming has found its medium-power applications for recharging portable electronic devices. The commercialized Cota system \cite{cota,cota_system} that can deliver power beam up to 30 feet without any line-of-sight transmission link is an example. Moreover, RF power beacon~\cite{K1207.5640Huang,Erol-Kantarci2014} has been advocated to power mobile devices through high frequency microwave (e.g., 60GHz~\cite{K.2015Huang}) in cellular networks. However, the practicability requires further experimental evaluation.

In the above two subsections, we have provided an overview of major wireless charging techniques and their applications. In the next two subsections, attention will be paid to the non-radiative propagation model and hardware design for non-radiative charging systems. The reader can refer to~\cite{X.LuSurvey} for more detailed information about recent advances in radiative charging system.

\subsection{Wireless Charging System Overview}

In this subsection, we present an overview of wireless charging system in the aspects of architectures, hardware designs and implementations. 

\subsubsection{Architecture}

\begin{table*} \small
\centering
\caption{\footnotesize Comparison of Hardware Implementations of Inductive Coupling.} \label{Hardware_comparison} 
\begin{tabular}{|l|c|p{2.6cm}|p{2.9cm}|p{2.9cm}|c| } 
\hline
\footnotesize Implementation &  Technique & Output &  Maximum Charging Efficiency  &  Maximum Charging Distance & Frequency \\   
\hline
Yoo {\em et al}~\cite{J.2010Yoo} (2010) &   0.18$\mu$m CMOS & 1.8V & 54.9$\%$ & 10mm & 13.56MHz  \\
\hline
Lee {\em et al}~\cite{M.2012Lee} (2012) &   0.5$\mu$m CMOS &  3.1V & 77$\%$ & 80mm & 13.56MHz    \\
\hline
Lee {\em et al}~\cite{Y.2013Lee} (2013) &   0.18$\mu$m CMOS & 3V & 87$\%$ & 20mm & 13.56MHz    \\
\hline
Lazaro{\em et al}~\cite{O.2013Lazaro} (2013) &  0.18$\mu$m CMOS & 1.5V & 82$\%$ & 11.35mm & 100-150kHz   \\
\hline 
Li {\em et al}~\cite{X.2015Li} (2014) &   0.13$\mu$m CMOS & 3.6V & 65$\%$ &    20mm & 40.68MHz  \\
\hline
\end{tabular}
\end{table*} 

\begin{table*} \small
\centering
\caption{\footnotesize Comparison of Hardware Implementations of Magnetic Resonance Coupling.} \label{Hardware_RC}
\begin{tabular}{|l|p{2.1cm}|p{2.1cm}|l|l|l| } 
\hline
\footnotesize Implementation &  Transmit Coil Diameter & Receive Coil Diameter & Charging Distance  &  Charging Efficiency & Frequency \\
\hline
Kurs {et al}~\cite{Kurs2007A} (2007) & 60$\times$60cm  & 30$\times$30cm & 75cm & 93$\%$ & 9.9MHz\\
\hline
Low {et al}~\cite{N.2009Low} (2009) & 21$\times$21cm  & 13$\times$13cm & 1cm & 75.7$\%$ & 134kHz\\
\hline
Wang {et al}~\cite{D.2012Wang} (2012) &  30$\times$30cm  & 30$\times$30cm &  5mm & 74.08$\%$  & 15.1MHz \\
\hline
Ahn {et al}~\cite{D.2013Ahn} (2013) & 35$\times$30cm & 31.5$\times$22.5cm & 20-31cm & 45-57$\%$  & 144kHz\\
\hline
Ali {et al}~\cite{T.2014Ali} (2014) & 13.6$\times$13.6 cm & 5$\times$5cm & 3mm & 88.11$\%$ & 22.2-22.4MHz \\
\hline
\end{tabular}
\end{table*} 

Figure~\ref{Block_diagram} shows a block diagram of a general non-radiative wireless charging system. The transmitter side consists of i) an AC/DC rectifier, which converts alternating current (AC) to direct current (DC); ii) a DC/DC converter, which alters the voltage of a source of DC from one level to another; and iii) a DC/AC inverter, which changes DC to AC. The receiver side is composed of i) an AC/DC rectifier, which converts high-frequency AC into DC, ii) a DC/DC converter, which tunes the voltage of the DC, and iii) a load for charging applications.

The wireless charging process works as follows. First, a power source is required to actuate the AC/DC rectifier. As the commercial AC worldwide operates either in 50Hz or 60Hz frequency~\cite{R.2014Kurtus}, which is too low to drive wireless charging, the charger increases the AC frequency by converting the AC to DC first, and then raising the voltage of DC and changing the DC back to high-frequency AC power. As the high-frequency AC that runs through the transmit loop coil creates a magnetic field around it, AC is induced at the receive loop coil separated away from the transmit coil by an air gap. The energy receiver then converts the induced AC to DC, and reshapes to the voltage required by the load. The battery of an electronic device can then be replenished at the load.      

Inductive coupling systems are generally based on four basic topologies, namely, series-series, series-parallel, parallel-series, and parallel-parallel~\cite{J.2010Moradewicz}. These topologies differ in the way of utilizing compensation capacitance in the circuit. Parallel-series and parallel-parallel regulate the inverter current flowing into the parallel resonant circuit based on an additional series inductor, which results in larger converter size and cost. Furthermore, these two topologies have varying resonant capacitance values depending on the coupling and quality factors~\cite{J.2010Moradewicz}. Therefore, series-series and series-parallel structures are more generally adopted. Performance comparison among these four compensation topologies can be found in~\cite{N.2013Jamal}. By contrast, the main types of the input port of magnetic resonance coupling system are series pattern and the parallel pattern circuits~\cite{M.2013Yang}. The series pattern and parallel pattern circuits should be adopted when the system operating efficiency is high and low, respectively, to achieve relatively higher value of induction coil.

\begin{figure}
\centering
\includegraphics[width=0.49\textwidth]{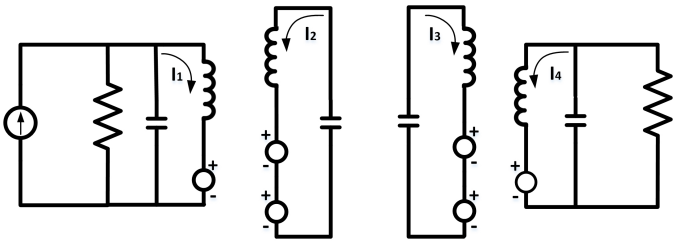}
\caption{The architecture of four-coil magnetic resonance coupling based wireless charging system.} \label{four-coil}
\end{figure}

The inductive coupling systems generally adopt the two-coil system architecture, as shown in Figure~\ref{Block_diagram}. By contrast, the system architectures being utilized by magnetic resonance coupling are more diverse. The recent progress has extended the magnetic resonance coupling application to a four-coil system with impedance matching~\cite{J.2010Chen,P.2011Duong}, relay resonator system~\cite{X.2012Zhang,F.2011Zhang} and domino-resonator systems~\cite{X.2012Zhong,K.2012Lee,W.2013Zhong}.   

The idea of four-coil system was first proposed in \cite{T.1998Boys} in 1998. The structure contains an excitation coil and a transmit resonator on the transmitter side, a receive resonator and a load coil on the receiver side, as shown in Figure~\ref{four-coil}. The utilization of the excitation coil and load coil involves two extra mutual coupling coefficients, i.e., the coefficient between the excitation coil and transmit resonator, and that between the receiving resonator and load coil. Compared with the two-coil system, the two extra coefficients introduce extra freedom in spreading the transfer distance. However, the overall transfer efficiency will not exceed 50$\%$ because of the independence matching requirement~\cite{S.2011Cheon}. The detailed circuit analysis of the four-coil system and optimization of independence matching to maximize charging power can be found in~\cite{S.2011Cheon}.


The relay resonator system is formed by adding an extra relay resonator between the transmit coil and receive coil. Optimization and experimental evaluations of such a system have been conducted with 115.6kHz~\cite{X.2015Zhong}, 1.25MHz~\cite{W.2011Kim}, 6.78MHz~\cite{Y.2012Kim}, 7MHz~\cite{F.2011Zhang}, and 13.56MHz~\cite{M.2011Kiani} operating frequencies.  
To further extend the transmission range of relay resonator system, domino-resonator systems can be formed by placing multiple adjacent resonator relays between the transmit coil and receive coil. The placement of the resonator relays is very flexible and can be made in various domino settings, such as straight line, circuit, curved and Y-shaped  patterns~\cite{X.2012Zhong,K.2012Lee,W.2013Zhong}.
The power paths can be split or combined, which allows a very malleable control of power transfer.

\subsubsection{Hardware Design and Implementation}
The intensity of the magnetic field can be characterized as a function of distance $d$ from the source as follows~\cite{I.2013Mayordomo}:
\begin{eqnarray}\label{density}
	H(d)=\frac{INr^{2}}{2\sqrt{(r^{2}+d^{2})^{3}}}	,
\end{eqnarray}
where $I$, $N$ and $r$ denote the current, the number of turns and the radius of the transmit coil, respectively.

From (\ref{density}), it is straightforward that increasing the number of turns and the radius of the transmit coil can amplify the intensity. However, the number of turns and the coil size cannot be enlarged without limit, because they need to be optimized by taking into account the transmission frequency and resistances~\cite{I.2013Mayordomo}. To capture the transferred energy from the transmit coil optimally, the receive coil should be designed with low impedance~\cite{W.2012Kim}.  

The power transfer efficiency of a non-radiative charging system is highly dependent on the mutual inductance between two coils, the quality factor Q, and load matching factor. Mutual inductance of a coil pair indicates how a variation in one coil influences the induced current in the other coil. The mutual inductance between a coil pair is proportional to the geometric
mean of the self-inductance of the two coils 
through a coupling coefficient~\cite{W.2005Nilsson}. The coupling co-efficiency that reflects the tightness of coupling is determined by the alignment and distance, the ratio of diameters, and the shape of two coils. 

The quality factor Q is defined as the ratio of the energy stored in the resonator over the energy provided by a generator~\cite{T.2009Imura}. Higher Q indicates a smaller rate of system energy loss during power transmission. Therefore, in a high Q power system, the oscillation/resonance decline slowly. The quality factor is affected by the self-inductance, resistance and intrinsic frequency, which mainly depend on the fabricated materials. The load matching factor mainly hinges on the distance. Since the resonance frequencies of a coil pair change as the gap varies~\cite{T.2011Imura}, load matching factor measures how tight the resonance frequencies are matched. To tune the load matching factor for maintaining  resonance frequency matching at varying distance  existing, literature has proposed various solutions such as coupling manipulation~\cite{P.2011Duong}, frequency matching~\cite{P.2011Sample}, impedance matching~\cite{C.2010Beh}, and resonator parameter tunning~\cite{2010I.Awai}.

In Table~\ref{Hardware_comparison} and Table~\ref{Hardware_RC}, we show some of the recently developed hardware implementation of IPT systems and magnetic resonance coupling systems, respectively. It is shown that 50$\%$-80$\%$ charging efficiency can be achieved within several centimeters charging distance for IPT systems. For magnetic resonance coupling systems, the charging distance extends to several decimeters with efficiency ranging from 50$\%$-90$\%$.

\subsection{Wireless Power Propagation Models}
The far-field RF propagation models are well known in literatures \cite{P.2006Barsocchi,P.2007Almers}. This subsection focuses on introducing the characterization of near-field magnetic wave propagation. We start with the fundamental of magnetic induction model in the basic single-input-single-output (SISO) setting. Then, the model is extended to multiple-input-single-output (MISO), single-input-multiple-output (SIMO) and multi-input multi-output (MIMO) configurations.
 
 \begin{figure} 
 \centering
 \subfigure [SISO Model] {
  \label{SISO}
  \centering
  \includegraphics[width=0.4 \textwidth]{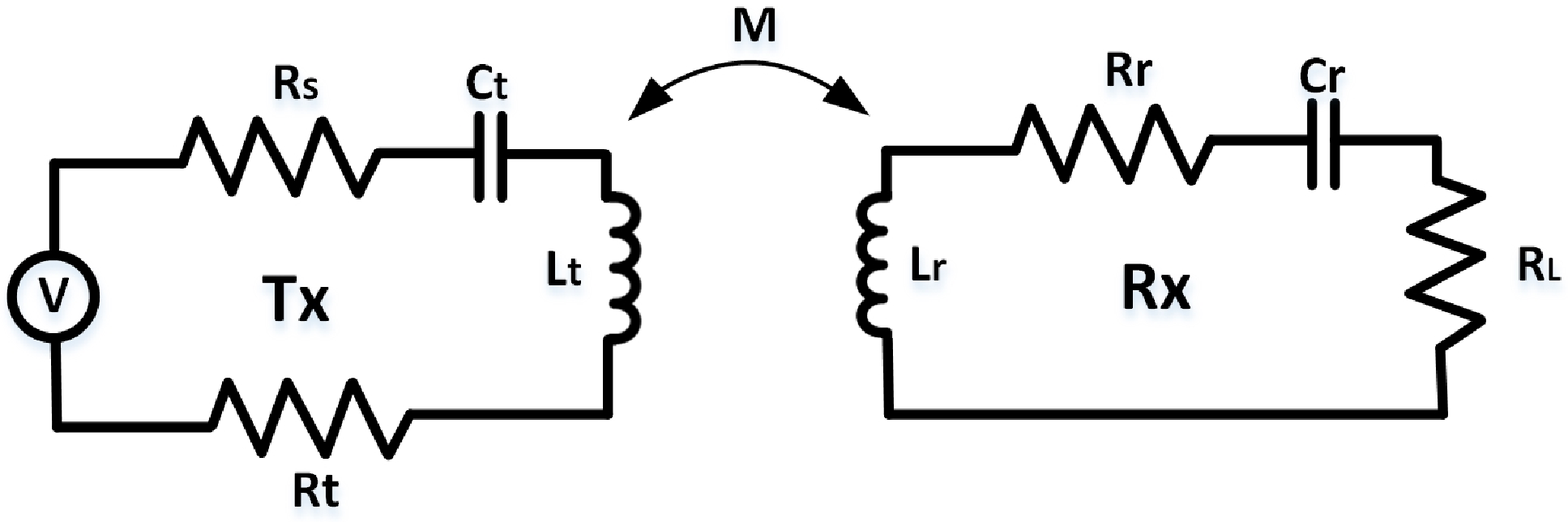}}  
  \centering
  \subfigure [MISO Model] {
   \label{MISO}
   \centering
   \includegraphics[width=0.39  \textwidth]{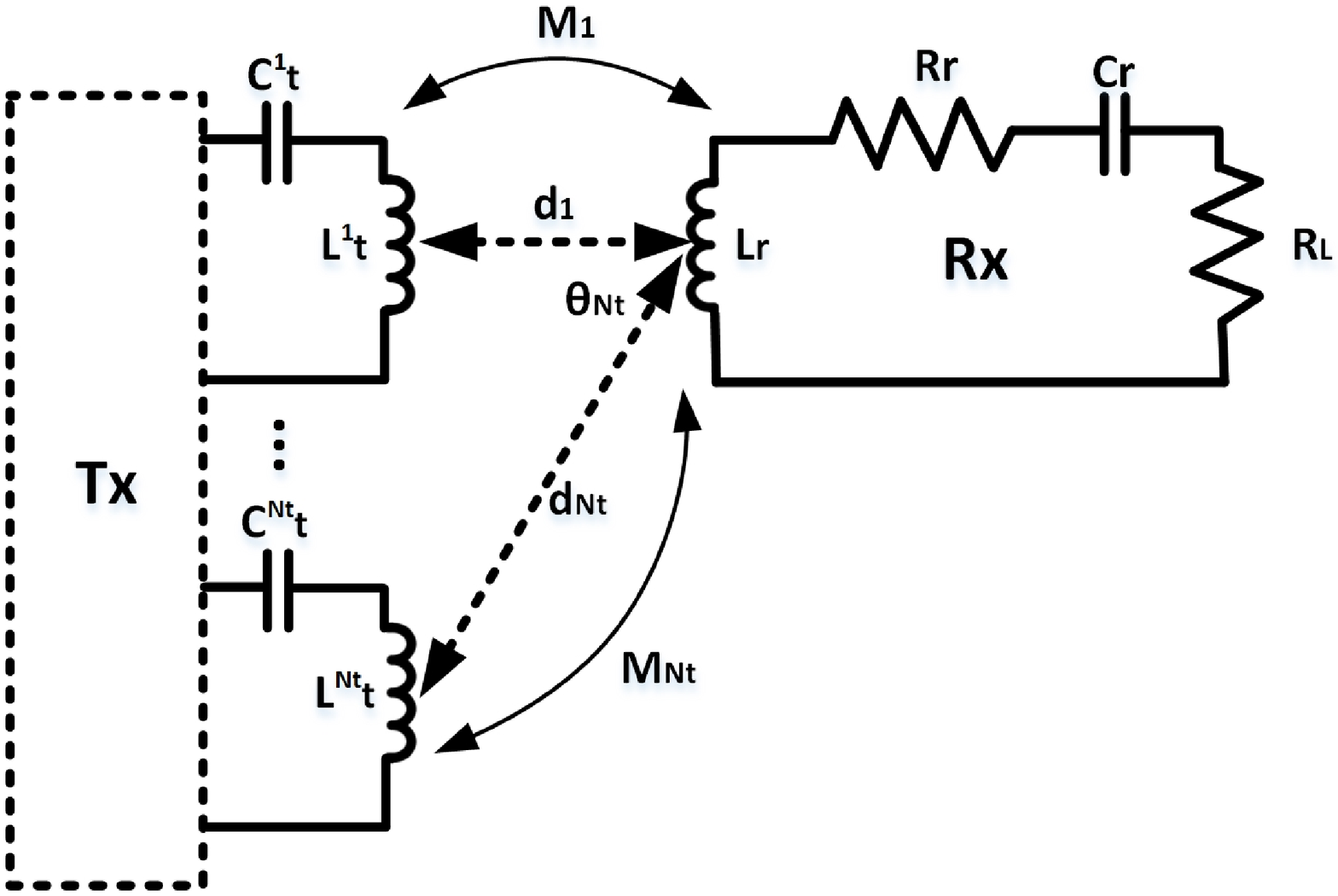}}   \\
   \centering
  \subfigure [SIMO Model] {
  \label{SIMO}
   \centering
 \includegraphics[width=0.39  \textwidth]{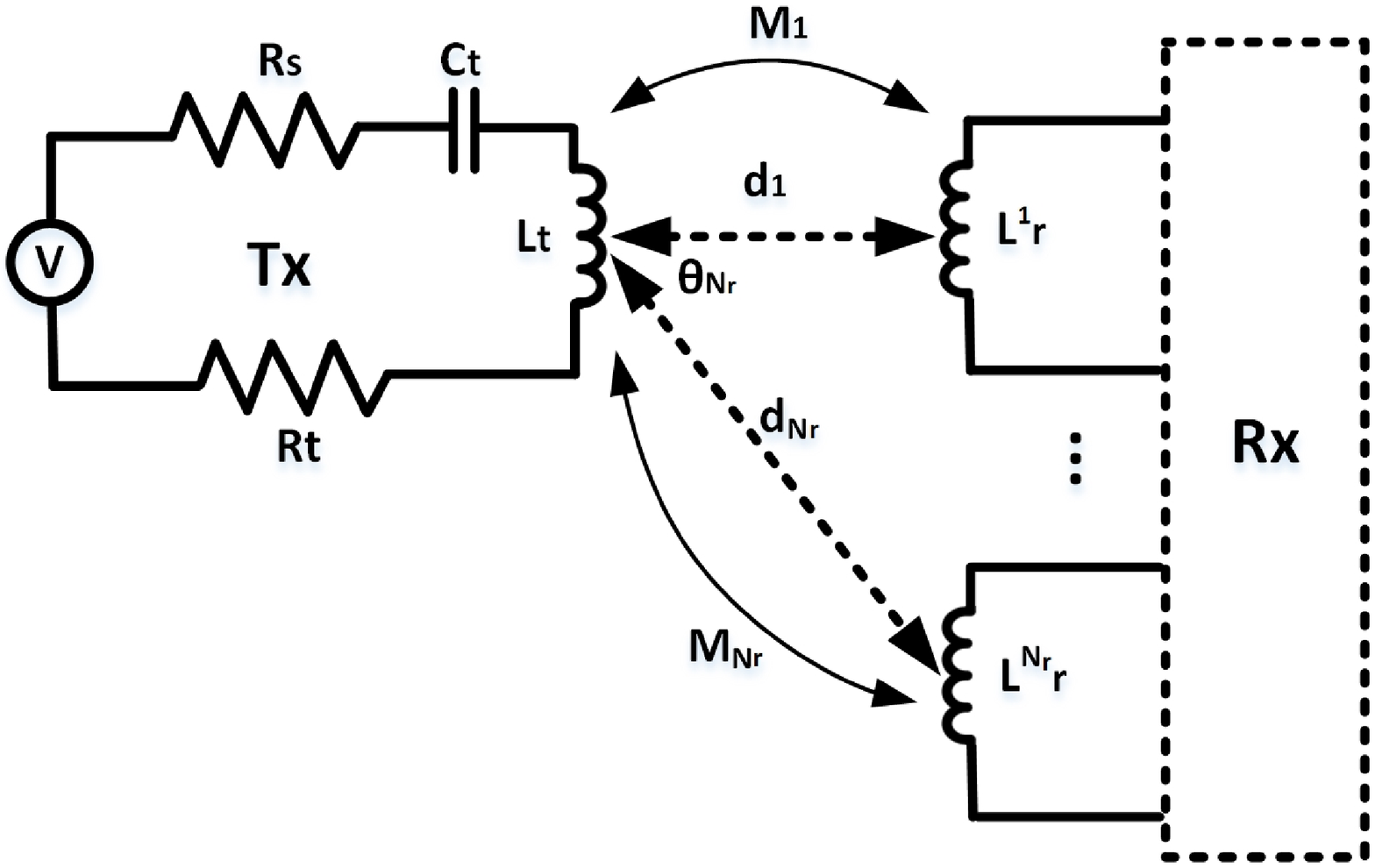}}
 \centering
  \subfigure [MIMO Model] {
  \label{MIMO}
   \centering
 \includegraphics[width=0.33  \textwidth]{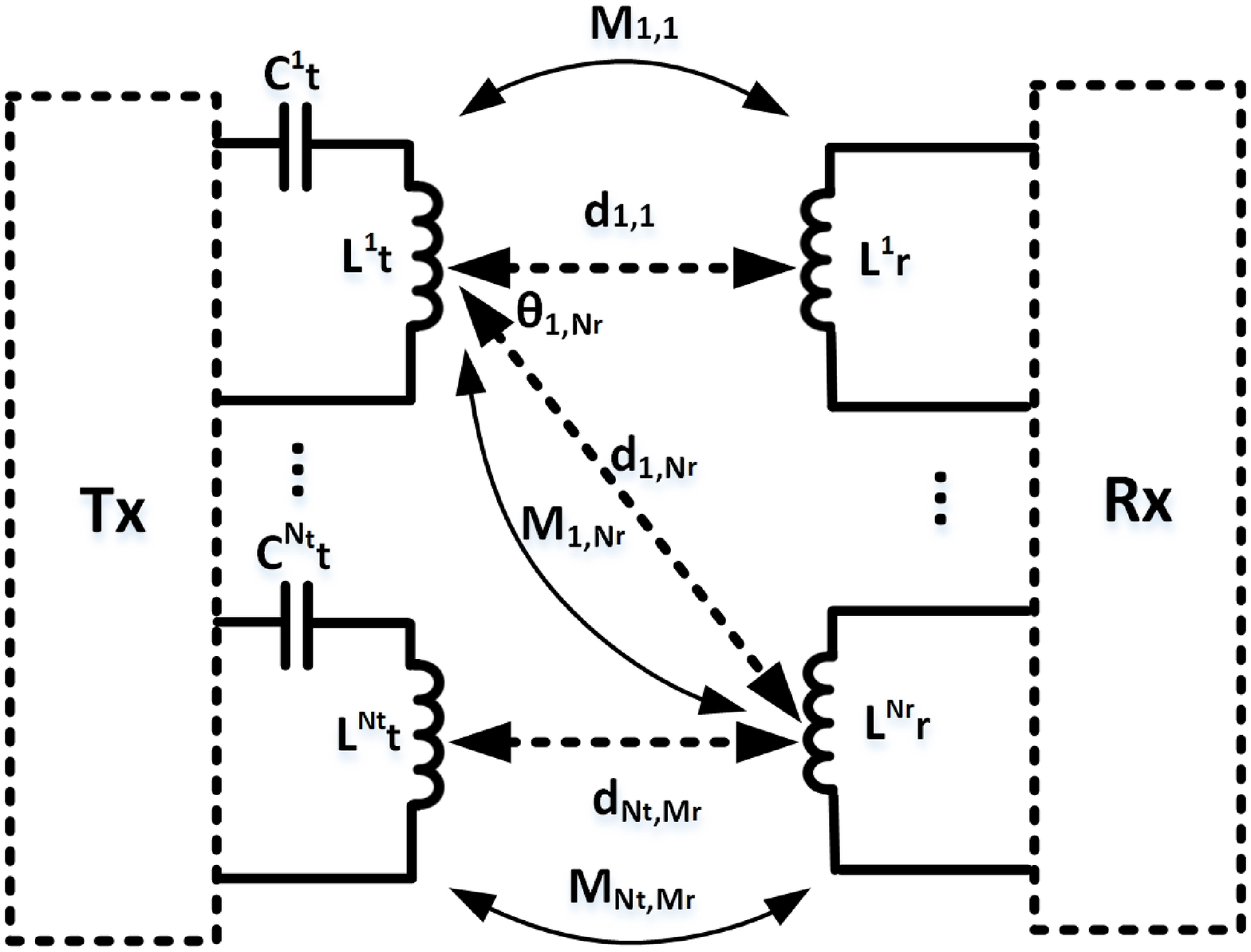}} 
 \centering
 \caption{Point-to-point transmission reference models.} 
 \label{P2P_model}
 \end{figure}

\subsubsection{SISO}

The SISO magnetic induction system is demonstrated in Figure~\ref{SISO}. Let $r_{t}$ and $r_{r}$ denote the radii of the coils of the transmitter and receiver, respectively. The distance between the two coils is represented by $d$. Let $\omega_{o}$ denote the resonance angular frequency that the two coils are coupling at. Then, $\omega_{o}=\frac{1}{\sqrt{L_{t}C_{t}}}=\frac{1}{\sqrt{L_{r}C_{r}}}$ where $L_{t}$ and $L_{r}$ are the self inductances of the two coils at the transmitter and receiver, respectively. $M$ is the mutual inductance, while $C_{t}$ and $C_{r}$ are two resonant capacitors. The resistances of the transmit coil and receive coil, are denoted by $R_{t}$ and $R_{r}$, respectively. The impedances at the source of the transmitter and the load of the receiver are denoted as $R_{S}$ and $R_{L}$, respectively. According to the Kirchoff's voltage law \cite{H.2013Nguyen}, the AC source voltage across the two coils can be expressed as follows:
\begin{eqnarray} \label{KVL}
(R_{S}+R_{t}+j \omega L_{t} +\frac{1}{j \omega C_{t}} ) I_{t} + j\omega M I_{r} = V_{S}	, \nonumber \\ 
j \omega M I_{t} + (R_{L} + R_{r} + j \omega L_{r}+\frac{1}{j \omega C_{r}}) I_{r} =0 	.
\end{eqnarray}

Simplified from (\ref{KVL}), the receive power at the load of the receiver can be obtained as follows~\cite{U.2012Azad}:
\begin{eqnarray}\label{P_r}
	P_{r} = P_{t}Q_{t}Q_{r}\eta_{t}\eta_{r}k^{2}(d) ,
\end{eqnarray}
where $P_{t}$ is the transmit power at the source of transmitter. 
$\eta_{t}$ and $\eta_{r}$ represent the efficiencies of the transmitter and receiver, respectively, which are given by 
\begin{eqnarray}
\eta_{t}=\frac{R_{S}}{R_{t}+R_{S}}, \hspace{3mm} \eta_{r}=\frac{R_{L}}{R_{r}+R_{L}}	.
\end{eqnarray}
$Q_{t}$ and $Q_{r}$ are the quality factors of the transmitter and receiver, given by
\begin{eqnarray}
Q_{t}=\frac{\omega L_{t}}{R_{t}+R_{S}}, \hspace{3mm} Q_{r}=\frac{\omega L_{r}}{R_{r}+R_{L}}	.
\end{eqnarray}
Moreover, $k(x)$ denotes the coupling coefficient factor between the two coils.

Coupling coefficient is a function of the mutual inductance, denoted as $M$, and the self inductance of transmit and receive coils, which can be estimated by the following expression, 
\begin{eqnarray}
k=\frac{M}{\sqrt{L_{t}L_{r}}}.
\end{eqnarray}

If the radius of transmit and receive coils as well as the charging distance between them are known, the coupling coefficient can also be expressed as the following function \cite{I.2011Agbinya,I.2013Agbinya}, 
\begin{eqnarray}\label{k_d}
k^{2}(d)=\frac{r^{3}_{t}r^{3}_{r}\pi^{2}}{(d^2+r^{2}_{t})^3}.
\end{eqnarray}

By inserting (\ref{k_d}) into (\ref{P_r}), the receive power in a SISO channel can be rewritten as follows:
\begin{eqnarray} \label{P_r2}
P_{r}=P_{t}Q_{t}Q_{r}\eta_{t}\eta_{r} \frac{r^{3}_{t}r^{3}_{r}\pi^{2}}{(d^2+r^{2}_{t})^3}.
\end{eqnarray}

\subsubsection{MISO}

Figure~\ref{MISO} shows the reference model for point-to-point transmission with MISO channel. 
Let $N_{t}$ represent the number of transmit coils. At resonant frequency, each coil of a charger is coupled with that of the energy receiver. The power delivered to the receiver from the charger's coil $n \in \{1,\ldots,N_{t}\}$ is given by~\cite{H.2013Nguyen}
\begin{eqnarray}
P^{n}_{r} = P^{n}_{t} Q^{n}_{t} Q_{r} \eta^{n}_{t} \eta_{r} k^{2}_{n}(d_{n})	,
\end{eqnarray}
where $P^{n}_{t}$, $Q^{n}_{t}$, and $\eta^{n}_{t}$ denote the transmit power, quality factor, and efficiency of the charger's coil $n$, respectively. $d_{n}$ represents the distance between the charger's coil $n$ and the receiver's coil.

The coupling efficiency between the charger and receiver is expressed by~\cite{H.2013Nguyen}
\begin{eqnarray}\label{coefficiency}
k^{2}_{n}(d)= \frac{r^{3}_{n}r^{3}_{r}}{\big ( d^{2}_{n} +r^{2}_{n}\big)^{3}} = \frac{r^{3}_{n}r^{3}_{r}}{\big ( (\frac{d_{1}}{cos \theta_{n}}  )^{2}+r^{2}_{n}\big)^{3}}	,
\end{eqnarray}
where $\theta_{n}$ is the angle between $d_{1}$ and $d_{n}$ as shown in Figure~\ref{MISO}. $r_{n}$ is the radius of the charger $n$'s coil. 

The aggregated receive power at energy receiver is additive, which can be calculated as follows: 
\begin{eqnarray}
P_{r}= Q_{r}\eta_{r}r^{3}_{r} \Bigg(P_{1}Q_{1}\eta_{1}\frac{r^{3}_{1}}{\left ((\frac{d_{1}}{cos\theta_{1}})^{2} + r^{2}_{1} \right)^{3}} +	\cdots \nonumber  \\ 
+ P_{N_{t}}Q_{N_{t}}\eta_{N_{t}} \frac{r^{3}_{N_{t}}}{\left ((\frac{d_{1}}{cos\theta_{N_{t}}})^{2} + r^{2}_{N_{t}} \right)^{3}}   \Bigg)	.
\end{eqnarray}

When the coils of the charger are considered to be identical, i.e., $P_{1}=P_{2}=\cdots=P_{N_{t}}=P_{T}$, $Q_{1}=Q_{2}=\cdots=Q_{N_{t}}=Q_{T}$, $\eta_{1}=\eta_{2}=\cdots=\eta_{N_{t}}=\eta_{T}$, and $r_{1}=r_{2}=\cdots=r_{N_{t}}=r_{T}$, the transferred power can be simplified as follows:
\begin{eqnarray}
P_{r}=P_{T}Q_{T}Q_{r}\eta_{t}\eta_{T}\pi^{2} r^{3}_{T} r^{3}_{r} 
\Bigg( \frac{1}{(1+r^{2}_{T}/d_{1}^{2})d_{1}^{6}} 
+ \cdots \nonumber  \\
 + \frac{cos^{6}_{\theta_{n}}}{(1+r^{2}_{T}cos^{2}_{\theta_{N_{t}}}/d^{2}_{n})d^{6}_{n}}	\Bigg)	.
\end{eqnarray}

\begin{figure*} 
\centering
\subfigure [Qi-compliant wireless power transfer model] {
 \label{WPT_Qi}
 \centering
 \includegraphics[width=0.85 \textwidth]{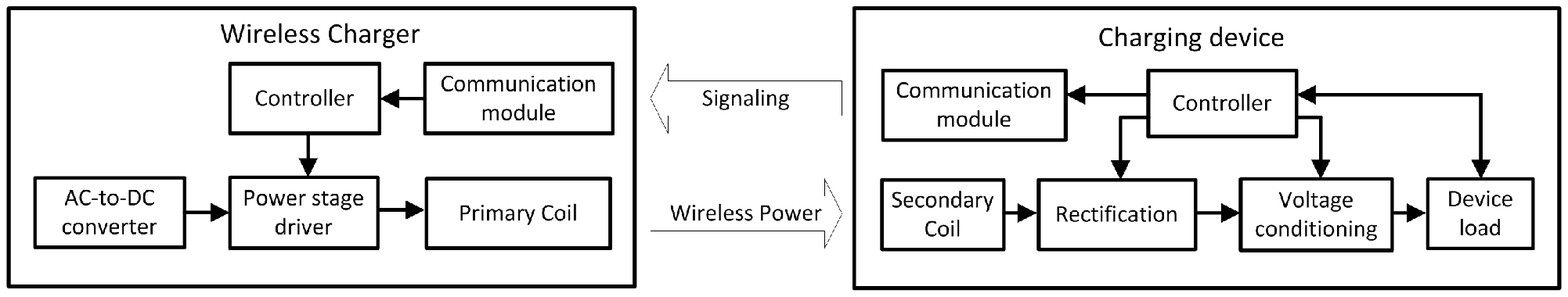}}  
 \centering
 \subfigure [A4WP-compliant wireless power transfer model] {
  \label{WPT_A4WP}
  \centering
  \includegraphics[width=0.85 \textwidth]{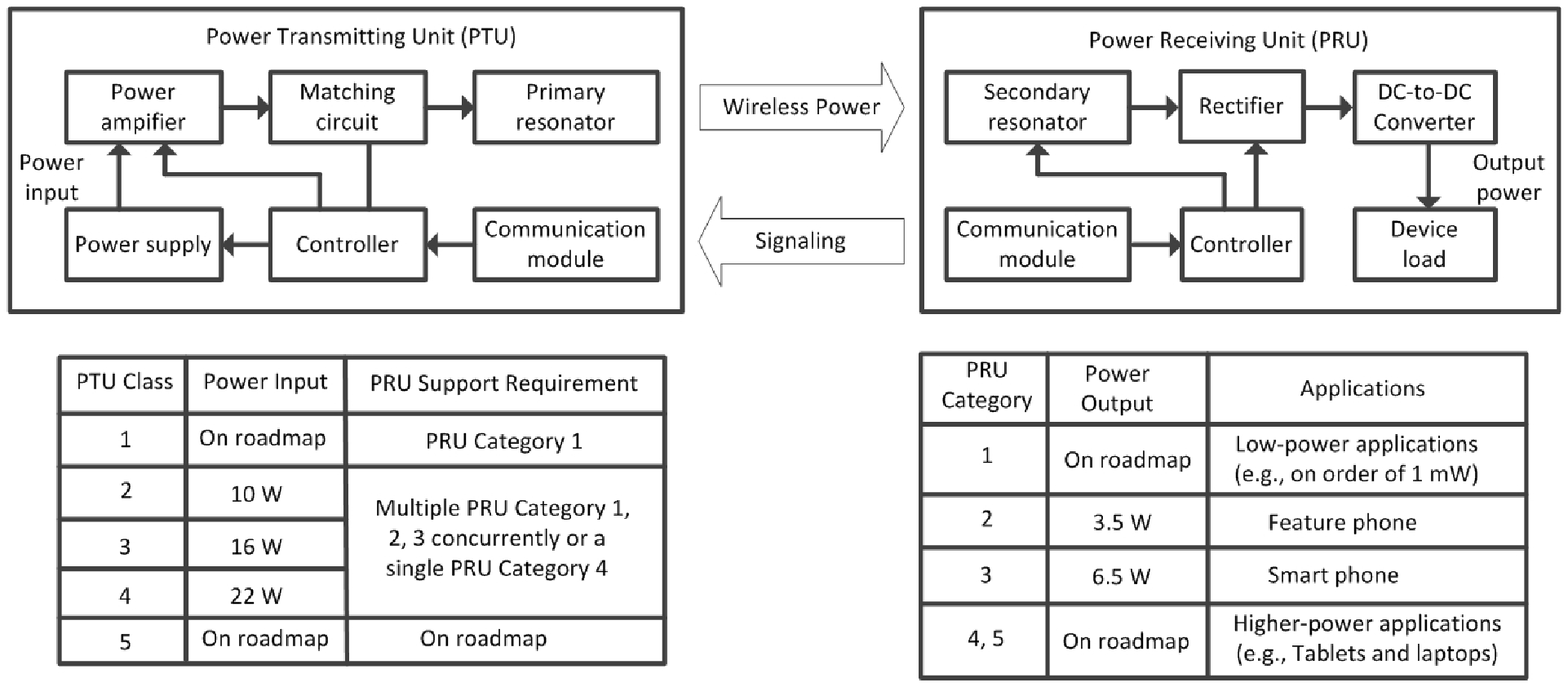}}   
  \centering
\caption{Reference models of near-field wireless power transfer protocol.} 
\label{WPT}
\end{figure*}  

\subsubsection{SIMO}
Figure~\ref{SIMO} demonstrates the reference model for point-to-point transmission with a SIMO channel. Let $N_{r}$ represent the number of coils at the energy receivers. Similar to the MISO system, at resonant frequency, the charger's coil is coupled with all the coils of energy receivers. Each receiver's coil captures a portion of energy from the charger. The receive power at the load of the receiver $m \in \{1,\ldots,N_{r}\}$ is given by
\begin{eqnarray}
P^{m}_{r}= P_{t} Q_{t} Q^{m}_{r} \eta_{t} \eta^{m}_{r} k^{2}_{m}(d_{m})	,
\end{eqnarray}
where $Q^{m}_{r}$ and $\eta^{m}_{r}$ denote the quality factor and efficiency of the charger's coil $m$, respectively. $d_{m}$ represents the distance between the charger's coil $m$ and the receiver's coil. 

Then, the total transferred power can be calculated as follows:
\begin{eqnarray}
P_{r}= P_{t} Q_{t} \eta_{t}  \big (  Q^{1}_{r} \eta^{1}_{r} k^{2}_{1}(d_{1})+\cdots +  Q^{m}_{r} \eta^{m}_{r} k^{2}_{m}(d_{m})+\cdots \nonumber \\ 
 +  Q^{N_{r}}_{r} \eta^{N_{r}}_{r} k^{2}_{N_{r}}(d_{N_{r}}) \big).
\end{eqnarray}
where $k^{2}_{m}(d_{m})$ is given by (\ref{coefficiency}).


\subsubsection{MIMO}

Let $k_{n,m}$ and $d_{n,m}$ denote coupling co-efficiency and distance between the transmit coil $n$ and receive coil $m$, respectively.
In the point-to-point MIMO transmission model, as shown in Figure~\ref{MIMO}, the receiver receives the power from each individual transmit coil separately. The crosstalk between the transmit coils and receive coils is small~\cite{H.2013Nguyen}. The receive power at the load of the receive coil $m \in \{1,\ldots,N_{r}\}$ from the transmit coil $n \in \{1,\ldots,N_{t}\}$  is given by
\begin{eqnarray}
P^{n,m}_{r}= P^{n}_{t} Q^{n}_{t} Q^{m}_{r} \eta^{n}_{t} \eta^{m}_{r} k^{2}_{n,m}(d_{n,m})	,
\end{eqnarray}
where $k^{2}_{n,m}(d_{n,m})$ is given by ($\ref{coefficiency}$).

The total transferred power can be derived as follows:
\begin{eqnarray}
P_{r}= \sum^{N_{t}}_{n=1} \sum^{N_{r}}_{m=1}  P^{n,m}_{r} .
\end{eqnarray}

\section{Wireless Charging Standards and Implementations}

Different wireless charging standards have been proposed. Among them, Qi and A4WP are two leading standards supported by major smartphone manufacturers. This section presents an overview of these two standards.

\subsection{International Charging Standards}

\subsubsection{Qi}

Qi (pronounced ``chee'') is a wireless charging standard developed by WPC~\cite{WPC}. A typical Qi-compliant system model is illustrated in Figure~\ref{WPT_Qi}. Qi standard specifies interoperable wireless power transfer and data communication between a wireless charger and a charging device. Qi allows the charging device to be in control of the charging procedure. The Qi-compliant charger is capable of adjusting the transmit power density as requested by the charging device through signaling. 

\begin{figure} 
\centering
\subfigure [Guided Positioning (Magnetic Attraction)] {
 \label{GPMA}
 \centering
 \includegraphics[width=0.157 \textwidth]{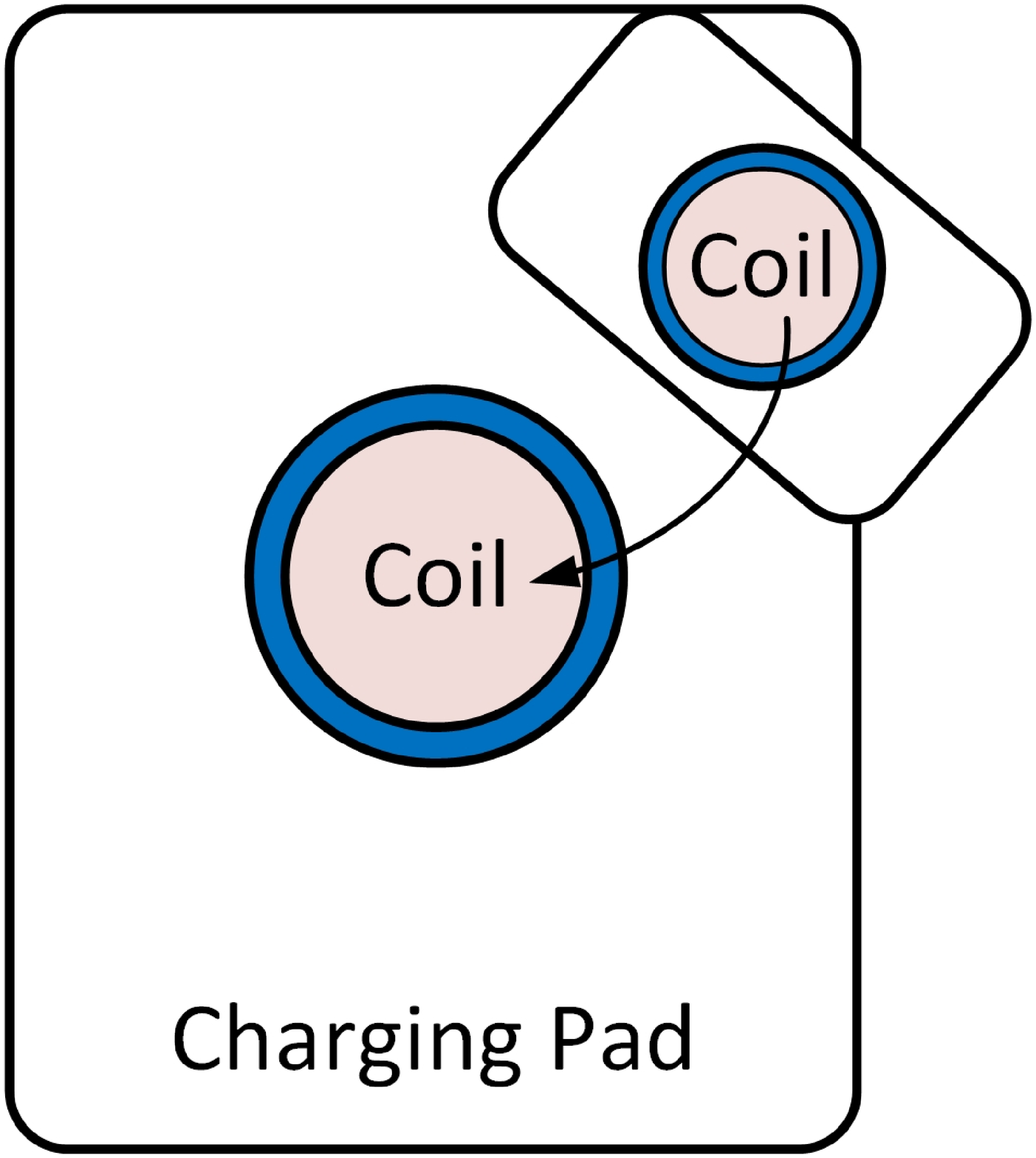}}  
 \centering
 \subfigure [Free Positioning (Moving Coil)] {
  \label{FPMC}
  \centering
  \includegraphics[width=0.13  \textwidth]{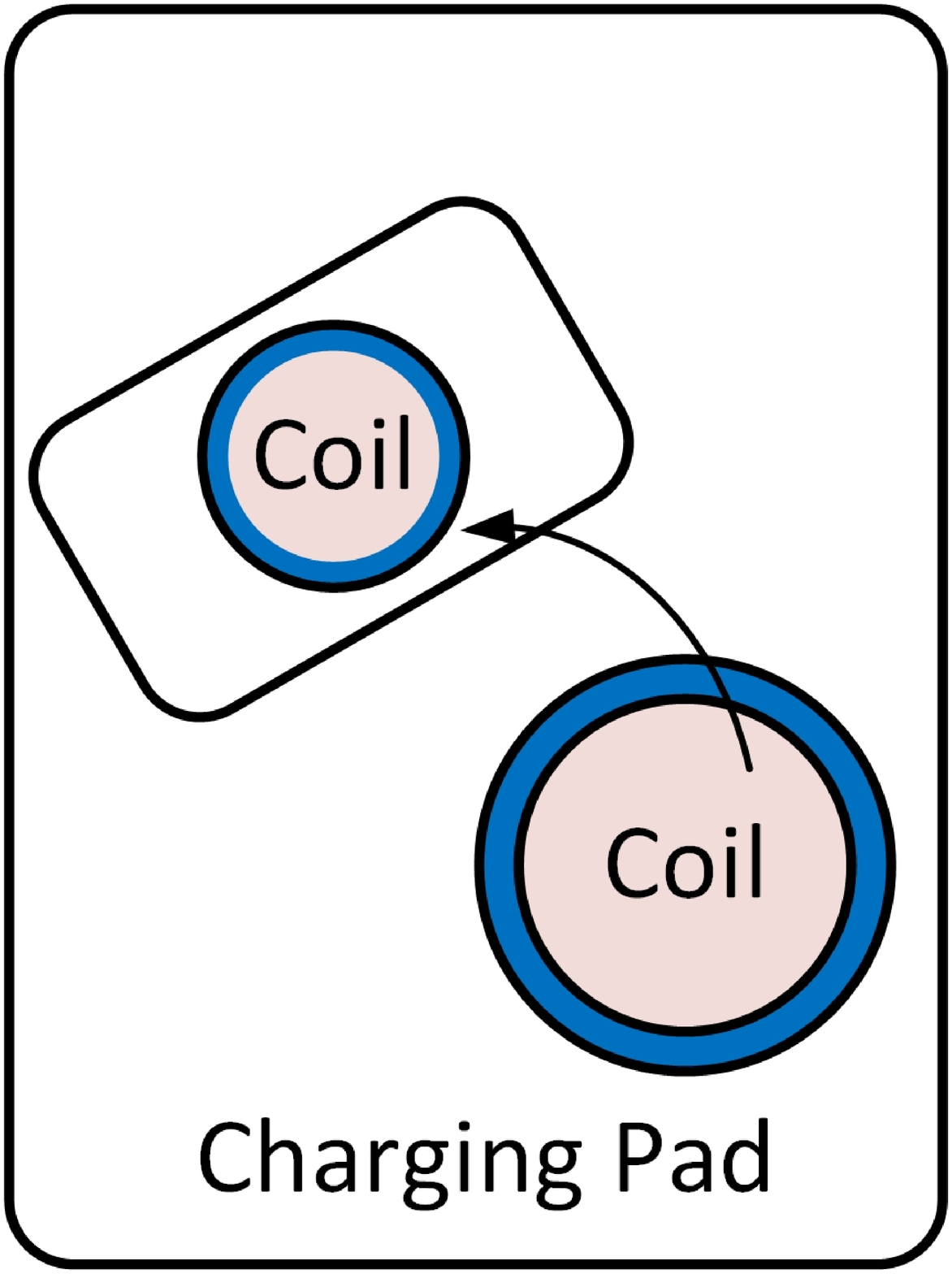}}   
  \centering
  \hspace{3mm}
 \subfigure [Free Positioning (Coil Array)] {
 \label{FPCA}
  \centering
\includegraphics[width=0.13  \textwidth]{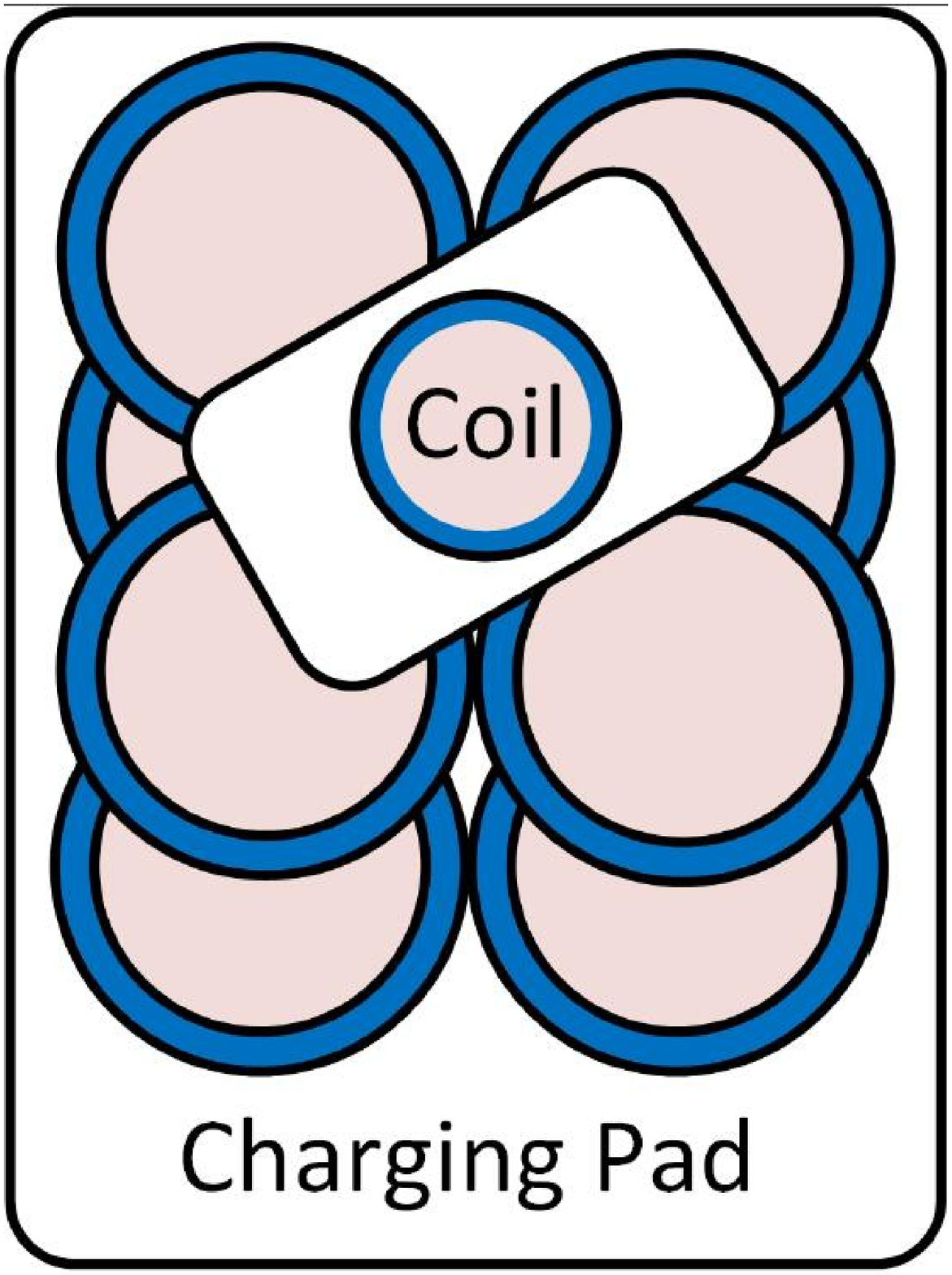}}\\
\centering
\caption{Models of wireless charging system.} 
\label{alignment_designs}
\end{figure}

Qi uses the magnetic inductive coupling technique, typically within the range of 40 millimetres. Two categories of power requirement are specified for Qi wireless charger, i.e.,  
\begin{itemize}
	\item Low-power category which can transfer power within 5W on 110 to 205kHz frequency range, and
	\item Medium-power category which can deliver power up to 120W on 80-300kHz frequency range.
\end{itemize}
Generally, a Qi wireless charger has a flat surface, referred to as a charging pad, of which a mobile device can be laid on top. As aforementioned, the tightness of coupling is a crucial factor in the inductive charging efficiency. To achieve tight coupling, a mobile device must be strictly placed in a proper alignment with the charger. Qi specifies three different approaches for making alignment~\cite{WPC_Qi}.
\begin{itemize}
	\item Guided positioning, shown in Figure~\ref{GPMA}, i.e., a one-to-one fixed-positioning charging, provides guideline for a charging device to be placed, for achieving an accurate alignment. The Qi specification guides the mobile device into a fixed location by using magnetic attractor. The advantage of this alignment approach is simplicity; however, it requires implementation of a piece of material attracted by a magnet in the charging device. Consequently, eddy-current-related power loss (and thus temperature rise) will be induced in the magnetic attractor \cite{X.2011Zhong}.    

	\item Free-positioning with movable primary coil, illustrated in Figure~\ref{FPMC}, is also a one-to-one charging that can localize the charging device. This approach requires a mechanically movable primary coil that tunes its position to make coupling with the charging device. This can be achieved by either inductive or capacitive means. The implementation of this alignment approach is simple if the charging pad is designed to accommodate only one device. However, the movable mechanical components tend to make the systems less reliable. Additionally, for multiple device charging, the motor control for the primary coils can be complicated and costly. 

	\item Free-positioning with coil array, demonstrated in Figure~\ref{FPCA}, allows multiple devices to be charged simultaneously irrespective of their positions.  The Qi specification endorses the ``vertical-flux" approach~\cite{R.2007Hui}, which utilizes the whole charger surface for power transfer without any restriction on the orientation of the secondary coil. For example, this free-positioning approach can be applied based on the three-layer coil array structure~\cite{R.2005Hui}. Compared with the above two approaches, this alignment approach offers more user-friendliness, at the expense of more costly and complex winding structure and control electronic element.
 
\end{itemize}

The Qi-compliant wireless charging model supports in-band communication. The data transmission is on the same frequency band as that used for the wireless charging. The Qi communication and control protocol is defined to enable a Qi wireless charger to adjust its power output for meeting the demand of the charging device and to cease power transfer when charging is finished. The protocol works as follows.
\begin{itemize}

\item Start: A charger senses the presence of a potential charging device.  

\item Ping: The charging device informs the charger the received signal strength, and the charger detects the response.  

\item Identification and Configuration: The charging device indicates its identifier and required power while the charger configures energy transfer.
 
\item Power Transfer: The charging device feeds back the control data, based on which the charger performs energy transfer. 

\end{itemize}

\subsubsection{A4WP}

A4WP aims to provide spatial freedom for wireless power~\cite{R.2013Tseng}. This standard proposes to generate a larger electromagnetic field with magnetic resonance coupling. To achieve spatial freedom, the A4WP standard does not require a precise alignment and even allows separation between a charger and charging devices. The maximum charging distance is up to several meters. Moreover, multiple devices can be charged concurrently with a different power requirement. Another advantage of A4WP over Qi is that foreign objects can be placed on an operating A4WP charger without causing any adverse effect. Therefore, the A4WP charger can be embedded in any object, improving the flexibility of charger deployment.

Figure~\ref{WPT_A4WP} shows the reference model for A4WP-compliant wireless charging. It consists of two components, i.e., power transmitter unit (PTU) and  power receiving unit (PRU). The wireless power is transferred from the PTU to the PRU, which is controlled by a charging management protocol. Feedback signaling is performed from the PRU to the PTU to help control the charging. The wireless power is generated at 6.78MHz Industrial Scientific Medical (ISM) frequency band. Unlike Qi, out-of-band communication for control signaling is adopted and operates at 2.4GHz ISM band.

\begin{itemize}

\item A PTU, or A4WP charger has three main functional units, i.e., resonator and matching circuit components, power conversion components, and signaling and control components. The PTU can be in one of following function states: {\em Configuration}, at which PTU does a self-check; {\em PTU Power Save}, at which PTU periodically detects changes of impedance of the primary resonator; {\em PTU Low Power}, at which PTU establishes a data connection with PRU(s); {\em PTU Power Transfer}, which is for regulating power transfer; {\em Local Fault State}, which happens when the PTU experiences any local fault conditions such as over-temperature; and {\em  PTU Latching Fault}, which happens when rogue objects are detected, or when a system error or other failures are reported.
 
\item The A4WP PRU comprises the components for energy reception and conversion, control and communication. The PRU has the following functional states: {\em Null State}, when the PRU is under voltage; {\em PRU Boot}, when the PRU establishes a communication link with the PTU, {\em PRU On}, the communication is performed; {\em PRU System Error State}, when there is over-voltage, over-current, or over-temperature alert; {\em PRU System Error}, when there is an error that has to shut down the power. 
\end{itemize} 
Figure~\ref{WPT_A4WP} also shows the classes and categories for the PTU and PRU (e.g., for power input and output, respectively). No power more than that specified shall be drawn for both PTU and PRU. 
 
Similar to the Qi standard, A4WP also specifies a communication protocol to support wireless charging functionality. A4WP-compliant systems adopt a Bluetooth Low Energy (BLE) link for the control of power levels, identification of valid loads, and protection of non-compliant devices. The A4WP communication protocol has three steps.
\begin{itemize}
\item Device detection: The PRU that needs to be charged sends out advertisements. The PTU replies with a connection request after receiving any advertisement. Upon receiving any connection request, the PRU stops sending advertisements. Then, a connection is established between the PTU and PRU. 

\item Information exchange: The PTU and PRU exchange their {\em Static Parameters} and {\em Dynamic Parameters} as follows. First, the PTU receives and reads the information of the PRU {\em Static Parameters} which contain its status. Then, the PTU specifies its capabilities in the PTU {\em Static Parameters} and sends them to the PRU. The PTU receives and reads the PRU {\em Dynamic Parameters} that include PRU current, voltage, temperature, and functional status. The PTU then indicates in the {\em PRU Control} to manage the charging process.

\item Charging control: It is initiated when {\em PRU Control} is specified and the PTU has enough power to meet the PRU's demand. The PRU {\em Dynamic Parameter} is updated periodically to inform the PTU with the latest information so that the PTU can adjust {\em PRU Control} accordingly. If a system error or complete charging event is detected, the PRU sends PRU alert notifications to the PTU. The PRU {\em Dynamic Parameter} includes the reason for the alert.
\end{itemize}
 
\begin{table*}\small 
\centering
\caption{\footnotesize Comparison of Different Wireless Charging Systems.} \label{WPT_systems}
\begin{tabular}{|l|l|l|c|}
\hline
\footnotesize {\bf System} & {\bf Source Power } & {\bf Frequency} & {\bf Effective Charging Distance}  \\ 
\hline
\hline
 RAVpower \cite{RAVPower} &  7.5W & 110 - 205kHz  &  8mm  \\
\hline
 Duracell Powermat \cite{Duracell2011} & 18W & 235 - 275kHz &  5mm \\
\hline
 Energizer Qi \cite{Energizer} &  22W & 110 - 205kHz &  11mm  \\
\hline
 Writicity WiT-2000M  \cite{WitricityCorp2000M}   & 12W &  6.78MHz &  20mm \\
\hline
  UW Prototype   \cite{P.2013Sample}  & 30W &  13.56MHz &   100mm \\
 \hline
 Writicity WiT-3300   \cite{WitricityCorp3300} & up to 3.3kW & 85kHz  & 150mm  \\
\hline
 MagMIMO  \cite{J.2014Jadidian}  & 20W &  1.0MHz &  400mm \\ 
\hline  
\end{tabular}
\end{table*} 
 
\subsection{Implementations of the International Charging Standards}

With the release of these international charging standards, research effort has been made on prototype studies. 
Due to the ease of implementation and early announcement, most of the existing implementations are based on the Qi standard. In the following, we review these hardware designs.   

In \cite{S2014Hached}, the authors proposed a Qi-compliant charger for implantable medical devices. The charger consists of a Bluetooth low-power communication module which allows remote control and supervision of the devices. In the system, the device's charging cycle control, real-time batteries and system status acquisition were performed remotely. When operated over 3W output power, the prototype was demonstrated to reach its maximum efficiency around 75$\%$. The authors in~\cite{S2013Miura} conducted a performance evaluation of bidirectional wireless charging between portable devices, under WPC Qi specifications. With an output power of 2.5W, 70$\%$ charging efficiency was achieved from the distance of 2mm. 

The authors in~\cite{M2013Galizzi} and \cite{V.2015Quang} built their prototype with integrated circuits. The authors in \cite{M2013Galizzi} introduced a Qi-compliant wireless charging system including a wireless power transmitter and a wireless power receiver. The power transmitter adopts a full-bridge resonant inverter and a full-bridge variable voltage regulator as the architecture. The prototype systems were implemented using an integrated circuit and discrete components. The experimental results showed that 70$\%$ charging efficiency was achieved at 5W output power for a 5mm charging distance. In \cite{V.2015Quang}, the authors presented the design of a fully integrated Li-ion battery charger in accordance to the Qi standard. With a constant current, the maximum and average charging efficiencies of $83\%$ and $79\%$ were achieved, respectively.

The authors in~\cite{Tiwari2013} and~\cite{X.2011Zhong} focused on the alignment control by presenting a design of a control unit and communication controller for guided positioning single receiver wireless charging platform. The control unit sets the response time values, the data exchanged between charger and receiver pair and the operating frequency using a serial communication interface. The function of the communication controller is to initiate, monitor and control wireless charging. Moreover, the authors introduced additional data processing and storage capability to make the design adaptive in terms of response time and the size of control data transfer. The implementation was shown to reduce the hardware design complexity and internal power consumption of both power transmitter and receiver.
The authors in~\cite{X.2011Zhong} introduced a design based on single-layer winding array to enable multiple-device simultaneous charging in a free-positioning manner. The proposed approach utilized the mathematical packing theory to localize the charging flux within the covered charging area, which enables the free placement of the devices (i.e., secondary coils). The measurements showed that energy efficiency in the range of $86\%-89\%$ was achievable for any position of the charging device. 
 
In \cite{P.2010Choi}, the authors compared four different power conversion methods, namely voltage control, duty-cycle control, frequency control and phase-shift control, for Qi-compliant wireless power transfer applications. The experiment demonstrated that the two phase-shift control approach outperforms the others, though the corresponding circuit is more costly. With the use of phase-shift control, an overall system efficiency of $72\%$ was achievable for 5W wireless charging.

In Table~\ref{WPT_systems}, we provide a comparison of the up-to-date wireless charging systems in terms of source power, frequency and effective charging distance.

\section{Static Wireless Charger Scheduling Strategies}


This section reviews the charging strategies for static chargers in WPCNs. In WPCNs, wireless devices communicate using only the harvested energy from wireless chargers. Typically, in WPCNs, two types of wireless chargers are considered. The first type is dedicated to providing wireless charging, referred to as an energy access point (E-AP). The second type can additionally support data communication and work as a data assess point (D-AP), referred to as a hybrid access point (H-AP). For the research efforts for WPCNs, there are two major directions. The first direction focuses on exclusive wireless charging, i.e., wireless power transfer and information transmission are separated, which is the focus of this section. The second direction is the research on SWIPT, in which wireless charging and information transmission are coupled to achieve some tradeoff. Substantial number of studies of SWIPT have been conducted in various contexts, e.g., point-to-point channels~\cite{I.FlintMay2015,X.LuMarch2015,I.FlintDecember2014,C.2015Liu}, broadcast channels~\cite{K.2015Lee,S.2015Timotheou,W.K.2014Ng}, relay channels~\cite{K.2015Xiong,H.2015Chen,Y.2015Huang,A.2013Nasir}, OFDMA channels \cite{K.2013Ng},  multi-antenna channels with additive white Gaussian noise (AWGN) or fading AWGN~\cite{X.2015Chen,R.2015Morsi,R.2015Zhang,W.K.Ng2014}, opportunistic channels~\cite{K.2015Ng,I.2014Krikidis,N.Zhao2015,M2015Tian} and wiretap channels~\cite{H.2015Xing,2015A.Khandaker,Q.2015Shi,Q.Zhang2015}. Additionally, cooperative SWIPT in distributed systems has also been investigated in~\cite{2015S.Lee,W.K.Ng2015}.  Our previous work in~\cite{X.LuSurvey} has provided a prodigious survey on this topic, and thus we omit the discussion in this paper.

\begin{figure*} 
\centering
 \subfigure [Wireless powered communication Network with hybrid access point] {
 \label{WPCN_integrated}
 \centering
\includegraphics[width=0.48  \textwidth]{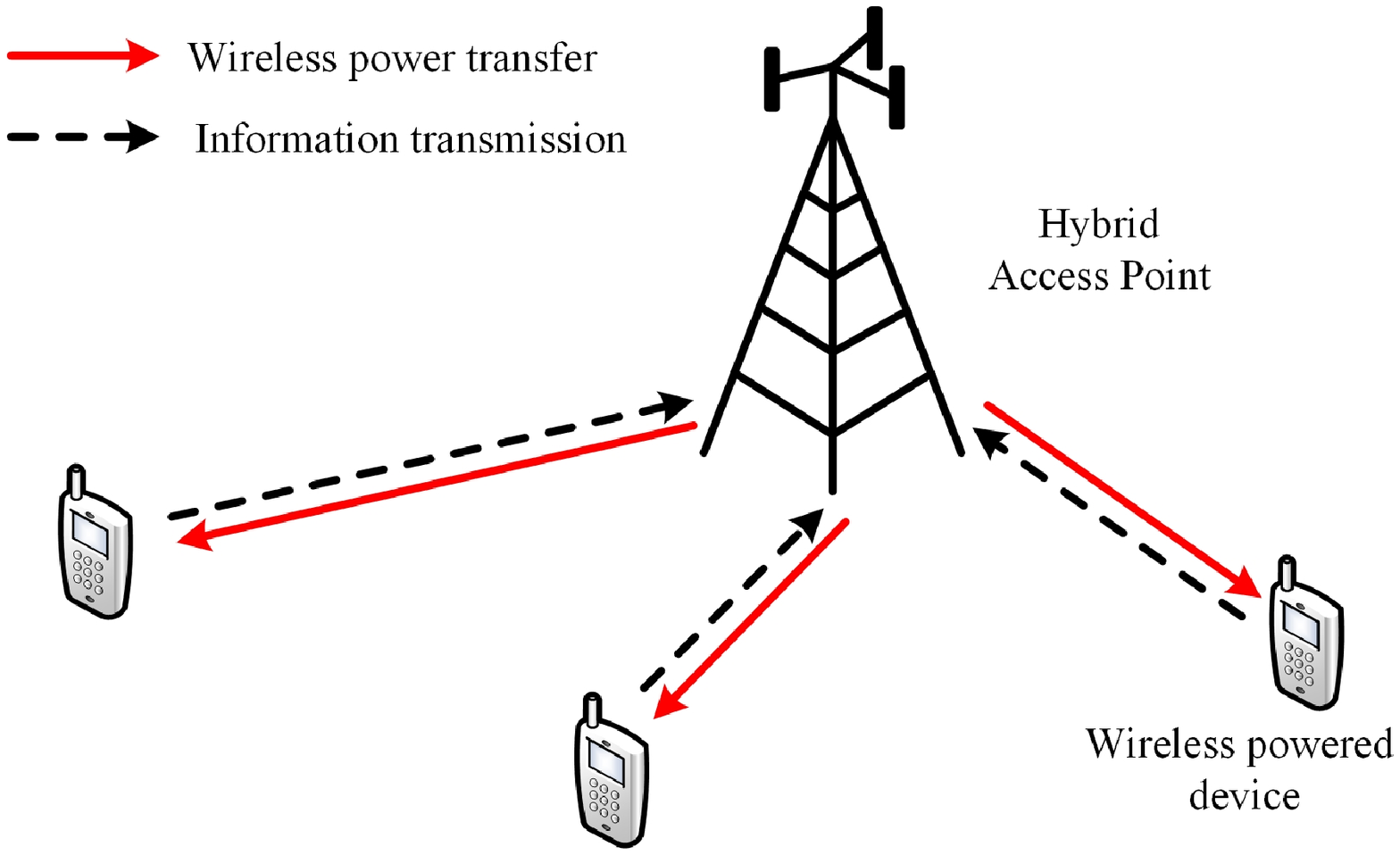}}
\centering
\subfigure [Wireless powered communication Network with separated energy access point and data access point] {
 \label{WPCN_separated}
 \centering
 \includegraphics[width=0.47 \textwidth]{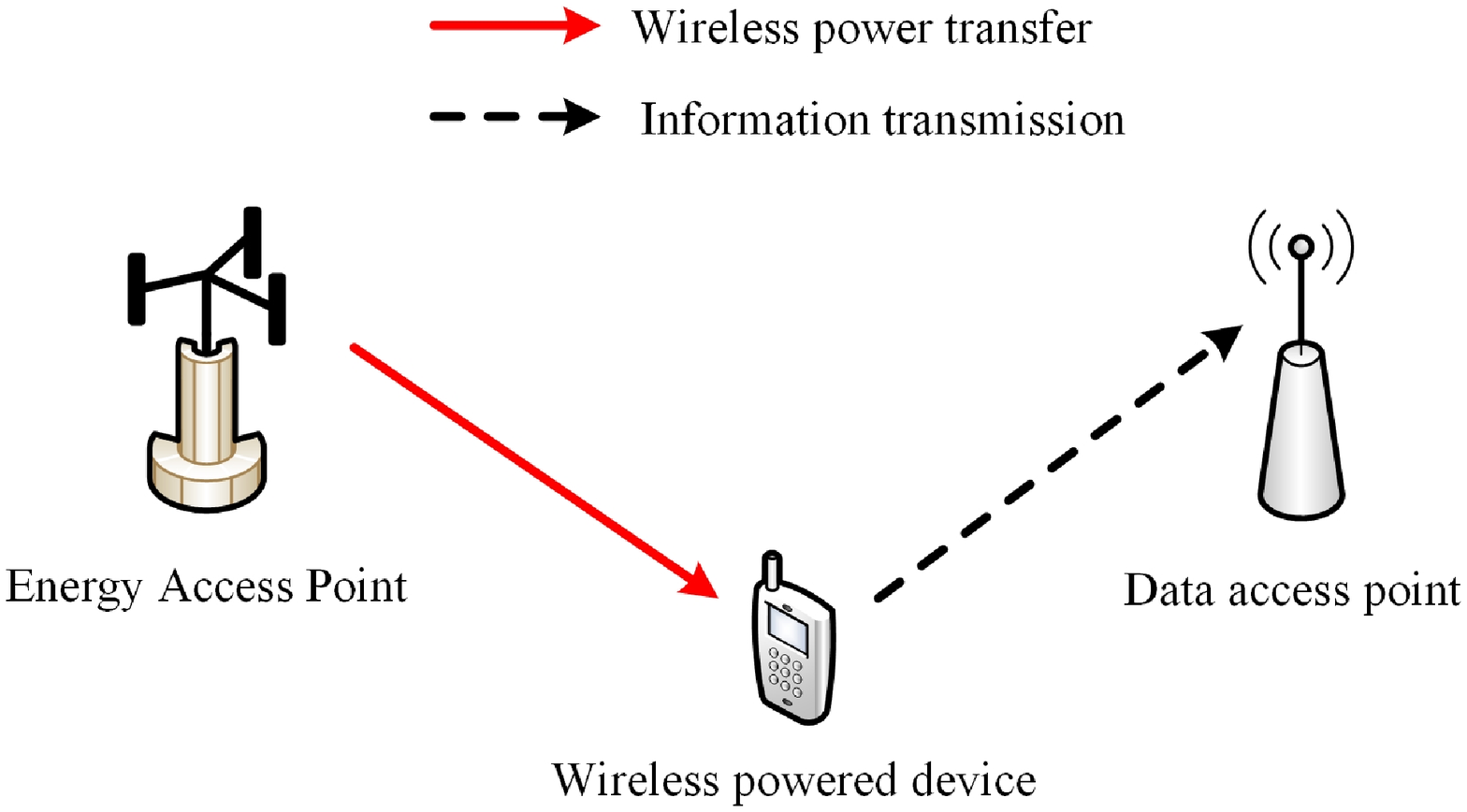}}  \\
 \centering 
 \subfigure [Wireless powered communication Network with relay and hybrid access point] {
 \label{WPCN_relay}
 \centering
\includegraphics[width=0.48  \textwidth]{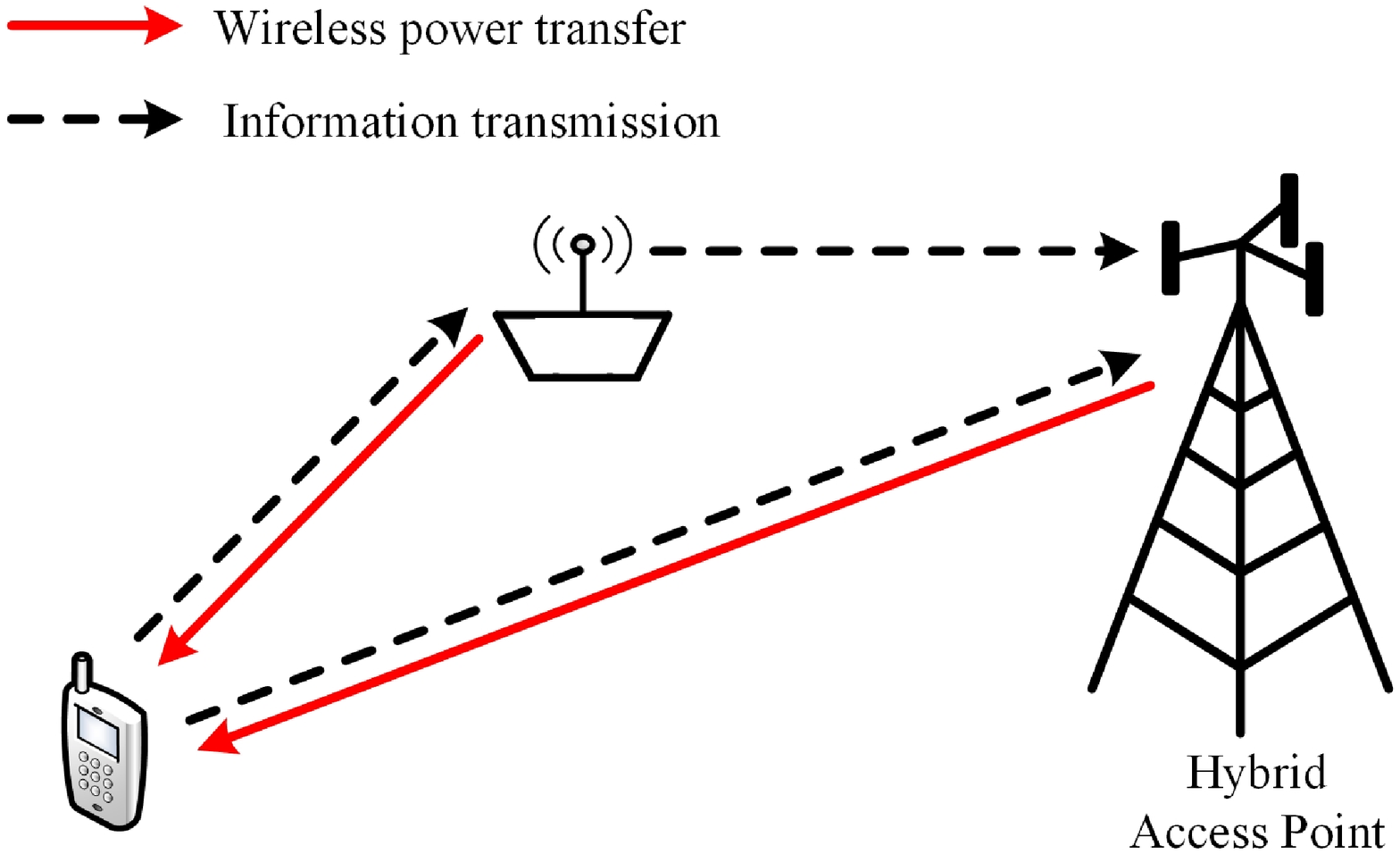}} 
\centering
 \subfigure [Wireless powered communication Network with multi-antenna energy access point] {
 \label{WPCN_beamforming}
 \centering
\includegraphics[width=0.48  \textwidth]{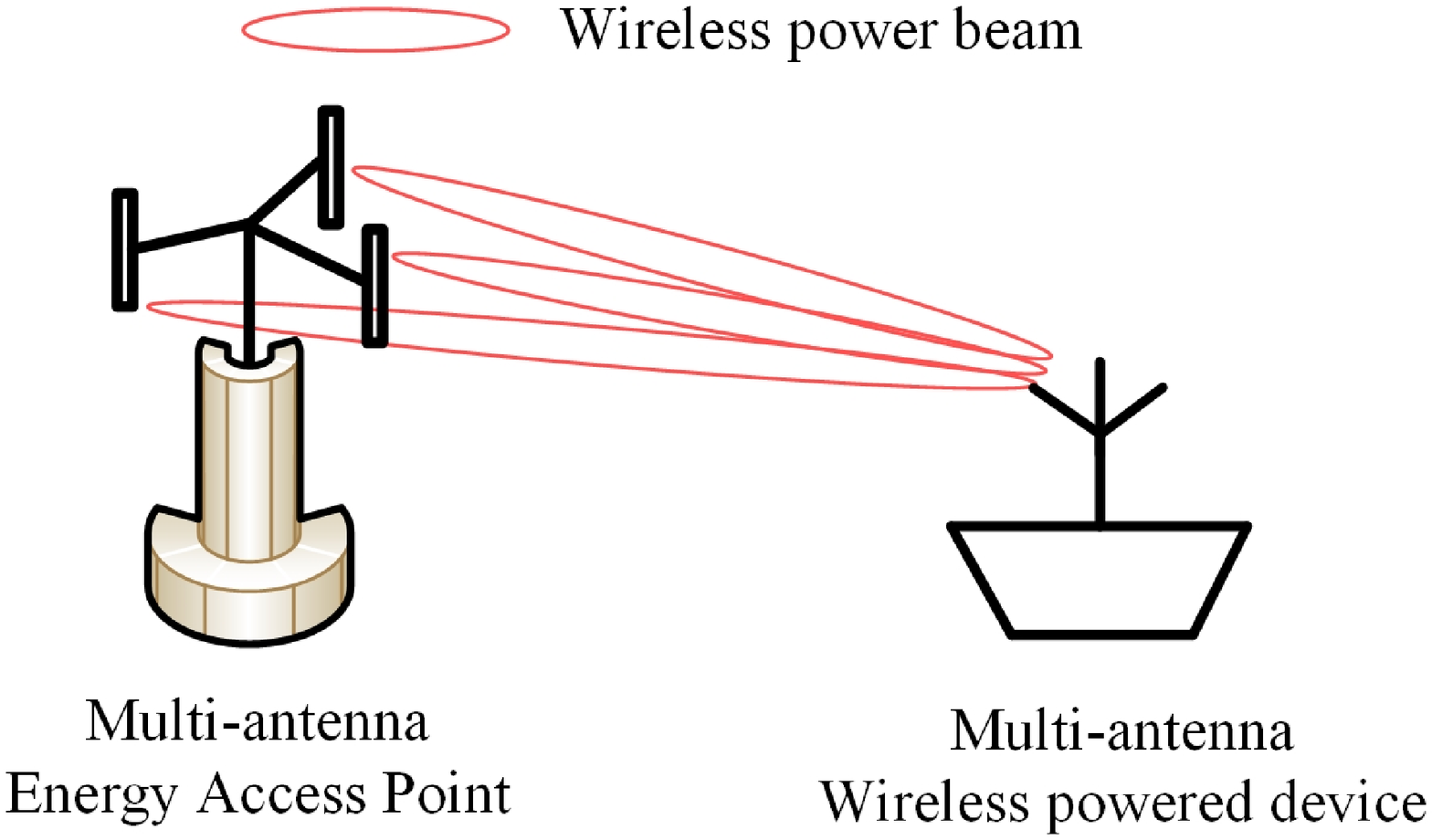}}\\
\centering
\caption{Reference models of wireless charging for wireless powered communication networks.} 
\label{static_charger_scheduling}
\end{figure*}

The existing literature considered four types of system models, as shown in Figure~\ref{static_charger_scheduling}, which are briefly described as follows.

\begin{itemize}

\item {\em WPCN with H-AP}: This system model (Figure~\ref{WPCN_integrated}) employs an H-AP to perform downlink wireless charging to and receive information transmission from a user device. 

\item {\em WPCN with dedicated E-AP}: In Figure~\ref{WPCN_separated}, downlink wireless charging and uplink information reception are conducted separately by an E-AP and a D-AP, respectively. 

\item {\em Relay-based WPCN with H-AP}:  This system model (Figure~\ref{WPCN_relay}) has a relay to facilitate the uplink transmission from a device to the H-AP. 

\item {\em WPCN with multi-antenna E-AP}: This system model (Figure~\ref{WPCN_beamforming}) adopts multiple antennas/coils at the E-AP to improve the charging efficiency by steering the energy beam spatially toward the direction of a device. The energy beamforming strategy is the main focus, while the information transmission is performed separately with wireless charging.

\end{itemize}
Note that, in WPCNs, a full-duplex H-AP allows for simultaneous wireless charging and information transmission in the downlink and uplink directions, respectively. By contrast, a half-duplex H-AP needs a coordination between wireless charging and information transmission of distributed devices in different time divisions. Moreover, a full-duplex device requires out-of-band wireless charging, which is performed on a frequency band different from that of information transmission. Half-duplex devices support in-band wireless charging, which overlaps with information transmission frequency band. 

In the following subsections, we review the charging strategies according to the types of WPCNs that they apply to.  

\subsection{Charging Strategies for Hybrid Access Point}

In the WPCN with an H-AP, the major issue is resource allocation to maximize the achievable throughput of the wireless powered devices.    


Both~\cite{H.2014Ju} and~\cite{2014H.Ju} aimed to maximize the weighted sum-throughput of the network. Reference~\cite{H.2014Ju} proposed a \emph{harvest-then-transmit} protocol which first schedules network devices to harvest energy from wireless charging in the downlink. Then the harvested energy is utilized to transmit individual information to the H-AP in the uplink based on time division multiple access (TDMA). With this protocol, the authors jointly optimized the time allocations for wireless charging and data gathering at the half-duplex H-AP based on the users' channel information and their average energy harvesting rate. By applying convex optimization techniques, the optimal time allocations were derived in closed-form expressions.   
However, the considered system revealed a doubly near-far problem that the users far from the H-AP receive less energy, and they require more power for uplink information transmission. To address this issue, the authors introduced a performance metric called common-throughput, which imposed the constraint that all network devices would be assigned with the same throughput regardless of their location. Moreover, an iterative algorithm based on a simple bisection search was developed to address the common-throughput maximization problem. It is demonstrated that the proposed iterative algorithm is useful to solve the doubly near-far problem, but incurs a cost of sum-throughput degradation.


Reference~\cite{2014H.Ju} extended~\cite{H.2014Ju} by employing a full-duplex H-AP. To address the maximization problem in the considered network, the authors jointly optimized the power allocation and time allocation at the H-AP in the downlink, as well as the time allocation for the users in the uplink. The problem was shown to be convex and non-convex for the cases considering perfect and imperfect self-interference cancellation, respectively. Accordingly, the authors obtained an optimal and a suboptimal solutions for joint time and power allocation. It was revealed that more charging power should be transferred during the time slots of the users with poorer channels and/or lower weights. The simulation results showed that the system with a full-duplex H-AP outperforms that of half-duplex ones when the self-interference can be  canceled effectively. 

Both~\cite{H.2014Ju} and~\cite{2014H.Ju} considered user devices with deterministic locations. The authors in~\cite{L.1409.3107Che} took a different approach to analyze the performance of randomly located devices by adopting a stochastic geometry approach. With the aim to maximize the system spatial throughput, the author developed a joint framework to optimize the uplink transmit power as well as the time partition between downlink energy transfer and uplink information transmission. Through the proposed framework, the authors characterized the successful information transmission probability. Moreover, the spatial throughput optimization problems were solved for both battery-free and battery-deployment cases.  
Numerical results illustrated the impacts of battery storage capacity on the system spatial throughput.

Rather than optimizing the multi-user scheduling, reference~\cite{H.2015Tabassum} analyzed the performance of state-of-the-art greedy and round-robin scheduling schemes jointly with the \emph{harvest-then-transmit} protocol. Closed-form expressions for the minimum power outage probability were derived. The authors then modified the analyzed schemes to improve the spectral efficiency on a given uplink channel with the zero power outage probability. The modified versions were illustrated to outperform the original ones in terms of fairness among user devices. However, this study only considered a single cell and ignored the impact of the accumulated RF energy from ambient cells.     


Different from the above three works, the system models in~\cite{L.2014Liu} and~\cite{G.1403.3991Yang} employed multiple antennas at the H-AP. The multi-antenna H-AP could control the energy transfer rate to different devices via tuning the energy beamforming weights. 
The objective of~\cite{L.2014Liu} is to maximize the minimum throughput of all devices.
To address the doubly near-far problem, the authors formulated a non-convex problem to optimize time allocation, the downlink energy beamforming, uplink transmit power allocation and receive beamforming jointly. The problem could be optimally solved using a two-stage algorithm, i.e., to obtain optimal downlink energy beamforming and time allocation. However, due to the high complexity of the two-stage algorithm, two suboptimal designs were introduced to mitigate the complexity of the proposed algorithm. It was revealed that the performance of the proposed suboptimal solutions approaches that of the optimal solution in terms of the max-min throughput. 

In~\cite{G.1403.3991Yang}, the authors designed a frame-based transmission protocol in a massive MIMO system with imperfect channel state information (CSI). Under the protocol, each time frame is divided into different phases. The access point first estimates downlink channels through exploiting channel reciprocity from the pilot signals sent by devices in uplink transmission. Next, RF energy is broadcast to all devices. Then, by using the harvested energy, the devices transmit their individual information to an access point simultaneously. The scheme maximizes the minimum rate among all devices through optimizing the time and energy allocation. Moreover, a metric, called massive MIMO degree-of-rate-gain, was defined as the asymptotic uplink rate normalized by the logarithm of the number of antennas at the access point. It was shown that the proposed transmission scheme is optimal with reference to the proposed metric. Moreover, the best possible fairness can be guaranteed by asymptotically obtaining a common rate for all devices.

\subsection{Charging Strategies for Dedicated Energy Access Point}

For WPCN with a dedicated E-AP, the focus is to control the wireless charging power of E-AP to achieve some optimization objective.

The studies in~\cite{F.2015Zhao}, \cite{X.2014Zhou} and~\cite{Q.2014Sun} aimed to  maximize the achievable throughput. 
The authors in~\cite{F.2015Zhao} considered the harvest-then-transmit protocol. An optimization framework to balance the time duration between energy harvesting and information transmission was developed under the energy, time and information error rate constraints. The solution was shown to be optimal. Unlike~\cite{F.2015Zhao}, reference~\cite{X.2014Zhou} studied a full-duplex mode system, in which the energy harvesting and data transmission were performed over two separated and time-varying channels. Under the assumption that the information of both channels was a priori known and the user's battery capacity was unlimited, the authors proposed an efficient algorithm to obtain optimal power allocation. The performance gap between the considered system and a conventional system with random energy harvesting rate with the same total power was examined by numerical simulation. However, this work only considered a single device, which left the power allocation to multiple devices an open issue.

The system model in~\cite{Q.2014Sun} extended that in~\cite{X.2014Zhou} by considering multiple devices and multiple antennas adopted at an E-AP.  
Considering TDMA, the system sum-throughput maximization problem was formulated as a non-convex problem that jointly optimizes time allocation and energy beamforming. 
The authors applied the semi-definite relaxation technique to reformulate a convex problem and proved the tightness as well as the global optimality of the semi-definite relaxation approximation. Furthermore, the authors devised a fast semi-closed form solution, which was numerically shown to reduce the implementation complexity substantially. 

Different from~\cite{X.2014Zhou} and~\cite{Q.2014Sun} where wireless charging is deterministic, reference~\cite{O.2012Bicen} assumed the E-APs and devices opportunistically access the same channel for wireless charging and information transmission. The power control of multiple E-APs with multiple wireless powered devices was considered in the system model. The authors proposed a power control strategy for the E-APs to minimize the energy consumption subject to a desired distortion level requirement at the D-AP. This energy consumption depends on the amount of information sent from the device under noise. Simulation results characterized the tradeoff among estimation distortion, number of E-APs and their power level.

\subsection{Charging Strategies for Relay with Hybrid Access Point}

The main concern in relay-based WPCN is to design an operation protocol to coordinate data transmission and wireless charging for throughput maximization.

The studies in~\cite{H1404.4120Chen} and~\cite{H.2014Chen} analyzed different cooperative strategies for the relay to improve network performance.  In~\cite{H1404.4120Chen}, the authors designed a harvest-then-cooperate protocol. The protocol schedules the user device and relay to first harvest energy and then perform information transmission in the uplink direction cooperatively. For the case of single relay with delay-limited transmission, the authors derived the approximate expression for the average system throughput under the proposed protocol over a Rayleigh fading channel in closed-form. For the case of multiple relay nodes, the approximate system throughput under the proposed protocol with two different relay selection schemes was derived. It was demonstrated by simulations that the proposed protocol outperforms the harvest-then-transmit protocol~\cite{H.2014Ju} in all evaluated cases.


Reference~\cite{H.2014Chen} further proposed two cooperative protocols, namely, energy cooperation and dual cooperation. The former allows a relay to cooperate with an H-AP for only downlink energy transfer. The latter instead lets the relay first cooperate with the H-AP for downlink energy transfer, and then assist the user device for uplink information transmission. The authors formulated the system throughput maximization problems by jointly designing the power allocation and time allocation. The optimal solutions for both problems were derived. Theoretical analysis revealed that, to maximize system throughput under the energy cooperation protocol, the strategy is to let an H-AP and relay always transmit with the peak power irrespective of the optimal time allocation. Furthermore, shown by simulation, when the signal-to-noise ratio (SNR) is high, the energy cooperation protocol achieved better performance than that of the dual cooperation protocol.

\subsection{Charging Strategies for Multi-antenna Energy Access Point}

The research efforts for a multi-antenna E-AP mainly deal with designing beamforming strategy and/or CSI feedback mechanism to improve the wireless charging efficiency. Note that the strategies in this category do not consider information transmission related issues.

The studies in~\cite{G.YangTSP}, \cite{Y.1403.7870Zeng} and \cite{Y.Zeng2015} investigated RF-based multi-antenna systems with the same objective to maximize the amount of transferred energy.
In~\cite{G.YangTSP}, the authors designed an adaptive energy beamforming scheme for a point-to-point MISO system with imperfect CSI feedback. The considered system operates on a frame-based protocol which first schedules the receiver to estimate channel via the preambles sent from the transmitter and feed the CSI estimation back to the transmitter. Then, the RF energy is transferred from the transmitter through beamforming. To maximize the harvested energy, the authors exploited the tradeoff between channel estimation duration and power transfer duration as well as allocation of transmit power. They first derived the optimal energy beamformers. Then, an optimal online preamble length and an offline preamble length were obtained for the cases with variable and fixed length preambles, respectively. The transmit power is allocated based on the channel estimation power and the optimal preamble length. 

In~\cite{Y.1403.7870Zeng}, considering a point-to-point MIMO energy beamforming system, the authors investigated the optimal design for a channel acquisition. Based on channel reciprocity, the E-AP estimates the channel status through dedicated reverse-link training signal from the user device. 
The study revealed the tradeoff between training duration and wireless power transfer in the energy beamforming system. Particularly, too short training duration lessens the precision of channel status estimation and therefore lowers the energy beamforming gain. On the other hand, too long training duration causes excessive energy consumption of user device, and thus decreases the time duration for energy transfer. Based on this tradeoff, the authors introduced an optimal training design to maximize the net energy of the user device, which is calculated by normalizing the energy used for channel training with the total harvested energy. However, this proposed design only applies for narrow-band flat-fading channels. 

Reference~\cite{Y.Zeng2015} extended \cite{Y.1403.7870Zeng} by considering more complicated wide-band frequency-selective fading channels, which offers additional frequency-diversity gain for energy transfer efficiency, compared to its narrow-band counterpart. To achieve both the diversity and beamforming gain, the authors introduced a two-phase channel training scheme. In the first phase, the E-AP selects a set of sub-bands with the largest antenna sum-power gains through the pilot signals sent from the user device. Then in the second phase, the E-AP estimates the MISO channels by the additional pilot signals only from the selected sub-bands. Numerical results demonstrated that the propose scheme is able to optimally balance the achievable frequency-diversity and energy-beamforming gains with energy constrained training. The authors further derived the closed-form expression of the amount of harvested energy at the user device under the proposed two-phase training scheme. This analytical result indicates that the amount of harvested energy is upper bounded by a constant value as the number of sub-bands approaches infinity. However, this study only considered independent channels, while leaving the case with co-related channels unexplored.

Different from the above studies~\cite{G.YangTSP,Y.1403.7870Zeng,Y.Zeng2015}, the research efforts in~\cite{J.2014Jadidian} focused on the coupling-based multi-coil system. Specifically, a near-field charging system, called MagMIMO, which performs wireless charging by beamforming the non-radiative magnetic field, was designed. The authors of~\cite{J.2014Jadidian} introduced the channel estimation scheme based on the measurement of the load that the receiver imposes on the transmitter circuit. This scheme differs from conventional communication system where a multi-antenna transmitter acquires the channel information either through feedback or from inferring the reciprocal channels by listening to some transmission from the receiver. 
Moreover, the authors also devised a protocol to allow MagMIMO to detect the presence of the device and its instantaneous load resistance according to the feedback information from the receiver. Experiment results demonstrated that the MagMIMO consumed comparable power to those of existing wireless chargers, such as Powermat and Energizer Qi, while enabling substantially longer effective charging distance. 


\subsection{Discussion and Summary }
In this section, we have reviewed the wireless charging strategies in four different types of networks, i.e., WPCN with H-AP, WPCN with dedicated E-AP, relay-based WPCN with H-AP and WPCN with multi-antenna E-AP. 
In Table \ref{Wireless_Charger_Scheduling_Strategies}, we summarize the reviewed literature of static wireless charger scheduling strategies.  

For the WPCN with an H-AP, charging strategies have been addressed for a half-duplex H-AP in SISO channels~\cite{2014H.Ju}, MISO channels~\cite{L.2014Liu} and MIMO channels~\cite{G.1403.3991Yang}, as well as a full-duplex H-AP in SISO channels~\cite{H.2014Ju}. Exploring charging strategy for multi-antenna WPCNs with the full-duplex H-AP can be one of the future directions. Moreover, most of the existing works consider TDMA-based schemes. However, it is interesting to analyze other multiple access schemes, e.g., OFDMA, to coordinate the uplink information transmission of user devices.

For WPCN with dedicated E-AP, existing literature has investigated single E-AP charging strategies for a single user device~\cite{X.2014Zhou} and multiple users~\cite{Q.2014Sun} with deterministic wireless charging. However, this strategies only apply for small-scale networks with limited number of devices. Charging strategies to coordinate multiple E-APs in deterministic channels are worth exploring. Moreover, multiple E-AP charging strategy for multiple user devices has been studied in opportunistic channels. Opportunistic wireless charging is explored for relay-based promising techniques to improve the spectrum efficiency. More research efforts can be made toward the analysis of charging throughput, system capacity as well as interference. 

For relay-based WPCN, network protocols have been proposed to address two cases in which the relay is wirelessly powered~\cite{H.2014Chen} and with wired power connection~\cite{H1404.4120Chen}. Compared to the former, the latter can further collaborate with the H-AP to perform downlink wireless charging to the user devices. The full-duplex relay and multi-antenna relay can be two directions to be further explored for relay-based WPCN. 

For WPCN with multi-antenna E-AP, beamforming strategies have been investigated for far-field multi-antenna system with MISO channel~\cite{G.YangTSP} and MIMO channel~\cite{Y.1403.7870Zeng}. However, existing literature only considered point-to-point charging. Energy beamforming for multiple energy receivers is a crucial issue to be addressed. Moreover, a near-field multi-coil system that generates energy beamforming with magnetic field has also been invented very recently in~\cite{J.2014Jadidian}. More research effort in analysis and experiment is required to understand the empirical performance in various network conditions.    

In addition, most of the existing studies only adopt theoretical analysis and numerical simulations. There is a need to design protocols for practical applications and perform experimental evaluations.  Existing work in \cite{S.2015Nikoletseas}, \cite{S.Nikoletseas2015} and \cite{Q.1504.00639Liu} performed realistic validation based on experimental setting of real devices. However,  the wireless communication requirements are not considered. Designing charging protocols associated with optimization of communication performance for real-world implementation is a critical research direction.

\begin{table*} \small 
\centering
\caption{\footnotesize Summary of Static Wireless Charger Scheduling Strategies.} \label{Wireless_Charger_Scheduling_Strategies}
\begin{tabular}{|p{1.5cm}|p{1.8cm}|p{1.2cm}|p{1cm}|p{5.5cm}|p{2.5cm}|p{1.7cm}|} 
\hline
\footnotesize  Literature & System model & \multicolumn{2}{ |c|  }{ Channel model }  &  Objective & Solution & Evaluation \\
\cline{3-4}
 & & Downlink & Uplink & & & \\
\hline
 \footnotesize  Ju {\em et al} \cite{H.2014Ju}  & WPCN with H-AP &  SISO & SISO & To maximize system throughput by jointly optimizing the time allocation and power allocation in a full-duplex system and a half-duplex system & Convex optimization & Theoretical analysis, numerical simulation  \\
\hline
 \footnotesize  Ju {\em et al} \cite{2014H.Ju}  & WPCN with H-AP & MISO & SIMO & To maximize the weight sum system throughput by jointly
 optimizing the time allocation and power allocation; to maximize the common-throughput
 by optimizing the time allocation & Convex optimization  & Theoretical analysis, numerical simulation  \\
\hline
 \footnotesize  Che {\em et al} \cite{L.1409.3107Che}   & WPCN with H-AP & SISO & SISO &  To maximize system spatial throughput by jointly optimizing time allocation and the power allocation  & Two search algorithms & Theoretical analysis, numerical simulation \\
\hline
 \footnotesize  Tabassum {\em et al}  \cite{H.2015Tabassum}  & WPCN with H-AP & SISO &　SISO &  To improve the spectral efficiency gains and fairness of conventional user scheduling algorithms & Two energy harvesting-constrained
 user scheduling schemes & Theoretical analysis, numerical simulation \\
\hline
 \footnotesize  Liu {\em et al}  \cite{L.2014Liu}  & WPCN with H-AP & MISO & SIMO & To maximize the minimum throughput among users via a
 jointly optimizing time allocation, power allocation and energy beamforming & Alternating optimization technique, two suboptimal designs based on convex optimization &  Numerical simulation  \\
\hline
 \footnotesize  Yang {\em et al} \cite{G.1403.3991Yang}  & WPCN with H-AP & MISO & SIMO & To maximize the minimum throughput among users by jointly optimizing time allocation and power allocation & Numerical search & Theoretical analysis, numerical simulation \\
\hline
 \footnotesize  Zhao {\em et al} \cite{F.2015Zhao} &  WPCN with separated E-AP and D-AP & SISO  & SISO & To maximize system throughput by balancing the time duration between the wireless power transfer phase and the information transfer phase  & Local searching & Numerical simulation  \\
\hline
 \footnotesize  Zhou {\em et al} \cite{X.2014Zhou}  &  WPCN with separated E-AP and D-AP & SISO & SISO & To maximize the achievable rate at the D-AP by jointly optimizing the power allocation at both links  & A water-filling algorithm & Theoretical analysis, numerical simulation \\
\hline
 \footnotesize  Sun {\em et al} \cite{Q.2014Sun} &   WPCN with separated E-AP and D-AP & MISO   & SISO  & To maximize the system throughput via joint time allocation and beamforming design & Successive convex approximation, convex optimization & Theoretical analysis, numerical simulation  \\
\hline
 \footnotesize  Chen {\em et al} \cite{H1404.4120Chen}  & WPCN with relay and H-AP & SISO & SISO & To design a communication protocol for WPCNs with relay and H-AP & An energy harvesting and data transmission protocol & Theoretical analysis, numerical simulation  \\  
\hline
 \footnotesize  Chen {\em et al} \cite{H.2014Chen} &  WPCN with relay and H-AP & SISO & SISO & To maximize the system throughput by optimizing the time and power allocation &  Non-convex optimization & Theoretical analysis, numerical simulation \\
\hline
 \footnotesize Yang {\em et al} \cite{G.YangTSP}  & WPCN with multi-antenna E-AP & MISO & Nil & To maximize the harvested energy in dynamic-length preamble and fixed-length preamble cases &  Dynamic optimization & Theoretical analysis, numerical simulation \\
\hline
 \footnotesize  Zeng {\em et al} \cite{Y.1403.7870Zeng} &  WPCN with multi-antenna E-AP & MIMO & Nil & To find optimal training design, including the number
 of receive antennas to be trained, training
 time and allocated power & Convex optimization  & Theoretical analysis, numerical simulation \\
\hline
 \footnotesize Zeng {\em et al} \cite{Y.Zeng2015} &   WPCN with multi-antenna E-AP & MISO & Nil & To maximize the net harvested energy
 at the user device & Convex optimization  & Theoretical analysis, numerical simulation \\
\hline
 \footnotesize   Jadidian {\em et al} \cite{J.2014Jadidian} &  WPCN with multi-antenna E-AP & MIMO & Nil & To maximize the power transfer efficiency & A beamforming design for 
 non-radiative magnetic field & System-level Simulation, Experiment \\
\hline
 
\end{tabular}
\end{table*}

\section{Mobile Wireless Charger Dispatch Strategies}


From this section, we will begin to review the network applications of wireless charging. The design issues can be broadly classified as mobile charger dispatching, static charger scheduling and wireless charger deployment, which will be introduced in a sequence in the following.

We firstly give a general introduction of the mobile charger dispatch problem. The problem is to schedule the travel of one or multiple mobile chargers so that they visit and recharge a collection of target devices, e.g., with wireless energy harvesting capability. The goal is to prolong the network lifetime. Typically, this problem is studied in the context of WRSNs~\cite{Y.2015Yang}.

Generally, there are five issues to be addressed in the design of charger dispatch problems:
\begin{itemize}

\item Given a number of distributed devices and their locations, we have to obtain the best charging locations for a mobile charger to visit so that wireless charging can cover all the devices.     

\item Given a number of charging locations for a mobile charger to visit, we have to determine an optimal travel path (sequence) for the charger to visit all the locations so that certain goal(s) can be achieved. 
 
\item Given a number of sojourn locations for a mobile charger to visit, we have to obtain an optimal charging duration for the charger to dwell in each location so that none of the devices is under-charged. 

\item Given a number of devices, their locations and data flow requirement, we have to obtain the best data flow rates and data routing paths for the devices so that the overall data gathering performance is optimized. 

\item In the context of collaborative energy provisioning with multiple chargers, we have to determine the minimum number of chargers to be deployed to meet a certain objective (e.g., minimum cost). 

\end{itemize}

The above five issues, respectively, involve the optimization of charging location, travel path, charging time, data rate and routing path, as well as the number of chargers. 

\begin{figure*} 
\centering
\subfigure [Separated wireless energy provisioning and data gathering  ] {
 \label{path_planning_separated}
 \centering
 \includegraphics[width=0.65 \textwidth]{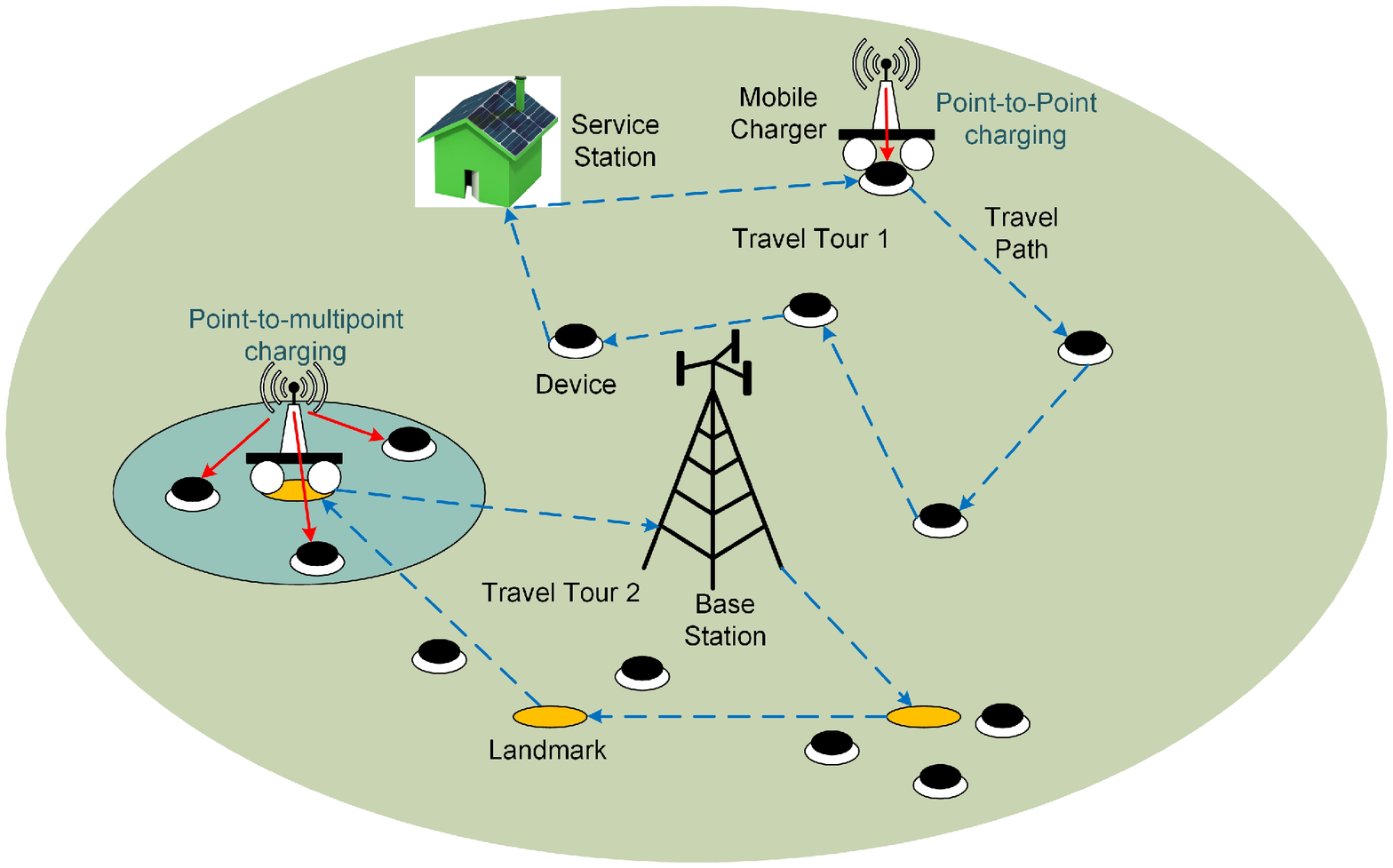}}  
 \centering
 \subfigure [Joint wireless energy provisioning and data gathering] {
 \label{path_planning_joint}
 \centering
\includegraphics[width=0.65  \textwidth]{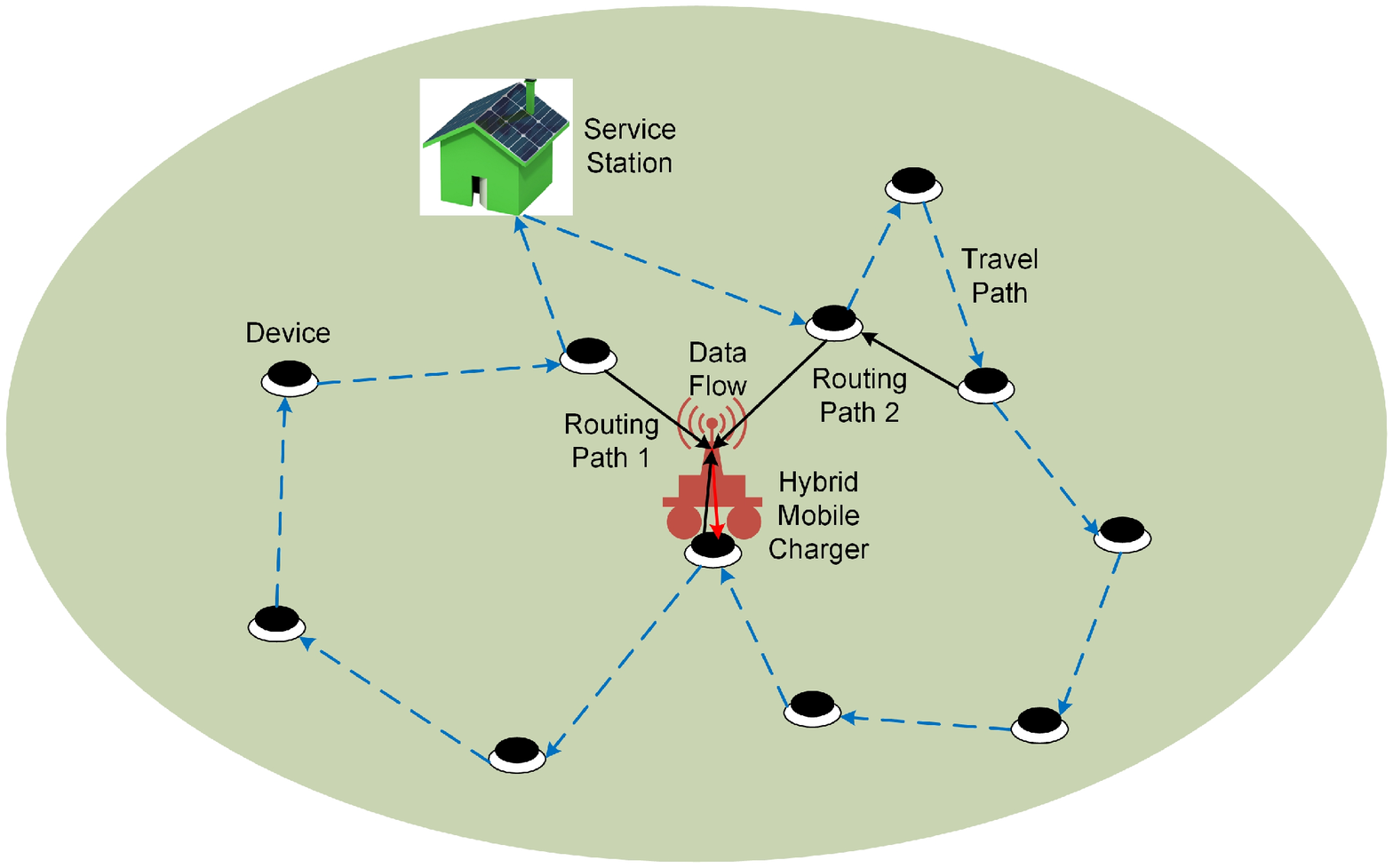}}\\
\centering
\caption{Reference models of mobile charger dispatch.} 
\label{path_planning}
\end{figure*}

In Figure~\ref{path_planning}, we demonstrate two typical system models considered in literature for the mobile charger dispatch planning. In the first model, as shown in Figure~\ref{path_planning_separated}, while wireless charging is performed by the mobile charger(s), data gathering is done by a data sink (or a base station). Thus, the data flow routing and energy consumption rate of network devices do not depend on the movement of the charger. Usually, a charger is sent out from a service station or a data sink. After each travel tour, the charger returns to the service station and receives energy to replenish its battery. A mobile charger can adopt point-to-point charging or point-to-multipoint charging technology, examples of which are shown as Travel Tour 1 and Travel Tour 2 in Figure~\ref{path_planning_separated}, respectively. For point-to-multipoint charging scenario, the charger can transfer energy to multiple target devices within its charging range concurrently at a selected landmark location (also referred to as an anchor point in~\cite{S2014Guo, M2014Zhao}). 

The second system model, as shown in Figure~\ref{path_planning_joint}, employs a hybrid charger that can perform both data collection/forwarding and wireless power transfer. Data can be forwarded to the hybrid charger when it visits a charging location, either in a single-hop or multi-hop fashion shown as Routing Path 1 and Routing Path 2 in Figure~\ref{path_planning_joint}. Conventionally, mobile data collection in wireless sensor networks has been extensively studied, referring to the literature survey in~\cite{D.2011Francesco}. Differently, in this second reference model, wireless energy provisioning and data gathering are jointly optimized. In this case, dynamic routing is required due to time-varying charger's location.

\begin{figure}
\centering
\includegraphics[width=0.5\textwidth]{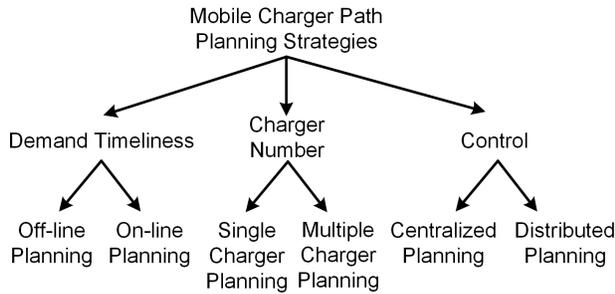}
\caption{Taxonomy of mobile charger dispatch strategies.} \label{taxonomy_planning}
\end{figure}

Figure~\ref{taxonomy_planning} illustrates the taxonomy of mobile charger dispatch strategies. From the perspective of timeliness of demand, the strategies can be classified as offline and online dispatch planning. Alternatively, the strategies can be classified as single-charger and multiple-charger strategies. Based on the control structure, they can be divided into centralized and distributed approaches. In the following subsections, we review the offline and online strategies. Within each subsection, we first present the works with a single charger and then the works with multiple chargers. Moreover, we summarize the strategies in tables and indicate whether each one is centralized or distributed.

\subsection{Offline Charger Dispatch Strategy}

Most of the existing works focus on an offline scenario, in which the energy replenishment scheduling is performed in a deterministic and periodic fashion. 


\subsubsection{Single-Charger Strategy}

The majority of the single-charger strategies~\cite{L2012Xie,L2013Fu,Z2013Qin,L.2014Shi,L2013Xie,LXie2013,L2014Xie} target on minimizing the total service time of the charger (including travel and charging), typically under: 1) the total time constraint for each duty cycle, 2) energy flow constraint that the charged power and consumed power balance at each charging node,  and 3) energy constraint that the energy level of each node always maintains above a certain threshold. This objective is equivalent to: i) the maximization of charger's vacation time~\cite{Z2013Qin}, ii) the maximization of the ratio of charger's vacation time to cycle time in~\cite{L2012Xie}, and iii) the minimization of the charger's energy consumption in~\cite{LXie2013} under the same constraints.   

The authors in~\cite{L2012Xie} first introduced the concept of renewable energy cycle where the residual energy level in a device exhibits some periodicity over a time cycle. Both the necessary and sufficient conditions for renewable energy cycle to achieve unlimited network lifetime were provided. Then, the authors theoretically proved that an optimal travel path for the charger to sustain the renewable energy cycle is the shortest Hamiltonian cycle (SHC). Typically, an SHC can be obtained by solving the well-known Traveling Salesman Problem (TSP)~\cite{M.1991Padberg}, which is non-deterministic polynomial-time hard (NP-hard) in general.
However, although it is NP-hard, the optimal travel path for a TSP with thousands of points can be solved quickly e.g., by applying the technique in~\cite{M.1991Padberg}  or the tool in~\cite{Concorde}. Based on the resulted optimal travel path, a non-linear optimization problem for joint charging duration and data flow routing was formulated and shown to be NP-hard. By adopting a piecewise linear approximation technique, the authors derived a feasible solution and validated its near-optimality through both theoretical proof and numerical results.

Similar to~\cite{L2012Xie}, the authors in~\cite{L.2014Shi} also developed a non-linear programming problem to optimize the travel path, charging duration and data flow routing jointly. The difference is that the flow routing was assumed to be invariable in~\cite{L2012Xie}, while dynamic time-varying flow routing was considered in~\cite{L.2014Shi}.  
By applying linearizion techniques, the authors reformulated the original problem as a linear programming (LP) model that can be solved within polynomial time. Simulation results demonstrated that compared with static data routing, the proposed strategy yields much larger objective value and incurs lower complexity. Moreover, any single static data routing may result in an infeasible solution, because some nodes would deplete their energy before being charged in the next cycle.

Unlike the above two studies, the authors in~\cite{L2014Xie,LXie2013,L2013Xie} further involved the selection of charging locations. As point-to-point charging was assumed in~\cite{L2012Xie}, the study in~\cite{L2014Xie} extended~\cite{L2012Xie} by investigating point-to-multipoint charging. The formulated  non-linear programming (NLP) problem was shown to be NP-hard. By applying discretization and a reformulation-linearizion-technique~\cite{D.1998Sherali}, the NLP was first converted to a mix-integer NLP (MINLP) and then a mixed-integer linear programming (MILP). The designed solution was proven to be near-optimal. The numerical results also highlighted a considerable performance gap between point-to-point and point-to-multipoint charging scenarios.

Different from~\cite{L2012Xie,L.2014Shi,L2014Xie}, which adopted a separated mobile charger and static base station for wireless energy provisioning and data gathering, the system models in~\cite{LXie2013} and~\cite{L2013Xie},
considered them jointly. In these references, the data generated from devices is forwarded toward a hybrid charger in a multihop fashion. As the location of the charger changes over time, the data flow routing needs to be optimized dynamically.
Extending from~\cite{L2014Xie} by considering a hybrid charger, the authors in~\cite{LXie2013} developed the time-dependent optimization problem because of the dynamic data flow routing. Interestingly, the authors considered the special case that involves only location-dependent variables. This special case has the same optimal objective value and offers a solution space completely enclosed in that for the original problem. Consequently, a near-optimal solution to the special case problem was proposed and proven to be near-optimal. However, this work assumed that the travel path for the charger was known a priori. 

Then, the study in~\cite{LXie2013} was further extended in~\cite{L2013Xie} by investigating the case with an unknown travel path. However, this increases the complexity of the problem substantially. The authors first addressed an ideal case assuming zero traveling time for the charger. By adopting the discretization and logic point representation techniques, a provably near-optimal solution was obtained for any level of accuracy. Based on this solution, the authors further obtained the travel path of the original problem by finding the shortest Hamiltonian cycle. This cycle connects all the logical points that have non-zero sojourn time in the ideal case. With this travel path, a feasible solution was further derived. Moreover, the performance gap between the feasible solution and optimal solution was theoretically characterized.  

The path planing strategies proposed in~\cite{Z2013Qin} and~\cite{L2013Fu} also assumed point-to-multipoint charging. Based on the finding in~\cite{L2014Xie}, the proposed strategy in~\cite{Z2013Qin} adopted the shortest Hamiltonian cycle as a charger travel path and focused on optimizing the charging duration of each stop through a dynamic optimization model. The authors in~\cite{L2013Fu} formulated an LP model to optimize the charger's charging location and the corresponding duration. It was shown that significant reduction of searching space for an optimal solution can be achieved by utilizing the smallest enclosing space~\cite{E.1991Welzl} and charging power discretization. To further reduce the complexity, a heuristic approach based on the k-means clustering algorithm, called Lloyd's algorithm~\cite{S.1982Lloyd}, was introduced to merge the charger stop locations while keeping the charging delay under a limit. The simulation showed that the heuristic approach reaches a close-to-optimal performance and largely outperforms a set-cover-based approach~\cite{N.2006Alon}, that maximizes the number of under-charged devices nearby each stop.

The authors in~\cite{J2014Wang} and~\cite{H2013Dai} considered target-oriented WRSNs, where wireless charging strategies were jointly optimized with sensor activation for target monitoring. In particular, as each target monitoring induces the same information irrespective of the number of sensors that captured it, sensor activation scheduling is required for coordinating the sensors so as to avoid redundant monitoring. In~\cite{J2014Wang}, the authors formulated the problem to maximize the average number of targets monitored. The problem was shown to be NP-complete. Consequently, a greedy algorithm and a random algorithm were designed to compromise between computation complexity and performance. The simulation showed that the greedy strategy achieves comparable performance with the random strategy when the charger moves slowly. It outperforms the random strategy with the increase of charger's velocity. However, the performance gaps between the optimal solution and the proposed algorithms were not investigated.

The studies in~\cite{H2013Dai} and~\cite{F.2011Jiang} dealt with the problem to optimize the quality of monitoring (QoM). The QoM was defined as average information gained per event monitored by sensor networks. The authors in~\cite{F.2011Jiang} introduced a simple strategy, namely Joint Periodic Wake-up (JPW), which jointly dispatches a mobile charger to visit and charge nearby sensors at points of interest (PoI) within a predefined charging duration. Moreover, the charger can control a duty cycle of the sensors. The performance evaluation showed the effectiveness of charging duration on the QoM performance, however, failed to quantify the performance gap between JPW and the optimal solution. 

The authors in~\cite{H2013Dai} considered the optimization problem to maximize QoM. As the formulation of the problem was shown to be NP-hard, the authors first proposed a relaxed problem which ignores the travel time of the charger. By reformulating the relaxed problem as a monotone submodular function maximization problem under a special sufficient condition, the first algorithm was designed to achieve 1/6 approximation for the relaxed problem. Then, based on the results obtained by the approximation algorithm, the second algorithm was introduced for the original problem. Both the order of approximation and time complexity of the two proposed approximation algorithms were theoretically derived. Compared with the Joint Periodic Wake-up algorithm in~\cite{F.2011Jiang} using simulations, the second algorithm was demonstrated to obtain considerable performance gain.

Both~\cite{M2014Zhao} and~\cite{S2014Guo} aimed to maximize network utility functions that characterize an overall data gathering performance. 
In~\cite{M2014Zhao}, the authors devised a two-step strategy for the joint design problem. The first step involves selection of a subset of sensors to be an anchor point, while the second step is to optimize data gathering when a mobile charger moves among the selected anchor point. Furthermore, the authors provided a selection algorithm to search for the maximum number of sensors with the least energy level as the anchor points, while keeping the tour length of the charger below a threshold. Next, the authors developed an NP-hard flow-level network utility maximization model, and devised a distributed algorithm to obtain a system-wide optimal solution 
(proved in \cite{P.1989Bertsekas}) in a distributed manner.  The simulation verified the convergence of the proposed strategy and its effectiveness under different topologies. However, the charging duration of devices was ignored. Additionally, energy consumption for data receiving and sensing was not taken into account. 

The authors in~\cite{S2014Guo} extended~\cite{M2014Zhao} by considering heterogeneous device's energy consumption and time-varying charging duration. The formulation of this problem, under flow conservation, energy balance, link and battery capacity and charging duration limit, was shown to be non-convex. By employing some auxiliary variables, the authors were able to convert the original formulation into a convex one, and decompose the problem into two levels of optimization. The decomposed optimization was solved by a distributed cross-layer strategy, which adaptively adjusts the device's optimal data, routing paths, instant energy provisioning status, and charging duration to maximize the network utility. The NS-2 simulation~\cite{NS2} showed the fast convergence of the proposed strategy and robustness to small level of node failure. It was shown to outperform the strategy proposed in~\cite{M2014Zhao} in terms of network utility and lifetime. One shortcoming of this strategy is that it neglects the energy constraint of the mobile charger.

The above single-charger strategies were all based on the assumption that the mobile charger has sufficient (or infinite) energy capacity to visit and charge an entire network, at least within each tour. However, a more realistic problem is to design charger dispatch strategies for a mobile charger with limited capacity. Thus, the authors in~\cite{Y2010Peng} and~\cite{K2012Li} took the energy constraint of the mobile charger into account. The aim of~\cite{Y2010Peng} was to find an optimal travel path that the network lifetime is maximized. The authors showed the NP-completeness of the developed charging problem and designed two heuristic algorithms to reduce the computation overhead. Under a given charger's battery capacity, the first one attempts to prolong the network lifetime as much as possible, while the second one improves the first one by employing binary search to find more suitable target network lifetime. 

Another strategy introduced in~\cite{K2012Li} was devised to maximize the number of devices that can be charged, subject to the constraint of charger's total energy consumption for both traveling and charging. Under the assumption of multiple-node charging, the strategy optimizes the charging location selection to reduce the tour length. The authors proved the NP-hardness of this problem and proposed heuristic solutions based on the meta-heuristic of particle swarm optimization (PSO)~\cite{M.2007Clerc}. Simulation results showed that the PSO-based solutions achieve a small gap between the heuristic and optimal TSP solution. However, the required number of iterations is significantly larger for the heuristic case.    
 
From the above reviewed literature, we observe that majority of the work dealt with energy provisioning of static devices with mobile charger(s). The key feature in this case is that two important factors, i.e., the charging delay of to-be-charged devices and travel distance of the charger, are interrelated. By contrast, reference~\cite{2014L.He} was a pioneering work that explores the dispatch planning for a mobile charger to replenish mobile devices. The difference lies in that the aforementioned two factors might conflict with each other. The authors designed a tree-based strategy to minimize the travel distance while maintaining charging delay within an acceptable level,
given the travel profile of all devices. Using a queue-based approach, the authors also identified an energy threshold according to which the device requires for energy transfer. Both analytical and simulation results indicated that the tree-based strategy approaches the optimal solution when the speed or the requested charging duration increases. The limitation of this strategy is that it works only when the routes of the network devices were preplanned.

\subsubsection{Multiple-Charger Strategy}

The multiple-charger strategy is to dispatch mobile chargers from a common or several distributed service stations to visit a collection of target devices collaboratively. Compared with the single-charger dispatch problem, multiple-charger dispatch further involves the coordination among mobile chargers.
Therefore, the design of multiple-charger strategy usually entails two steps: minimization of the number of chargers given a charging coverage requirement, and scheduling of the optimal dispatch planning given the minimum number of chargers. 

The majority of the multiple-charger strategies consider point-to-point charging. References~\cite{R.2014Beigel} and~\cite{J.2014Wu} investigated a one-dimensional (1D) linear WRSN with negligible charging time. Both of the works aim to minimize the number of chargers for maintaining the operation of the networks. The authors in~\cite{R.2014Beigel} first provided an optimal solution with linear complexity in searching for the minimum number of chargers and corresponding dispatch planning in a homogeneous charging scenario, i.e., the charging frequency for all devices is identical. Then, for heterogeneous charging with different charging frequencies, the authors designed a greedy algorithm which is shown to have a factor of two optimal solutions by both mathematical proof and simulation. However, the chargers were assumed to have infinite battery capacity. Additionally, the proposed solutions were only examined in a small network up to 10 devices. Different from~\cite{R.2014Beigel}, the study in~\cite{J.2014Wu} assumed the energy limit on the chargers. The authors first discussed different approaches when each sensor is allowed to be charged by a single charger, and jointly by multiple chargers, as well as when mobile chargers are enabled to charge each other. Then, an optimal solution to minimize the number of chargers was proposed for the case that allows inter-charger charging. This solution was also shown to achieve the maximum ratio of energy consumed for charging and that for traveling. However, the proposed strategies were restricted for linear and ring topologies.

As opposed to the above two studies, references~\cite{H.Dai2014},~\cite{W.2014Xu} and~\cite{W.2014Liang} considered two-dimensional (2D) WRSNs with energy-constrained chargers. In~\cite{H.Dai2014}, the formulation to minimize the number of chargers in a 2D network was proven to be NP-hard. To solve this problem, an approximation algorithm was first proposed for a relaxed version of the original problem, i.e., by removing a linear constraint. Then, based on the obtained results from the relaxed problem, the authors devised two approximation algorithms for the original problem and derived the order of approximation for both algorithms. Simulation results demonstrated the advantage of the two approximation algorithms over a baseline algorithm; however, they still have considerable performance gap compared with the optimal solution. Additionally, one of the shortcomings of this study was that it only applies for the case where the energy consumption rate of all devices are identical.   

Reference~\cite{W.2014Xu} aimed to minimize the sum of traveling distance of all chargers. This can be formulated as a \emph{q-root} TSP which is to find $q$ closed tours covering all locations such that the total length of the $q$ tours is minimized. Due to the NP-hardness of this problem, the authors proposed an approximation algorithm with a provable 2-approximation ratio under the assumption that the energy consumption rates of all devices are fixed. The basic idea is to find $q$-root trees with the minimum distance and to transform each tree into a closed tour with the length of each tour being not more than twice of the corresponding tree. Then, for the case with heterogeneous energy consumption rates, a heuristic algorithm was developed. The simulation demonstrated the superiority of the proposed algorithm over a greedy baseline algorithm. However, again the performance gap compared to the optimal solution was unknown. 

In the similar context, reference~\cite{W.2014Liang} also developed a \emph{q-root} TSP to schedule multiple chargers, while further aiming to minimize the number of deployed chargers. The considered problem was solved by a two-step design. The authors first introduced a tree decomposition algorithm similar to that in~\cite{W.2014Xu} with a provable 5-approximation ratio to find $q$ closed tours. Then, by bounding the total distance of each tour, an approximation algorithm that invokes the first algorithm was proposed to minimize the number of chargers. Evaluated in the networks with linear and random distributions of energy consumption rate, the proposed strategy was shown to achieve a $40\%$ performance gap with the optimal solution. However, a drawback of this strategy is that its complexity grows exponentially with the number of to-be-charged devices.

Reference~\cite{C2013WangCoordination} considered a time constraint for each charger's travel duration.  With the aim to minimize the total traveling cost while inducing no node outage, a multiple TSP with deadlines was formulated and shown to be NP-hard. To reduce computational overhead, the authors devised a heuristic algorithm which selects the nodes to recharge according to the weighted sum of travel time and residual lifetime of sensor nodes. Furthermore, the complexity of the heuristic algorithm was derived. The simulation results validated the effectiveness of the proposed algorithm, however, ignored performance gap from the optimal solution.

The authors in~\cite{2015I.Farris} attempted to provide solutions to handle joint energy replenishment and data gathering in large-scale WISPs.
In the first approach, the data is stored in the RFID tags temporarily and later collected and forwarded to the data sink through the readers. In the second approach, data is forwarded to the readers on a real-time fashion. 
Both approaches first cluster the WISP nodes based on the energy constraints of the system and then optimize the movement tour for the involved RFID readers under the energy and time constraints of the WISP nodes.
It is shown that the proposed approaches always guarantee a feasible solution. Moreover, the second approach offers better delay performance.

While all the above multiple-charger strategies adopted centralized control, the focus of~\cite{A2013Madhja} is to investigate distributed control with local information. The aim was to explore the tradeoff between the charging performance and the amount of information available. The authors proposed two distributed strategies, in both of which each charger chooses the travel path to move based on the information about the status of its neighboring chargers.  The difference lies in that the first strategy assumes no network information, while the second one operates with local knowledge. Simulation illustrated that the first distributed strategy achieved comparable performance with its centralized counterpart, which was inferior to the second distributed strategy. From this algorithm-related observation, the authors claimed that for the situation with limited network knowledge, the coordination among the mobile chargers may be less crucial than the design of the travel path.

\subsection{Online Charging Dispatch Strategy}

Most of the literature introduced in the previous subsection was based on the assumption that the mobile charger operates with perfect global knowledge. However, in practice, the acquisition of global knowledge incurs large communication overhead and considerable power consumption. Moreover, operated based on a priori information, the offline strategies are vulnerable to any change of network condition. Consequently, in real systems where variation and uncertainty in network demand normally exist, the offline charger dispatch strategies lack adaptability and suffer from substantial performance degradation. To address this issue, an online strategy can be designed for real-time charging. In other words, the online strategy allows a mobile charger to receive new charging requests at any time instant, and the strategy constructs and adjusts the charger's travel path in an on-demand basis. In the following, we review the online charger dispatch strategies. 





The majority of research efforts on online strategy focused on the single-charger dispatch strategies, within which,~\cite{X.2014Ren} and~\cite{L.He2014} work in a centralized fashion. Reference~\cite{X.2014Ren} was to maximize the network charging throughput per travel tour. The offline formulation of this problem under the assumption that all charging requests were known in advance was first shown to be NP-hard and solved by an offline approximation algorithm. Then, for the online version with one-by-one arrived charging requests, a naive strategy was proposed to re-plan iteratively the travel path by always serving the request with the smallest processing time, which is the sum of traveling time and charging time. Furthermore, the case with point-to-multipoint charging was also analyzed. The authors introduced a cluster-based algorithm. The algorithm groups the requesting sensors into different clusters according to their locations. The charger then evaluates the cluster with a newly defined metric called a charging gain, and uses the heuristic algorithm to serve the cluster with the highest charging gain. Nevertheless, both of the proposed algorithms are highly location-biased, which provides little chance of energy transfer for the devices far away from the charger. 

In~\cite{L.He2014}, the authors devised an energy synchronized charging (ESync) protocol with an aim to reduce both travel distance and charging delay. Considering on-demand energy provisioning, a collection of nested TSP tours is constructed by only involving the devices with low residual energy. To further optimize the travel tour, the concept of energy synchronization is adopted to harmonize the charging sequence of the devices. The travel tour construction is dynamically adjusted based on the request sequence to synchronize the devices in each charging round. The efficiency of ESync in reducing the traveling distance and charging delay was verified by both experiment and simulation. 

Different from~\cite{X.2014Ren} and~\cite{L.He2014}, the authors in~\cite{M2013Angelopoulos},~\cite{L.HeTMC} and~\cite{L.2014Jiang} concentrated on the design of distributed strategies.
Reference~\cite{M2013Angelopoulos} considered the energy provisioning for a circular network with devices uniformly distributed at random. Different from the above centralized online strategies, the authors proposed a distributed and adaptive strategy that requires only local information. Under the assumption that all sensors have the same data rate, the charger tries to choose the travel path that the charger's battery depletes at the fastest rate, clearly influenced by the adopted data routing protocol. Additionally, a partial charging scheme that determines the amount of energy to transfer was shown to be optimal in the number of alive devices. In general, an algorithm based on global information should outperform its counterpart relying on local information. The proposed strategy in this study was shown by simulation to even outperform some strategies relying on global information in some cases. However, the charging duration was neglected in this study.

The strategy in~\cite{L.HeTMC} was based on the nearest-job-next with preemption discipline that takes both spatial and temporal properties of the incoming charging requests into consideration. The basic idea is to trigger the re-selection of the next to-be-charged node upon either the charging completion of a device or the arrival of a new charging request. The charger then chooses the spatially closest requesting node to serve. The performance bounds of throughput as well as charging delay were analyzed. Both numerical and system-level simulations showed that the proposed strategy outperforms the first-come-first-serve discipline. However, similar to~\cite{X.2014Ren}, the proposed strategy is location-biased which results in unfairness for wireless power distribution. Another drawback is that the proposed strategy was evaluated only in terms of charging throughput and delay. Its performance in other metrics, such as charging coverage and performance gap between optimal solutions, were not analyzed.  

Reference~\cite{L.2014Jiang} explored online multiple-charger strategy. The authors aimed to maximize the charging coverage with on-demand scheduling in an event monitoring WRSN, and proved that this problem is NP-complete. Then, the two metrics were introduced. The first metric was incremental effective coverage (IEC) which was defined to represent the set of point of interests~\cite{H.2010Tan}. The second metric is trail covering utility (TCU) which was the average coverage utility during the charging time of the sensor. Three greedy heuristic algorithms that serve to-be-charged devices based on maximum IEC, maximum average TCU, and maximum average TCU with multiple chargers were proposed. The first two algorithms were evaluated to achieve comparable performance in terms of charging coverage. For the third algorithm, simulation characterized the tradeoff between charging coverage and the number of chargers deployed. However, the third algorithm lacks efficient coordination strategies among multiple chargers. As every charger just broadcasts the information to all the other chargers after device charging was completed, it may result in invalid travel distance of the other chargers.
    
\subsection{Discussion and Conclusion}



In Table~\ref{offline_path_planning}, we summarize the reviewed offline dispatch strategies. The reviewed literature is compared in terms of the number of chargers applied, the energy constraint of the charger(s), optimization variables in the proposed strategies, charging patterns (point-to-point or point-to-multipoint charging), control methods (centralized or distributed) and evaluation methods. As for evaluation methods, there are four typical approaches, namely, numerical simulation, system-level simulation, theoretical analysis and experiment. As we can observe from Table~\ref{offline_path_planning}, most of the existing works rely on centralized control to schedule mobile charger(s). Distributed algorithms have been less studied, especially for multiple-charger dispatch strategies. Moreover, all the existing multiple-charger dispatch strategies employ point-to-point charging. Future work may incorporate landmark selection to reduce the length of travel tours for multiple chargers. 

Table~\ref{online_path_planning} presents the summary of the online dispatch strategies. Specifically, we compare the related literature in terms of objectives, number of chargers applied, energy constraint of the charger(s), charging patterns, control methods and evaluation methods. 
Only reference~\cite{L.2014Jiang} has provided the solution for multiple chargers. However, as aforementioned, efficient coordination among chargers is missing. How to coordinate chargers for online strategy, especially with distributed control, is challenging. Moreover, how to manage multiple charging requests by utilizing point-to-multipoint charging can be a future direction for online algorithms.   

\begin{table*}    \footnotesize
\centering
\caption{\footnotesize Summary of the offline mobile charger dispatch strategies, where ``charging pattern (CP)'', ``point-to-point charging (PPC)'', ``point-to-multipoint charging (PMC)'', ``numerical simulation (NS)'', ``system-level simulation (SS)'' and ``theoretical analysis (TA)'', respectively} \label{offline_path_planning}
\begin{tabular}{|p{2.2cm}|p{1cm}|p{1.3cm}|p{0.6cm}|p{1.0cm}|p{1.0cm}|p{0.9cm}|p{0.9cm}|p{0.6cm}|p{0.6cm}|p{1.3cm}|p{1.2cm}|}
\cline{1-12}
   & \multicolumn{2}{ |c|  }{\multirow{1}{*}{Charger} } & \multicolumn{6}{ c| }{Optimization Variable} &  & &\multicolumn{1}{ |c|  }{\multirow{1}{*}{Performance} }  \\ 
\cline{2-9}
\multicolumn{1}{ |c|  }{Literature}  & Number & Energy constraint & Travel Path & Charging location & Charging duration & Charger number & Data routing & Data rate & CP & Control & Evaluation \\ 
\cline{1-12} 
\cline{1-12} 
Zhao \emph{et al}~\cite{M2014Zhao}    & Single  & \multicolumn{1}{ |c|  }{No} &  \multicolumn{1}{ |c|  }{\checkmark} &  \multicolumn{1}{ |c|  }{\checkmark}  &   &  &  \multicolumn{1}{ |c|  }{\checkmark}  & \multicolumn{1}{ |c|  }{\checkmark} & PPC & Distributed & \multicolumn{1}{ |c|  }{TA, NS}  \\ 
\cline{1-12}  
Guo \emph{et al}~\cite{S2014Guo}    & Single  & \multicolumn{1}{ |c|  }{No} & \multicolumn{1}{ |c|  }{\checkmark} & \multicolumn{1}{ |c|  }{\checkmark} & \multicolumn{1}{ |c|  }{\checkmark}  &    &  \multicolumn{1}{ |c|  }{\checkmark} & \multicolumn{1}{ |c|  }{\checkmark} & PPC & Distributed & \multicolumn{1}{ |c|  }{TA, SS}   \\ 
\cline{1-12}
Xie \emph{et al}~\cite{L2012Xie}  & Single & \multicolumn{1}{ |c|  }{No} & \multicolumn{1}{ |c|  }{\checkmark}  & & \multicolumn{1}{ |c|  }{\checkmark}  &   &  \multicolumn{1}{ |c|  }{\checkmark} & & PPC  & Centralized &  \multicolumn{1}{ |c|  }{TA,NS}  \\
\cline{1-12}                                         
 Shi \emph{et al}~\cite{L.2014Shi}   & Single & \multicolumn{1}{ |c|  }{No} &  \multicolumn{1}{ |c|  }{\checkmark}&  & \multicolumn{1}{ |c|  }{\checkmark}  &  &  \multicolumn{1}{ |c|  }{\checkmark} & &  PPC  & Centralized & \multicolumn{1}{ |c|  }{TA, NS}  \\  
 \cline{1-12}  
 Xie \emph{et al} ~\cite{L2014Xie}   & Single & \multicolumn{1}{ |c|  }{No} &  \multicolumn{1}{ |c|  }{\checkmark} &  \multicolumn{1}{ |c|  }{\checkmark} & \multicolumn{1}{ |c|  }{\checkmark} &  & \multicolumn{1}{ |c|  }{\checkmark}  & & PMC & Centralized & \multicolumn{1}{ |c|  }{TA, NS}   \\
 \cline{1-12} 
  Xie \emph{et al} ~\cite{LXie2013} & Single & \multicolumn{1}{ |c|  }{No} &    & \multicolumn{1}{ |c|  }{\checkmark} & \multicolumn{1}{ |c|  }{\checkmark} &  & \multicolumn{1}{ |c|  }{\checkmark} &  & PMC & Centralized & \multicolumn{1}{ |c|  }{TA, NS}  \\ 
\cline{1-12}Xie \emph{et al}~\cite{L2013Xie} 	  & Single & \multicolumn{1}{ |c|  }{No} & \multicolumn{1}{ |c|  }{\checkmark} & \multicolumn{1}{ |c|  }{\checkmark} & \multicolumn{1}{ |c|  }{\checkmark} &  & \multicolumn{1}{ |c|  }{\checkmark} &  & PMC & Centralized &  \multicolumn{1}{ |c|  }{TA, NS}  \\ 
\cline{1-12} Qin \emph{et al}~\cite{Z2013Qin}  & Single & \multicolumn{1}{ |c|  }{No} & \multicolumn{1}{ |c|  }{\checkmark} & & \multicolumn{1}{ |c|  }{\checkmark} &   &    & & PMC & Centralized &  \multicolumn{1}{ |c|  }{NS}   \\ 
\cline{1-12}
 Fu  \emph{et al}~\cite{L2013Fu} & Single & \multicolumn{1}{ |c|  }{No} & &  \multicolumn{1}{ |c|  }{\checkmark} &   \multicolumn{1}{ |c|  }{\checkmark} &   & &  & PMC & Centralized &  \multicolumn{1}{ |c|  }{NS}  \\ 
 \cline{1-12}
Wang \emph{et al}~\cite{J2014Wang}    & Single & \multicolumn{1}{ |c|  }{No} &  \multicolumn{1}{ |c|  }{\checkmark}& \multicolumn{1}{ |c|  }{\checkmark} &  \multicolumn{1}{ |c|  }{\checkmark} &  &   & &  PPC  & Centralized & \multicolumn{1}{ |c|  }{NS}  \\ 
\cline{1-12} 
Dai \emph{et al}~\cite{H2013Dai}  & Single & \multicolumn{1}{ |c|  }{No} & \multicolumn{1}{ |c|  }{\checkmark} & \multicolumn{1}{ |c|  }{\checkmark} & \multicolumn{1}{ |c|  }{\checkmark} & & & & PPC & Centralized & \multicolumn{1}{ |c|  }{TA, NS}  \\ 
\cline{1-12} 
Jiang \emph{et al}~\cite{F.2011Jiang} & Single & \multicolumn{1}{ |c|  }{No} &  &   & \multicolumn{1}{ |c|  }{\checkmark} & & & & PPC & Centralized & \multicolumn{1}{ |c|  }{TA, NS}  \\ 
\cline{1-12} 
Peng  \emph{et al}~\cite{Y2010Peng}  & Single & \multicolumn{1}{ |c|  }{Yes} & \multicolumn{1}{ |c|  }{\checkmark} &  & \multicolumn{1}{ |c|  }{\checkmark}  &  &  &  &  PPC & Centralized & Experiment \\ 
\cline{1-12} 
 Li  \emph{et al}~\cite{K2012Li}  & Single & \multicolumn{1}{ |c|  }{Yes} & \multicolumn{1}{ |c|  }{\checkmark} &   & \multicolumn{1}{ |c|  }{\checkmark} &  &  & & PMC & Centralized &  \multicolumn{1}{ |c|  }{TA, NS}  \\ 
\cline{1-12} He \emph{et al}~\cite{2014L.He}  & Single & \multicolumn{1}{ |c|  }{No} & \multicolumn{1}{ |c|  }{\checkmark} & \multicolumn{1}{ |c|  }{\checkmark} &   &  &  & & PPC & Centralized &  \multicolumn{1}{ |c|  }{TA, NS}  \\ 
\cline{1-12} Beigel \emph{et al}~\cite{R.2014Beigel}  & Multiple & \multicolumn{1}{ |c|  }{No} &  & &  \multicolumn{1}{ |c|  }{\checkmark} & \multicolumn{1}{ |c|  }{\checkmark} &  & &  PPC & Centralized & \multicolumn{1}{ |c|  }{TA, NS}  \\

\cline{1-12}  Wu \emph{et al}~\cite{J.2014Wu}   & Multiple & \multicolumn{1}{ |c|  }{Yes} &  & &   & \multicolumn{1}{ |c|  }{\checkmark} &  & & PPC & Centralized & \multicolumn{1}{ |c|  }{TA}  \\ \cline{1-12} 
Dai \emph{et al}~\cite{H.Dai2014}   & Multiple & \multicolumn{1}{ |c|  }{Yes} & \multicolumn{1}{ |c|  }{\checkmark} & &   & \multicolumn{1}{ |c|  }{\checkmark} &   & & PPC & Centralized & \multicolumn{1}{ |c|  }{TA, NS}  \\

\cline{1-12}  Liang \emph{et al}~\cite{W.2014Liang} & Multiple & \multicolumn{1}{ |c|  }{Yes} &  & &  &  \multicolumn{1}{ |c|  }{\checkmark} &  & & PPC & Centralized & \multicolumn{1}{ |c|  }{TA, NS}  \\ 
\cline{1-12} 
Wang \emph{et al}~\cite{C2013WangCoordination} & Multiple & \multicolumn{1}{ |c|  }{No} & \multicolumn{1}{ |c|  }{\checkmark} & \multicolumn{1}{ |c|  }{\checkmark} & \multicolumn{1}{ |c|  }{\checkmark} &
\multicolumn{1}{ |c|  }{\checkmark} & & & PPC & Centralized & \multicolumn{1}{ |c|  }{TA, SS}  \\ 
\cline{1-12}
Farris \emph{et al}~\cite{2015I.Farris}      & Multiple & \multicolumn{1}{ |c|  }{No} &  \multicolumn{1}{ |c|  }{\checkmark} &  &   & \multicolumn{1}{ |c|  }{\checkmark}   &  &  & PPC  &  Centralized & \multicolumn{1}{ |c|  }{NS}  \\

\cline{1-12}
 
Madhja \emph{et al}~\cite{A2013Madhja}      & Multiple & \multicolumn{1}{ |c|  }{Yes} &  \multicolumn{1}{ |c|  }{\checkmark} &  &   &   &   &  & PPC  &  Distributed & \multicolumn{1}{ |c|  }{NS}  \\
\cline{1-12}

\end{tabular}
\end{table*}

\begin{table*} \small 
\centering
\caption{\footnotesize Summary of the Online Charger Dispatch Strategies.} \label{online_path_planning}
\begin{tabular}{|p{2cm}|p{1cm}|p{1.3cm}|p{3.5cm}|p{2.1cm}|p{2cm}|p{3cm}|}
\hline
\footnotesize  & \multicolumn{2}{ |c|  }{\bf Charger }  &  &  &  &  \\
\cline{2-3}
\multicolumn{1}{ |c| }{\bf Literature}  & Number & Energy constraint & \multicolumn{1}{ |c|  }{\bf Objective} & \multicolumn{1}{ |c|  }{\bf Charging pattern} & \multicolumn{1}{ |c|  }{\bf Control} & \multicolumn{1}{ |c|  }{\bf Performance Evaluation } \\ 
\hline
\hline
He \emph{et al} ~\cite{L.He2014}  & Single & \multicolumn{1}{ |c|  }{Yes}   &  To mitigate the limit of TSP-based solutions  &  Point-to-point charging & Centralized  & Theoretical analysis, experiment, system-level simulation  \\
\hline
Ren \emph{et al}~\cite{X.2014Ren}   &  Single & \multicolumn{1}{ |c|  }{No}   & Maximization of charging throughput & Point-to-point charging & Centralized  & Theoretical analysis, numerical simulation  \\
\hline
Angelopoulos \emph{et al}~\cite{M2013Angelopoulos}  &  Single & \multicolumn{1}{ |c|  }{No}  & To balance the tradeoff between information knowledge and achieved performance &  Point-to-point charging & Distributed & Numerical simulation \\
\hline
He \emph{et al} ~\cite{L.HeTMC}  & Single & \multicolumn{1}{ |c|  }{Yes}  & To increase charging throughput and latency over first-come-first-serve principle   &  Point-to-point charging & Distributed & Theoretical analysis, numerical simulation \\
\hline
Jiang \emph{et al}~\cite{L.2014Jiang}  & Multiple & \multicolumn{1}{ |c|  }{No} &  Maximization of charging coverage  & Point-to-point charging & Distributed & Numerical simulation \\
\hline
\end{tabular}
\end{table*}

\section{Wireless Charger Deployment Strategies}





\begin{figure*} 
\centering
\subfigure [Point Provisioning Scenario] {
 \label{point_provisioning}
 \centering
 \includegraphics[width=0.42 \textwidth]{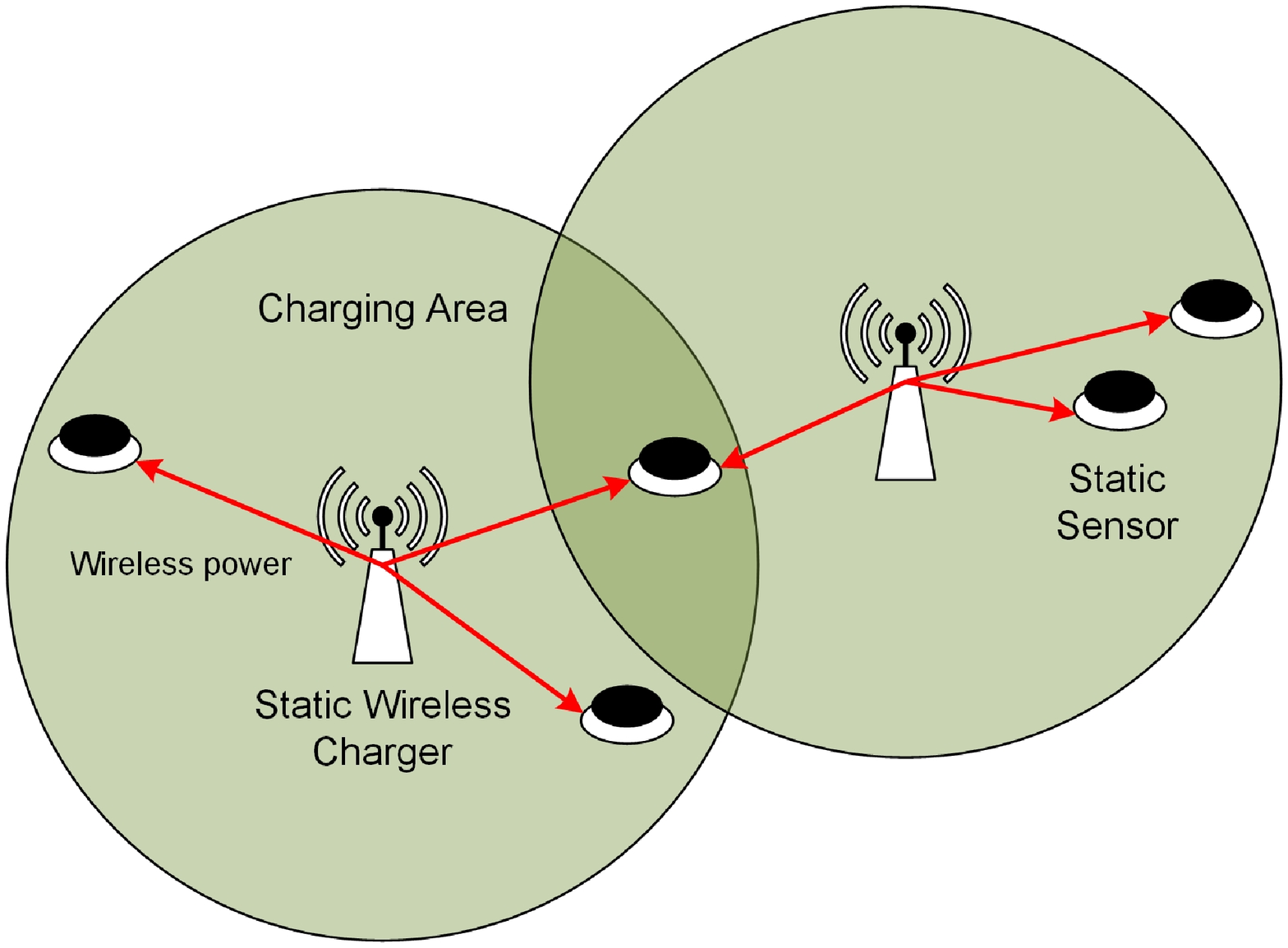}}  
 \centering
 \subfigure [Path Provisioning Scenario] {
  \label{path_provisioning}
  \centering
  \includegraphics[width=0.45 \textwidth]{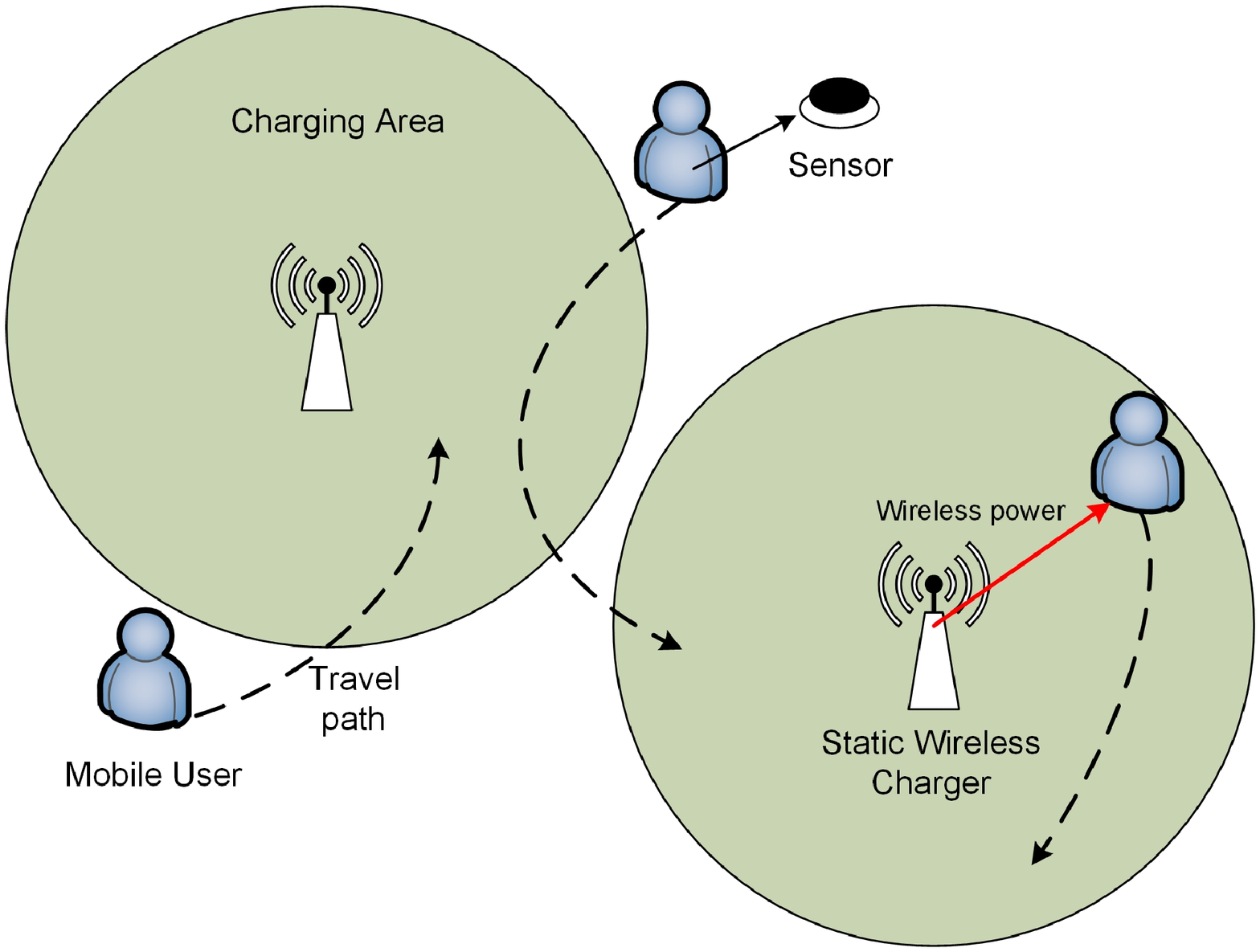}}   
  \centering
 \subfigure [Multihop Provisioning Scenario] {
  \label{multihop_provisioning}
  \centering
  \includegraphics[width=0.43 \textwidth]{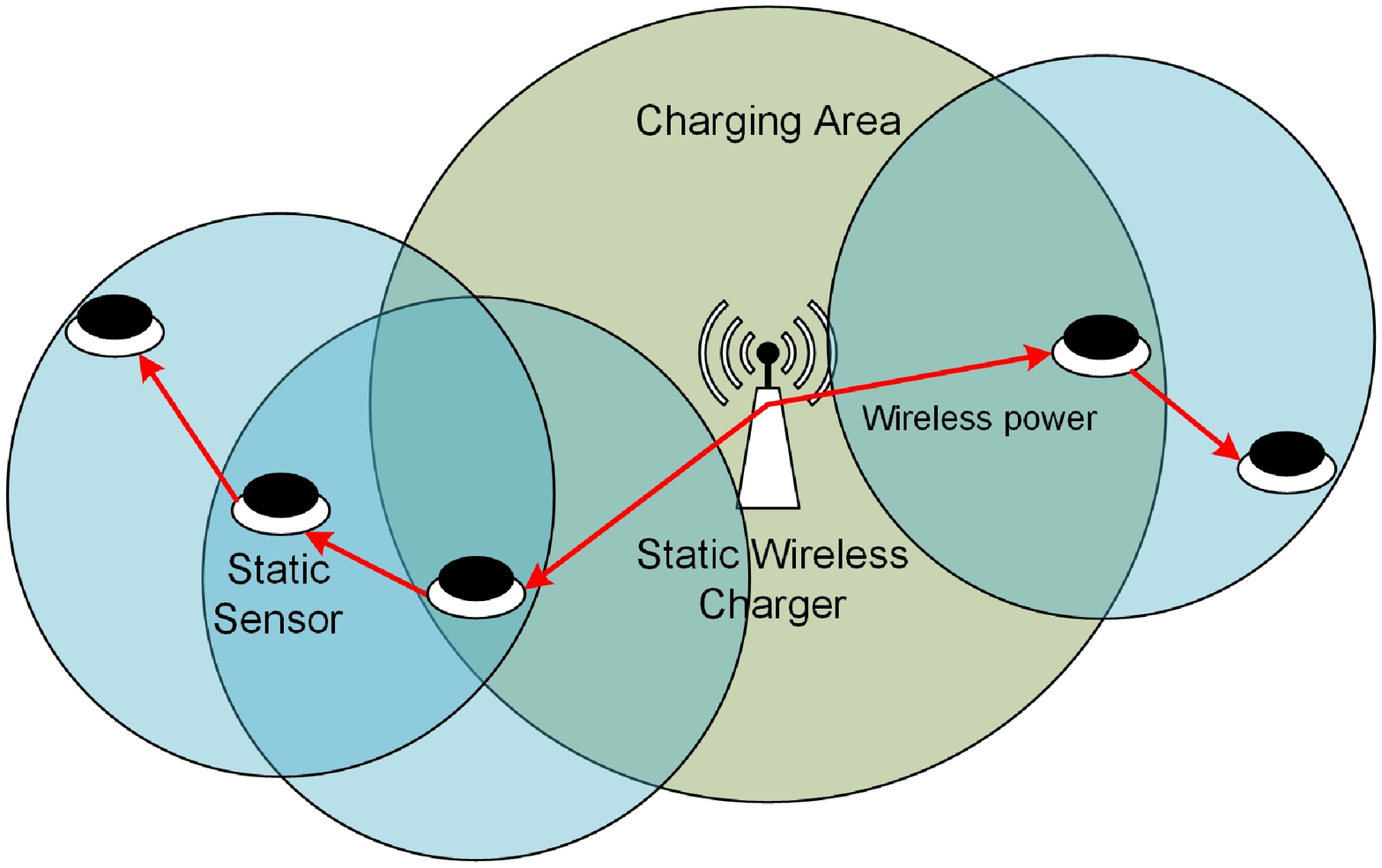}}   
  \centering  
 \subfigure [Landmark Provisioning Scenario] {
  \label{landmark_provisioning}
  \centering
\includegraphics[width=0.42  \textwidth]{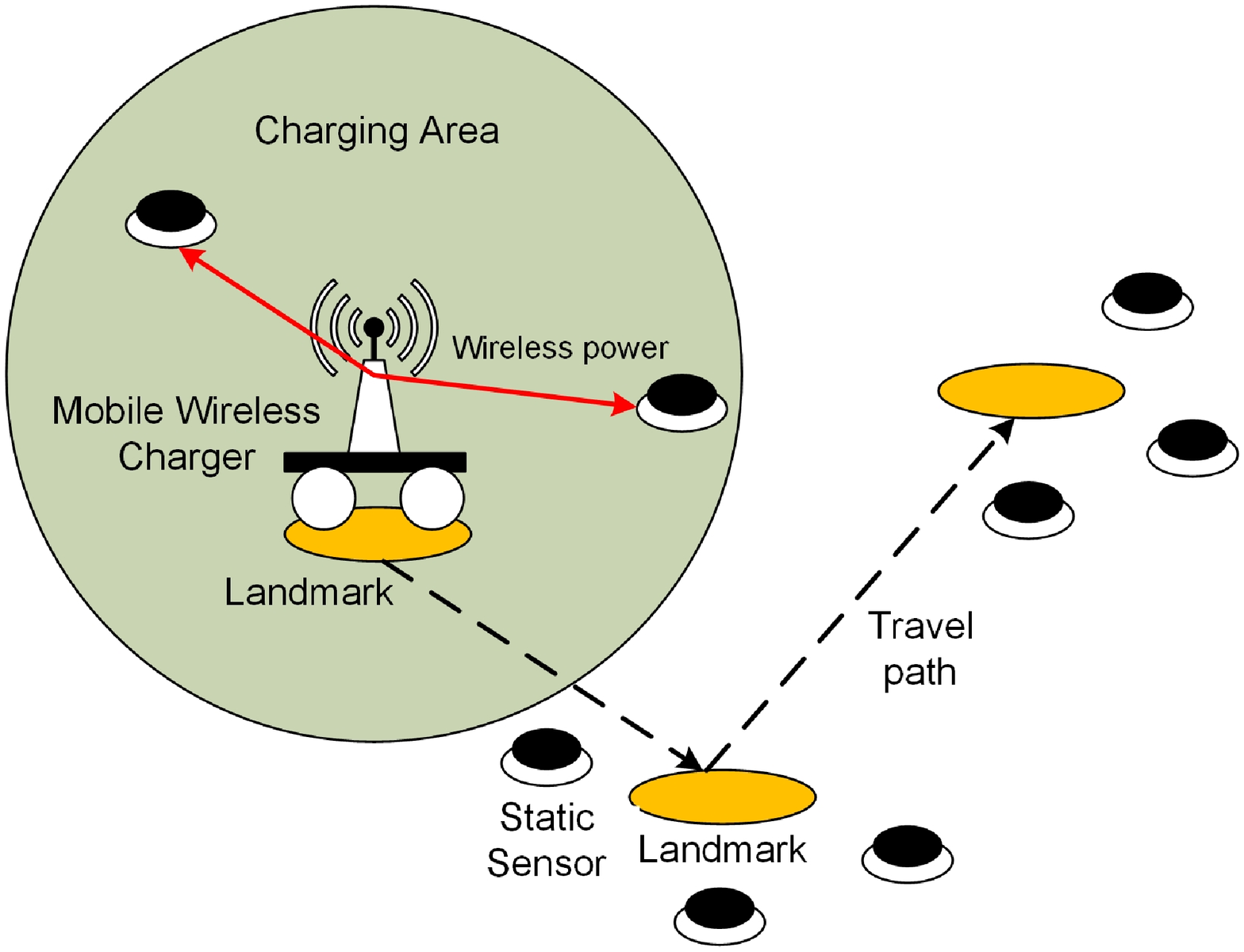}}\\
\centering
\caption{Reference models of wireless charger deployment scenario.} 
\label{deployment_scenario}
\end{figure*}

Wireless charger deployment involves planning of charger placement to support the sustainable operation of a wireless network. The deployment problems can be divided into two types: placement of static chargers and mobile chargers. As aforementioned, since the effective coverage range is only few meters for coupling-based wireless chargers, and tens of meters for RF-based chargers, the placement of static chargers is suitable and practical only in small areas. In a large network, a full-coverage static charger deployment is costly and incurs high overhead~\cite{C2012Chiu}. As shown in Figure~\ref{deployment_scenario}, the existing literature addresses wireless charger deployment strategies in four different scenarios.
\begin{itemize}
\item \emph{Point Provisioning}~\cite{S2013He} deals with the placement of static chargers to support static devices with wireless power.   

\item \emph{Path provisioning}~\cite{S2013He} aims to deploy static chargers to charge mobile devices (e.g., wearable or implanted sensors by human) in their travel paths.

\item \emph{Multihop provisioning} determines the locations to place static chargers in a static network, where the devices are also enabled with wireless power transfer function and can share energy with each other.

\item \emph{Landmark provisioning} involves two steps: selection of landmarks for the mobile chargers to visit by turns, and clustering landmarks as groups to deploy mobile chargers. The landmarks are the locations to park the charger to provide concurrent charging for multiple static devices in the vicinity.
\end{itemize}  

The first three scenarios are concerned with static charger deployment, while the last one requires mobile charger deployment. In the following two subsections, we review the strategies under these scenarios.

\subsection{Static Wireless Charger Deployment}


The majority of the existing works \cite{Erol-Kantarci2014,J.2014Liao,Y.2014Pang, T.2015He,H.2014Dai,HDai2014} focused on the deployment problem for point provisioning scenario.
The study in~\cite{J.2014Liao} investigated a WRSN where the wireless chargers are placed based on grid points at a fixed height. Each wireless charger is equipped with 3D RF-based beamforming and provides a cone-shape charging space, called charging cone. To minimize the number of chargers, the authors devised a node based greedy cone selecting (NB-GCS) algorithm and a node pair based greedy cone selecting (PB-GCS) algorithm. The former and latter generated charging cones on a node-by-node and pair-by-pair basis, respectively. It was shown by simulations that PB-GCS performs better than NB-GCS in terms of the number of chargers. Their performance gap increases with the number of sensor nodes. However, NB-GCS has significantly lower complexity, especially when the number of nodes is large. Compared with the system model in~\cite{J.2014Liao} where only a wireless charger serves as the energy source, reference~\cite{Erol-Kantarci2014} further examined this charger deployment problem where randomly deployed base stations coexist.  By adopting ILP, the authors explored the situations with BSs performing SWIPT in the first case and only transmitting information in the second case. The simulation claims that the first case results in fewer chargers and clearly outperforms the second case in terms of the transferred power.

The problem investigated in~\cite{T.2015He} was to deploy a finite number of wireless chargers next to the same number of bottleneck sensors in order to maximize the flow rate of the network. The authors first formulated an MILP to determine the routing and the set of bottleneck sensors to be charged. Then, a heuristic charger deployment scheme was also proposed and was demonstrated to obtain on average 85.9$\%$ of the optimal solution generated by the MLIP.

Reference~\cite{Y.2014Pang} dealt with the problem to provide charging coverage for a set of sensors with minimum number of wireless chargers. To reduce the complexity of the optimization problem, the authors devised an approximation solution based on a network partition algorithm to choose the deployment locations for wireless chargers. Moreover, the order of approximation has been theoretically characterized, under the condition that all the target sensors are evenly distributed. The authors also introduced a shifting strategy to prove the performance lower bound of the proposed partition algorithm. However, there was no simulation evaluation to examine its performance.

The focus of \cite{H.2014Dai} and \cite{HDai2014} was to study a safe wireless charging strategy under electromagnetic radiation regulation.
In~\cite{H.2014Dai}, the authors investigated an equivalent problem to the point provisioning problem. That is, given a set of deployed chargers, how to select the ones to be turned on so that nowhere on the planar field exposes electromagnetic radiation exceeding a limit. As the radiation limit applies everywhere, it incurs an infinite number of constraints. The authors demonstrated that searching for the optimal activation set of chargers to maximize the overall charging throughput, under the imposed constraints, is NP-hard in general.
By applying constraint conversion and constraint reduction techniques, the authors showed that the original problem can be transformed into two traditional problems, namely multidimensional 0/1 knapsack problem~\cite{A.2004Freville} and Fermat-Weber problem~\cite{P.1975Hurter}. Then an approximation algorithm with provable near optimality was proposed as a solution, which was shown to outperform a PSO-based heuristic algorithm by around $35\%$. However, the proposed solution is essentially centralized, which results in high complexity with the increase in number of chargers.

The study in~\cite{HDai2014} extended~\cite{H.2014Dai} by considering the transmit power of the chargers to be adjustable instead of on/off operation. The objective was to maximize the charging utility which is in proportional to the total charging power. Similar to~\cite{H.2014Dai}, this problem imposed infinite constraints. The authors first reformulated the optimization problem into a conventional LP problem by utilizing an area demonetization technique. To reduce the complexity of the LP problem, a distributed redundant constraint reduction approach was proposed to reduce the number of constraints. The authors further devised a distributed approximation algorithm to solve the optimization problem.  Experiment with a Powercaster testbed illustrated that around 40$\%$ average performance gain can be achieved by the proposed distributed algorithm over the centralized solution in~\cite{H.2014Dai}.

Reference~\cite{S2013He} considered both point and path provisioning problems for a wireless identification and sensing platform (WISP). In this platform, RFID tags are recharged wirelessly by RFID readers. Both problems possess the same objective to minimize the number of chargers. Under the assumption that the recharging power from multiple RFID readers is additive, the authors derived the lower bound on the number of readers required for both scenarios. It is shown by simulation that, compared with traditional triangular deployment approach in sensing disc model~\cite{X.2006Bai}, the proposed approach for point provisioning resulted in dramatic reduction in the number of chargers. Furthermore, the proposed approach for point provisioning was shown to achieve near-optimal performance, while the proposed approach for path provisioning achieved practically close to optimal performance. The study in~\cite{S2013He} focused on a full-coverage scheme, which is suitable only for a small network. In a large network, it may incur too much cost and overhead.


To develop a cost-effective charger deployment in large networks, reference~\cite{C2012Chiu} exploited the idea of partial coverage in path provisioning scenario, based on the observation that human movement has some degree of regularity~\cite{C.2010Song}. The objective is to design a mobility-aware deployment scheme maintaining a desirable survival rate with limited number of chargers. The authors formulated the mobility-aware charger deployment (MACD) problem for the maximum survival rate in a grid-based map, where the grid points are the potential locations to place chargers. It was proven that the MACD problem is NP-hard. Then, the authors designed a low-complex MACD algorithm based on a greedy approach. The simulation illustrated that, compared with the full-coverage scheme in~\cite{S2013He}, the proposed MACD algorithm manages to achieve the same survival rate with significant less number of chargers by making effective use of the end-devices' mobility regularity. 

The above work only considered one-hop wireless charging systems, where all wireless power is directly transferred from chargers. Reference ~\cite{T2013Rault} offered a solution to multi-hop provisioning, where each node could also transmit energy to its neighbors. In this context, the authors formulated a problem to minimize the number of chargers of fixed capacity as an MILP. It has the constraint on the maximum number of hops for energy transfer. Compared with a single-hop charging approach, by simulation, the proposed solution was shown to render much less number of required chargers, especially when the charger capacity is large. However, for multi-hop charging, there exists a tradeoff between charging efficiency and number of hops. The analysis of this tradeoff was missing in this work.

\subsection{Mobile Wireless Charger Deployment}  

The authors in~\cite{Erol-Kantarci2012Suresense} proposed a three-step scheme, called SuReSense, to address the deployment problem for multiple mobile wireless chargers in a WRSN. First, an integer linear programming (ILP) problem was formulated to minimize the number of landmarks based on the location and power demand of the sensors. Next, the landmarks are organized into clusters based on their proximity to docking stations which replenish mobile chargers. Finally, each mobile charger visits the landmark following the shortest Hamiltonian cycle. Compared with the scheme that the wireless charger visits each sensor individually according to the shortest Hamiltonian cycle, the simulation results showed that SuReSense is able to achieve shorter path length, especially when the power demand is low. 

The following works~\cite{M2012Erol-Kantarci} and~\cite{Erol-Kantarci2012DRIFT} based on~\cite{Erol-Kantarci2012Suresense} focused on the landmark selection for different objectives.  Reference~\cite{M2012Erol-Kantarci} considered the profit maximization problem~\cite{T.La2011Porta} in a WRSN with mission assignment~\cite{L.2014Porta}. The authors developed an ILP model, called mission-aware placement of wireless power transmitters (MAPIT), to optimize the number of devices charged from each landmark. It was demonstrated by simulation that the profit can be improved by confining the number of the landmarks. Moreover, the profit decreases with the increase of number of missions, because to complete more missions, the nodes require to be charged from more landmark locations.  

Both~\cite{Erol-Kantarci2012Suresense} and~\cite{M2012Erol-Kantarci} only considered the case that all the sensors are identical in priority. However, this may not be the general case in some environments. For example, the sensors in critical areas need to perform more precise monitoring and thus require more robust power previsioning. To address this concern, the study in~\cite{Erol-Kantarci2012DRIFT} proposed the strategy, called differentiated RF power transmission (DRIFT), to extend~\cite{Erol-Kantarci2012Suresense} by considering different priorities of the sensor nodes. The ILP model was developed with the objective to maximize the power delivered to the high priority nodes from each landmark. The simulation demonstrated that DRIFT allows the high priority node to receive significant higher power. However, SuReSense generates lower path length for the mobile charger. Furthermore, the authors demonstrated that there exists a tradeoff between power reception efficiency and the path length.
 
\begin{table*}  \small 
\centering
\caption{\footnotesize Summary of the Wireless Charger Deployment Strategies.} \label{Deployment_Strategies}
\begin{tabular}{|p{1.4cm}|p{1.5cm}|p{4cm}|p{4cm}|p{2.1cm}|p{2.5cm}|}
\hline
\footnotesize {\bf Literature} & {\bf Scenario } & {\bf Objective} & {\bf Constraint } & {\bf Solution} & {\bf Performance Evaluation } \\ \hline
\hline
  Erol-Kantarci \emph{et al} \cite{Erol-Kantarci2014} & Point provisioning & 1) Joint maximization of power transferred while keeping the number of BSs and energy transmitter at minimum; 2) Joint maximization of power transferred while keeping the number of energy transmitter at minimum & Location limit and number limit of energy transmitters and some deployment requirements & Centralized solution based on ILP  & Theoretical analysis, numerical simulation \\
\hline
 Chiu \emph{et al} \cite{C2012Chiu}  & Point provisioning &  Minimization of the number of chargers & Network charging coverage requirement & Two centralized greedy algorithms & Theoretical analysis, numerical simulation \\
 \hline
 He \emph{et al}\cite{T.2015He} & Point provisioning &  Maximization of system flow rate & The number of wireless chargers & Centralized solution based on MILP & Theoretical analysis,  numerical simulation  \\
\hline
  Peng \emph{et al}  \cite{Y.2014Pang}  & Point provisioning & Minimization of the number of chargers &   Network charging coverage requirement  & A centralized approximation solution  & Theoretical analysis \\
 \hline
 Dai \emph{et al}  \cite{H.2014Dai} & Point provisioning & Maximization of charging throughput &  Electromagnetic radiation limit & A centralized approximation solution  &  Theoretical analysis, numerical simulation, experiment \\
 \hline
  Dai \emph{et al} \cite{HDai2014} & Point provisioning & Maximization of charging throughput &  Electromagnetic radiation limit & A distributed approximation solution  & Theoretical analysis, numerical simulation, experiment \\
\hline
 He \emph{et al} \cite{S2013He}  & Point provisioning, path provisioning &  Minimization of the number of chargers  &  Average charging rate requirement & Centralized solution based on non-linear optimization  & Theoretical analysis, system-level simulation\\
\hline
 Liao \emph{et al} \cite{J.2014Liao}  &  Path provisioning &  Maximization of survival rate  & Limitation on number of chargers  & A centralized heuristic greedy algorithm  & Theoretical analysis, system-level simulation\\
\hline
  Rault \emph{et al}  \cite{T2013Rault}  & Multi-hop provisioning &  Minimization of the number of chargers & Network coverage requirement,   maximum  limit of hop number for energy transfer & Centralized solution based on mixed ILP & Numerical simulation \\
\hline
 Erol-Kantarci \emph{et al} \cite{Erol-Kantarci2012Suresense}  &  Landmark  provisioning & Minimization of the number of landmarks & Total energy replenishment demand, capacity limit of the chargers & Centralized solution based on ILP &  Numerical simulation \\
\hline
 Erol-Kantarci \emph{et al} \cite{M2012Erol-Kantarci} & Landmark  provisioning & Maximization of mission profit & Energy replenishment demand, capacity limit of the charger  & Centralized solution based on ILP &  Numerical simulation \\
\hline
 Erol-Kantarci \emph{et al} \cite{Erol-Kantarci2012DRIFT} & Landmark  provisioning & Maximization of the power delivered to the high priority nodes & Maximum number of landmarks, transmission range limit and power requirement of high priority nodes, capacity limit of the charger  & Centralized solution based on ILP &  Numerical simulation \\
\hline
\end{tabular}
\end{table*} 
 
\subsection{Discussion and Summary}



Table~\ref{Deployment_Strategies} summarizes the existing wireless charger deployment strategies. Clearly, multi-hop provisioning has been less investigated, only in~\cite{T2013Rault}. Additionally, it is important to study a system when each device can harvest energy from multiple transmitters. As for the deployment scenarios, none of the existing works considers the deployment of mobile chargers in mobile networks. Mobile charger deployment strategies based on the mobility pattern of user devices can be studied. 

Moreover, we observe that the deployment problems are formulated mostly as optimization problems with different objectives and constraints. All the solutions consequently need global information such as devices' battery capacity, location, and even hardware specification parameters and velocity (e.g., in~\cite{C2012Chiu}). Collecting these information incurs tremendous communication overhead. Though some of the proposed solutions (e.g., in~\cite{C2012Chiu} and~\cite{Erol-Kantarci2012Suresense}) claimed to be of low-complexity and scalable for large networks, its feasibility and practicability in deploying them in real systems have to be evaluated. Alternatively, decentralized approaches based on local information that relax the communication requirement can be one of the important future directions. Moreover, most of the proposals were evaluated by numerical simulation. Only references~\cite{S2013He} and~\cite{J.2014Liao} have provided system-level simulation. There is the need for future research to conduct more assessment through system-level simulations and real experiments to understand the empirical performance.

\section{Open Research Issues and Future Directions}
\label{sec:futureissues}

In this section, we first summarize some open issues with regard to both wireless charging technologies and data communication in wireless charging systems. Then, we envision several novel paradigms
emerging with the advance of wireless charging technologies.

\subsection{Open Research Issues}

This subsection first discusses some technical issues in wireless charging, then highlights some communication challenges. 

\subsubsection{Open Issues in Wireless Charging}

{\em Inductive coupling:} The increase of wireless charging power density gives rise to several technical issues, e.g., thermal, electromagnetic compatibility, and electromagnetic field problems~\cite{Y2013Hui}. This requires high-efficiency power conversion techniques to reduce the power loss at an energy receiver and battery modules with effective ventilation design.

{\em Magnetic resonance coupling:} Magnetic resonance coupling-based techniques, such as Witritiy and MagMIMO, have a larger charging area and are capable of charging multiple devices simultaneously. However, they also cause increased electromagnetic interference with a lower efficiency compared with inductive charging. Another limitation with magnetic resonance coupling is the relatively large size of a transmitter. The wireless charging distance is generally proportional to the diameter of the transmitter. Therefore, wireless charging over long distance typically requires a large receiver size.

{\em Near-field beamforming:} For multi-antenna near-field beamforming, the computation of a magnetic-beamforming vector on the transmission side largely depends on the knowledge of the magnetic channels to the receivers. The design of channel estimation and feedback mechanisms is of paramount importance. With the inaccuracy of channel estimation or absence of feedback, the charging performance severely deteriorates. Additionally, there exists a hardware limitation that the impedance matching is optimally operated only within a certain range~\cite{J.2014Jadidian}.

{\em Localization for RF-based energy beamforming:}
As aforementioned, energy beamforming is able to enhance the power transfer efficiency. However, the energy transmitter needs to know the location of the energy receiver to steer the energy beam to. Localization needs to make real-time spatial estimations for two major parameters, i.e., angle and distance. Self-detection and localization of to-be-charger devices is challenging especially for mobile WPCNs. Additionally, similar to near-field beamforming, channel estimation is also very critical in the design of beamforming vectors.

{\em Heating Effect:} A metallic or ferromagnetic material can absorb some of the near-field energy if it locates in a proximity of any wireless charger. The induced voltage/current on the material can 
cause temperature rise. As metallic material is an essential part of electronic devices, the resultant
temperature rise lowers charging efficiency and can render bad user experience. Although both Qi and A4WP have the mechanisms to avoid safety issues such as severe over-temperature, system power
loss is still inevitable and can be considerable especially if the device is large in size. Moreover, foreign objects may be another factor to cause power loss. How to mitigate the heating effect to diminish power loss is challenging.

{\em Energy conversion efficiency:} Wireless charging requires electricity energy to be transformed from  AC to electronmagnetic waves and then to DC. Each conversion adds the loss in overall energy, which leads to a normally wireless charging efficiency hovering between $50\%$ and $70 \%$. Efforts toward hardware improvement of energy conversion efficiency are instrumental to achieve highly efficient wireless charging.

\subsubsection{Open Issues in Data Communication}

To improve the usability and efficiency of the wireless charger, their data communication capability can be enhanced. 

{\em Duplex communication and multiple access:} The current communication protocols support simplex communication (e.g., from a charging device to charger). However, there are some important procedures which require duplex communication. For example, the charging device can request for a certain charging power, while the charger may request for battery status of the charging device. Moreover, the current protocols support one-to-one communication. However, multiple device charging can be implemented 
with multiple access for data transmission among charging devices and a charger has to be developed and implemented.

{\em Secure communication:} The current protocols support plain communication between a charger and a charging device. They are susceptible to jamming attacks \cite{D.NiyatoJammingICC,D.NiyatoJammingWCNC}  (e.g., to block the communication between the charger and the charging device), eavesdropping attacks (e.g., to steal charging device's and charger's identity) and man-in-the-middle attacks (e.g., malicious device manipulates or falsifies charging status). The security features have to be developed in the communication protocols, taking unique wireless charging characteristics (e.g., in-band communication in Qi) into account. 

{\em Inter-charger communication:} The protocols support only the communication between a charger and charging device (i.e., intra-charger). Multiple chargers can be connected and their information as well as charging devices' information can be exchanged (i.e., inter-charger). Although  the concept of wireless charger networking has been proposed in~\cite{X.1410.8635Lu}, there are some possible improvements. For example, wireless chargers can be integrated with a wireless access point, which is called a hybrid access point, to provide both data communication and energy transfer services.

\subsection{Future Directions}


In this subsection, we discuss several emerging paradigms which are anticipated in  wireless powered communication networks.

\subsubsection{Wireless Charger Network}
Similar to wireless communication networks that provide data service, a wireless charger network can be built to deliver energy provisioning service to distributed users. The wireless charger network that connects a collection of distributed chargers through wired or wireless links allows to exchange information (e.g., include availability, location, charging status, and cost of different chargers) to schedule the chargers. Such scheduling can either be made in a distributed or centralized manner to optimize certain objectives (e.g., system energy efficiency, total charging cost). A wireless charger network can be a hybrid system based on several charging techniques to satisfy heterogeneous charging and coverage requirement. For instance, the system may utilize short-range near-field chargers (e.g., inductive-based) to charge static devices that have high power demand, mid-range near-field chargers (e.g., resonance-based) to charge devices having no line-of-sight charging link and relax the coil alignment requirement. Furthermore, a far-field charger (e.g. Powercaster and cota system) can be employed to cover remote devices with low-power requirement and some local movement requirement, (e.g., wearable devices, MP3, watches, Google glasses, and sensors in smart building).  

\subsubsection{Green Wireless Energy Provisioning}

With the increasing deployment of wireless powered devices, how to provision wireless energy for large-scale networks in an eco-friendly way becomes an emerging issue. As reviewed above, both static and mobile charger scheduling strategies have been developed for power replenishment. However, these strategies could incur more pollution and energy consumption, if the power sources and charging techniques for wireless chargers are not appropriately adopted. For example, the vehicle equipped with wireless chargers for mobile energy provisioning will produce considerable amount of CO$_{2}$ emission. Moreover, due to the propagation loss and thus low transfer efficiency, a static RF-based charger powered by the electric grid could cause more consumption of conventional fuels, like coal, to harm the environment.  
Currently, how to perform green wireless energy provisioning remains an open issue and has been ignored by the majority of existing studies.
One promising solution is to equip renewable energy sources, e.g., solar, for wireless chargers. However, renewable energy could be unpredictable, thus hard for the chargers to deliver reliable wireless charging services. Significant relevant issues can be explored in this direction.

\subsubsection{Full-Duplex Self-energy Recycling Information Transmitter}

Full-duplex based wireless information transmitter~\cite{Y.2015Zeng} can be equipped with multiple antennas to transmit information and receive energy simultaneously in the same frequency band. Conventionally, a full-duplex system suffers from the self-interference as part of the transmitted RF signals is received by the transmitter itself. Self-interference is undesirable because it degrades the desired information signal. However, with the capability of harvesting RF energy, self-interference can facilitate energy saving. In particular, part of the energy used for information transmission can be captured by the receive antenna(s) for future reuse, referred to as self-energy recycling. This paradigm benefits both energy efficiency and spectrum efficiency. Moreover, it can be widely applied to a multi-antenna base station, access point, relay node, and user devices.   

\subsubsection{Millimeter-wave Enhanced Wireless Powered Cellular Network}

Millimeter-wave cellular communications~\cite{S.2014Rangan} that operates on frequency ranging from 30 to 300GHz have become a new frontier for next-generation wireless systems. Due to high frequencies, millimeter-wave cellular communication is a natural system to facilitate wireless energy beamforming. For a multi-antenna transmitter, the beamforming efficiency increases by increasing the frequency. Moreover, frequency is a key factor that affects the physical size of a rectenna based microwave power conversion system~\cite{G.2008Chattopadhyay}. At high frequency ranges, the required size of the antennas is small, which consequently renders a small form factor for the system. Moreover, a small form factor helps to advance beamforming by enabling a larger number of antennas to be placed in an array, which further helps to mitigate charging power attenuation. Thus,  a millimeter-wave RF transmitter is desired to be utilized for RF-based wireless charging and SWIPT.

\subsubsection{Near-field SWIPT System}
As afore-introduced, SWIPT (see~\cite{X.LuSurvey} and references therein) has been broadly investigated in RF-based wireless communication systems. With the emerging of coupling-based chargers, magnetic induction communication \cite{Z.July2010Sun} can also be incorporated in near-field charging system to induce SWIPT.  
Near-field communication based on magnetic field can achieve significant capacity gain compared with RF-based communication.
A hardware design and implementation were reported in \cite{N.2009Miura} that an inductive coupling based chip can deliver 11Gbps for a distance of 15$\mu$m in 180nm complementary metal-oxide semiconductor (CMOS).
Therefore, SWIPT-compliant near-field chargers have great potentials in high-speed data offloading in next generation communications. Being backhauled with high-speed Internet connections,  SWIPT-compliant near-field chargers can be integrated into cellular systems for seamless data service during charging.

\section{Conclusion}
 
Wireless power technology offers the possibility of removing the last remaining cord connections required to replenish portable electronic devices. This promising technology has significantly advanced during the past decades and introduces a large amount of user-friendly applications. In this article, we have presented a comprehensive survey on the paradigm of wireless charging compliant communication networks. Starting from the development history, we have further introduced the fundamental, international standards and network applications of wireless charging in a sequence, followed by the discussion of open issues and envision of future applications.  

The integration of wireless charging with existing communication networks creates new opportunities as well as challenges for resource allocation. This survey has shown the existing solutions of providing seamless wireless power transfer through static charger scheduling, mobile charger dispatch and wireless charger deployment. Among those studies, various emerging issues including online mobile charger dispatch strategies, near-field energy beamforming schemes, energy provisioning for mobile networks, distributed wireless charger deployment strategies, and multiple access control for wireless power communication networks are less explored and require further investigation. Additionally, the open issues and practical challenges discussed in Section VIII can be considered as main directions for future research.

\section*{Acknowledgements}

This work was supported in part by the National Research Foundation of Korea (NRF) grant funded by the Korean government (MSIP) (2014R1A5A1011478), Singapore MOE Tier 1 (RG18/13 and RG33/12) and MOE Tier 2 (MOE2014-T2-2-015 ARC 4/15), and the U.S. National Science Foundation under Grants US NSF CCF-1456921, CNS-1443917, ECCS-1405121, and NSFC 61428101.

\end{document}